\documentclass{raa}
\usepackage{graphicx,times}             
\usepackage[round]{natbib}
\usepackage{amssymb,amsmath}
\usepackage{multirow}
\usepackage{longtable}
\usepackage{subcaption}
\bibpunct{(}{)}{;}{a}{}{,}

\usepackage[pagebackref=true]{hyperref}

\begin{document}

  \title{CHES: a space-borne astrometric mission for the detection of habitable planets of the nearby solar-type stars
}

   \volnopage{Vol.0 (20xx) No.0, 000--000}      
   \setcounter{page}{1}          

   \author{Jiang-Hui Ji  
      \inst{1,2}
   \and Hai-Tao Li
      \inst{3,4}
   \and Jun-Bo Zhang
      \inst{5,6}
   \and Liang Fang
     \inst{4,5,7}
   \and Dong Li
     \inst{8}
   \and Su Wang
     \inst{1,2}
   \and Yang Cao
     \inst{3}
   \and Lei Deng
     \inst{8}
   \and Bao-Quan Li
     \inst{3,4}
   \and Hao Xian
     \inst{4,5,6}
   \and Xiao-Dong Gao
     \inst{5}
   \and Ang Zhang
     \inst{4,5,6}
   \and Fei Li
     \inst{8}
   \and Jia-Cheng Liu
     \inst{9}
   \and Zhao-Xiang Qi
     \inst{10}
   \and Sheng Jin
     \inst{1,2}
   \and Ya-Ning Liu
     \inst{3}
   \and Guo Chen
     \inst{1,2}
   \and Ming-Tao Li
     \inst{3,4}
   \and Yao Dong
     \inst{1,2}
   \and Zi Zhu
     \inst{9}
   \and CHES consortium
   }

   \institute{CAS Key Laboratory of Planetary Sciences, Purple Mountain Observatory, Chinese Academy of Sciences, Nanjing 210023, China; {\it jijh@pmo.ac.cn}\\
        \and
             School of Astronomy and Space Science, University of Science and Technology of China, Hefei 230026, China\\
        \and
             Key Laboratory of Electronics and Information Technology for Space Systems, National Space Science Center, Chinese Academy of Sciences, Beijing 100190, China\\
        \and
            University of Chinese Academy of Sciences, Beijing 100049, China\\
        \and
            Institute of Optics and Electronics, Chinese Academy of Sciences, Chengdu 610209, China\\
        \and
            Key Laboratory on Adaptive Optics, Chinese Academy of Sciences, Chengdu 610209, China\\
        \and
            The Key Laboratory of Science and Technology on Space Optoelectronic Precision Measurement, Chinese Academy of Sciences, Chengdu 610209, China\\
        \and
            Innovation Academy for Microsatellites of Chinese Academy of Sciences, Shanghai 201306, China\\
        \and
            School of Astronomy and Space Science, Nanjing University, Nanjing 210046, China\\
        \and
            Shanghai  Astronomical  Observatory,  Chinese Academy of Sciences, Shanghai 200030,  China\\
\vs\no
   {\small Received 2022 month day; accepted 2022 month day}}

\abstract{The Closeby Habitable Exoplanet Survey (CHES) mission is proposed to discover habitable-zone Earth-like planets of the nearby solar-type stars ($\sim 10~\mathrm{pc}$ away from our solar system) via micro-arcsecond relative astrometry. The major scientific objectives of CHES are: to search for Earth Twins or terrestrial planets in habitable zones orbiting 100 FGK nearby stars; further to conduct a comprehensive survey and extensively characterize the nearby planetary systems.
The primary payload is a high-quality, low-distortion, high-stability telescope. The optical subsystem is a coaxial three-mirror anastigmat (TMA) with a $1.2 \mathrm{~m}$-aperture, $0.44^{\circ} \times 0.44^{\circ}$ field of view and $500 \mathrm{~nm}-900 \mathrm{~nm}$ working waveband. The camera focal plane is composed of 81 MOSAIC scientific CMOS detectors each with $4 \mathrm{~K} \times 4 \mathrm{~K}$ pixels. The heterodyne laser interferometric calibration technology is employed to ensure micro-arcsecond level (1 $\mu$as) relative astrometry precision to meet the requirements for detection of Earth-like planets. CHES satellite operates at the Sun-Earth L2 point and observes the entire target stars for 5 years.
CHES will offer the first direct measurements of true masses and inclinations of Earth Twins and super-Earths orbiting our neighbor stars based on micro-arcsecond astrometry from space. This will definitely enhance our understanding of the formation of diverse nearby planetary systems and the emergence of other worlds for solar-type stars, and finally to reflect the evolution of our own solar system.
}

   \authorrunning{J.-H. Ji, et al. }            
   \titlerunning{CHES: a space-borne astrometric mission for habitable planets}  

   \maketitle

\section{EXECUTIVE SUMMARY}           
\label{sect:intro}
\subsection{CHES's major scientific goals}
To discover and explore habitable planets in our Galaxy, will provide essential answers to \emph{"Are we alone in the universe?", "Is the Earth unique?", "How planets evolve the cradle of life?"} or \emph{"Is our solar system special?"}. The in-depth understanding of the formation and evolution of planetary systems do rely on the detection of the diverse exoplanets (especially habitable planets), which is of great significance to enriching human beings' exploration of the unknown worlds, shedding light on the origin and evolution of life, and recognizing our status in the universe.

The CHES mission is proposed to discover Earth-like planets around the nearby solar-type stars via ultra-high-precision relative astrometry. The key scientific goals are:

(1) to search for the terrestrial planets in habitable zones orbiting 100 FGK stars $\sim$ 10 pc (the nearby solar-type stars);

(2) further to conduct a comprehensive survey and extensive characterization on the nearby planetary systems.

CHES will offer the first direct measurements of true masses and inclinations of \emph{Earth Twins} (with orbit, mass and environment similar to Earth) and super-Earths orbiting our neighbor stars based on micro-arcsecond astrometry from space. This will definitely enhance our understanding of the formation of diverse nearby planetary systems and the emergence of other worlds for the nearby solar-type stars, and finally to reflect the origin and evolution of our own solar system.

\subsection{Scientific instruments}
The payload is a high-quality, low-distortion, high-stability telescope with the optical subsystem, camera subsystem and on-board calibration subsystem. The optical subsystem is a coaxial three-mirror anastigmat (TMA) with a $1.2 \mathrm{~m}$-aperture, $0.44^{\circ} \times 0.44^{\circ}$ field of view and $500 \mathrm{~nm}-900 \mathrm{~nm}$ working waveband. The camera focal plane is composed of 81 MOSAIC scientific CMOS detectors each with $4 \mathrm{~K} \times 4 \mathrm{~K}$ pixels. The on-board calibration subsystem consists of a metrology assembly. A heterodyne laser interferometric calibration technology is employed to ensure micro-arcsecond level (1 $\mu$as) relative astrometry precision that is required to detect the habitable \emph{Earth Twins} orbiting our neighboring stars.

The mission orbit of the CHES satellite travels about the L2 point of the Sun and the Earth. The satellite is designed to have a lifespan of 5 years, during which the entire target stars will be extensively stared at and observed.

\subsection{Scientific additional benefits}
CHES will produce fruitful achievements not only in the Earth-like planets but also for cosmology, dark matter and black holes with micro-arcsecond accuracy in relative astrometry, which helps us better understand the philosophy, life and planet.
\clearpage
\begin{center}
\begin{longtable}{|l|l|}
\hline Science case & Habitable Exoplanets orbiting the nearby solar-type stars \\
\hline Science & $>$ To discover habitable Earths about nearby solar-type stars \\
objectives &$>$ To conduct a comprehensive survey and census on the nearby planetary systems\\
& $\quad$ \emph{Extended: cosmology, dark matter and black holes} \\
\hline Overview & $>$ Spacecraft at L2 for 5 years \\
& $>$ Optical telescope $(500 \mathrm{~nm}-900 \mathrm{~nm})$ \\
& $>$ Micro-arcsecond astrometry $(1~\mu \mathrm{as})$ \\
& $>$ Point and stare strategy to enable relative astrometry \\
\hline What makes & $>$ Ultra-high-precision relative astrometry simply reachable from\\
CHES unique? & $\quad$ space: 0.3 $\mu$as (habitable-zone Earths about the sun-like stars at 10 pc)\\
&  $>$ To obtain true masses and orbital architecture (inclinations,\\
& $\quad$ etc.) of habitable-zone terrestrial planets\\
&  $>$ To conduct the census and characterization of nearby planetary systems\\
\hline Primary  & $>$ Nearby F, G, K stellar systems (100 stars $\sim$ 10 pc)\\
targets & $\quad$ \emph{Extended: ultra-faint dwarf galaxies, neutron stars in X-ray binaries, etc.}\\
\hline Scientific & $>$ Coaxial three-reflection TMA system\\
Payload  & $>$ Primary mirror: $D=1.2 \mathrm{~m}$ diameter\\
& $>$ Long focal length: $f=36 \mathrm{~m}$\\
& $>$ FOV: $0.44^{\circ} \times 0.44^{\circ}$, with 6 to 8 reference stars\\ & $>$ Focal plane with
   81 scientific CMOS detectors ($4 \mathrm{~K} \times 4 \mathrm{~K}, \geq 50 \mathrm{fps}$ )\\
& $>$ Nyquist sampling of the PSF\\
& $>$ Metrology calibration of the Focal Plane Array(FPA): relative\\
& $\quad$ positions of pixels at the micro-pixel level for each detector,\\
& $\quad$ geometrical parameters of FPA\\
\hline Spacecraft & $>$ Spacecraft dry mass with margin: 1,558 kg. Launch Mass: 2,930\\
  & $\quad$  kg, fuel mass (990 kg + 382 kg)\\
 & $>$ Attitude Control System: pointing accuracy of 0.07  $\mathrm{arcsec}$,\\
& $\quad$ pointing stability of 0.0036  $\mathrm{arcsec} / 0.02 \mathrm{sec}$\\
& $>$ Propulsion system: orbital maneuver engines: $490 \mathrm{~N}+12 \times 10 \mathrm{~N}$,\\
& $\quad$ attitude control thrusters: $12 \times(1 \mu \mathrm{N} \sim 50 \mu \mathrm{N})+12 \times 20 \mathrm{mN}$\\
& $>$ Thermal Control System: working temperature: $20 \pm 5^{\circ} \mathrm{C}$ and\\
& $\quad$ temperature stability of $45 \mathrm{mK}$ for payload;\\
& $\quad$ working temperature: $-15 \sim+45^{\circ} \mathrm{C}$ for other equipments\\
& $>$ Telecommand: X-band, communication rate: $20 \mathrm{Mbps}$\\
\hline Launcher and & $>$ CZ-3C: GTO $(200 \mathrm{~km} \times 35,958 \mathrm{~km})$.\\
operations & $>$ Orbital maneuver to Halo orbit at L2.\\
& $>$ Launch in 2025\\
& $>$ Nominal mission: 5 yrs.\\
& $>$ Launch site: Xichang.\\
\hline
\end{longtable}
\end{center}

\section{SCIENTIFIC GOALS}
\label{sect:Obs}
\begin{figure}
   \centering
   \includegraphics[width=10cm]{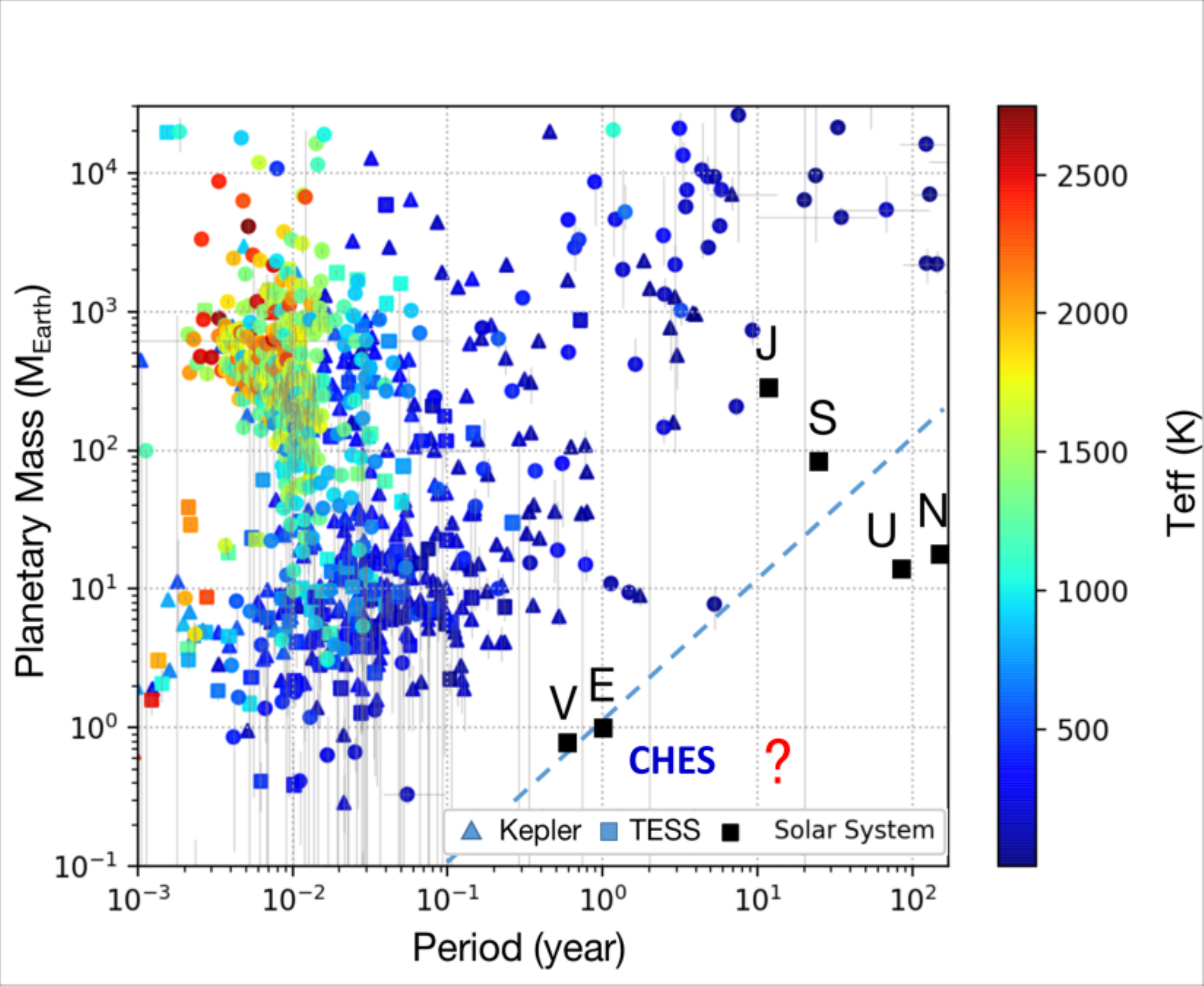}
   \caption{ Mass and orbital period distribution of exoplanets.}
   \label{fig:2-1}
   \end{figure}

As of today, more than 5,000 exoplanets have been detected (https://exoplanetarchive.ipac.caltech.edu) since the first planet around the main-sequence star 51 Peg was discovered \citep{MayorQueloz1995}. The statistical investigations on a large sample of exoplanets show that the planetary systems are actually complex and diverse far beyond our imagination.  Figure \ref{fig:2-1} shows the planetary mass versus orbital period of exoplanets \citep{Borucki2010, Borucki2011, Batalha2013, Huang2018}. Unlike their siblings of the solar system (marked up by the filled squares with capital letters), these planets can be classified into hot Jupiters, cold Jupiters, warm Neptunes, super-Earths, and terrestrial planets, where the color of the legends represents their surface temperature. However, the real Earth-mass planets in habitable-zone remain undetected. Therefore, the discovery of habitable-zone planets of solar-type stars and the characterization of rocky planets have become one of the most significant frontiers in the study of exoplanets.

Exploring the Earth-like planets in habitable zone (i.e., Earth Twins or ET 2.0) will definitely provide clues to the essential scientific questions such as \emph{"Are we alone in the universe"} or \emph{"How planets become the cradle of life?"}. The primary scientific goals of CHES will discover and explore terrestrial planets in the habitable zones orbiting nearby solar-type stars, and further conduct characterization of rocky planets, using ultra-high-precision relative astrometry.

\subsection{Science objective 1: search for terrestrial planets in the habitable zone}
To search for terrestrial planets in the habitable zone orbiting 100 $ \mathrm{FGK}$ stars in the solar system's nearest neighbors ( $\sim$ 10 pc), and discover terrestrial planets in the habitable zone of the nearby stars.

As is well-known, a number of space-borne missions of the detection of exoplanets have been carried out since 2000, e.g., the Gaia satellite (Global Astrometric Interferometer for Astrophysics) \citep{LindegrenPerryman1996, GaiaCollaboration2016} with astrometric method, the Kepler Telescope \citep{Koch1998, Borucki2010}, TESS (Transiting Exoplanet Survey Satellite) \citep{Ricker2015}, and CHEOPS (the CHaracterising ExOPlanet Satellite) \citep{Broeg2013}, etc. involved in transit method. As above-mentioned, the Kepler mission contributed more than half of the discovery of exoplanets \citep{Borucki2011, Batalha2013}, which gives an new insight into the orbital architecture, formation and dynamical evolution for the planetary systems.

\begin{figure}
   \centering
   \includegraphics[width=12cm]{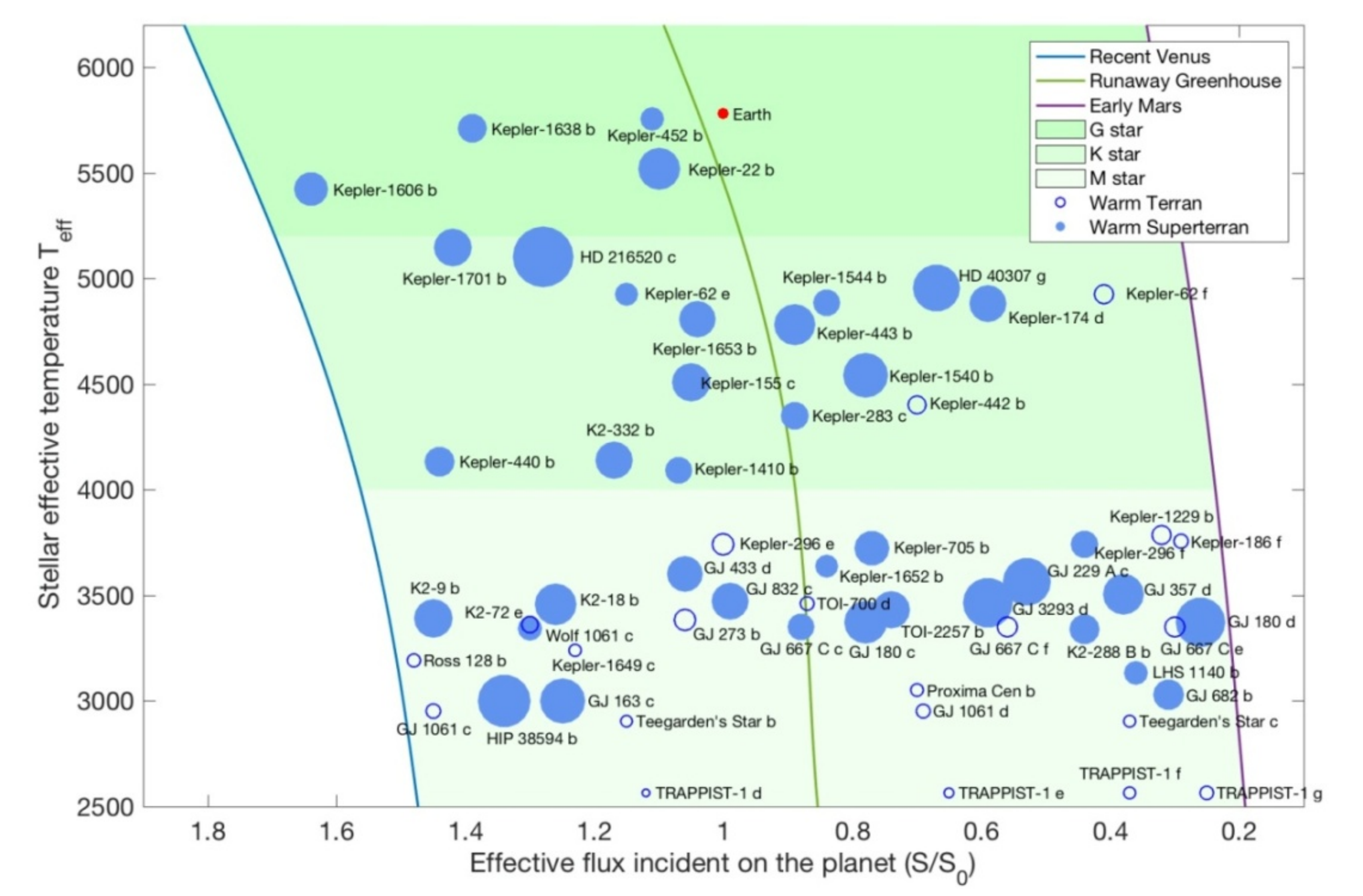}
   \caption{Distribution of known habitable-zone planets.}
   \label{fig:2-2}
   \end{figure}

Future missions aiming at the exoplanets include the Chinese Space Station Telescope (CSST) equipped with the coronagraph using direct imaging \citep{Gao2015}, the ROMAN Space Telescope (The Nancy Grace Roman Space Telescope) \citep{Green2012} with microlensing, and PLATO (the PLAnetary Transits and Oscillations of stars) \citep{Ragazzoni2016}, and ARIEL (the Atmospheric Remote-sensing Infrared mission Exoplanet Large-survey) \citep{Tinetti2016}, etc. Undoubtedly, these missions will considerably enable us to have an in-depth understanding of exoplanets, however as Figure \ref{fig:2-2} shown, the habitable-zone Earth-like planets orbiting the nearby solar-type stars are rarely found, where the Earth is indicated by red dot whereas the blue circles stand for warm terrestrial planets and super Earths, and the three separated curves, respectively, indicate the habitable border of the Recent Venus, Runaway Greenhouse and Early Mars  \citep{Kopparapu2013}.

CHES will specifically target the terrestrial planets in habitable zones using ultra-high-precision relative astrometric method in comparison with the transit detection such as Kepler, TESS, CHEOPS, ARIEL, PLATO and several astrometric space mission concepts such as SIM \citep{Catanzarite2005}, NEAT \citep{Malbet2012}, STEP \citep{Chen2014}, Theia \citep{TheTheiaCollaboration2017}, MASS \citep{Nemati2020}. CHES will be the first time to directly focus on the Earth-mass planets in the habitable zone about nearby solar-type stars. The scientific questions to be solved include whether there are planets in the habitable zone orbiting the nearby stars of our solar system, and how these planets distribute, and what is the probability of the occurrence of habitable-zone planets. Therefore, searching the habitable-zone planets in the solar system's neighbors by CHES can offer a significant clue to their distribution of the nearby solar-type stars, especially those terrestrial planets in the habitable zone.

\subsubsection{Terrestrial planets in the habitable zone}
As of March 2022, there are 38 terrestrial planets in the habitable zone discovered, most of which are around M-type stars \citep{Gillon2017,Luger2017,Gunther2019}, mainly detected through transit photometry and radial velocity. But the transit method can only measure the size of the planet rather than its true mass, and the radial velocity measurement can simply unveil the minimum mass of the planets (all of which are much larger than Earth), so it is not yet possible to confirm whether these planets are habitable-zone terrestrial planets.

The planetary candidates in the habitable zone discovered by Kepler are generally far away (1000-3000 light-years) \citep{Borucki2013}, which is difficult to verify and further identify through other observational means. Most of the habitable-zone planets observed are moving around M Dwarf stars with close-in orbits \citep{Luger2017,Tsiaras2019}, which may have strong ultraviolet radiation, and the atmospheric stability of the surrounding planets is uncertain. Therefore, whether these terrestrial planets in the habitable zone are habitable remains still controversial. Finally, some terrestrial planets in the habitable zone lack key information such as mass and size due to the limitation of observation, so it is still controversial whether they are terrestrial planets. Therefore, generally speaking, so far, the real \textit{Earth Twins} have not been found yet.

   \begin{figure}
   \centering
   \includegraphics[width=10cm]{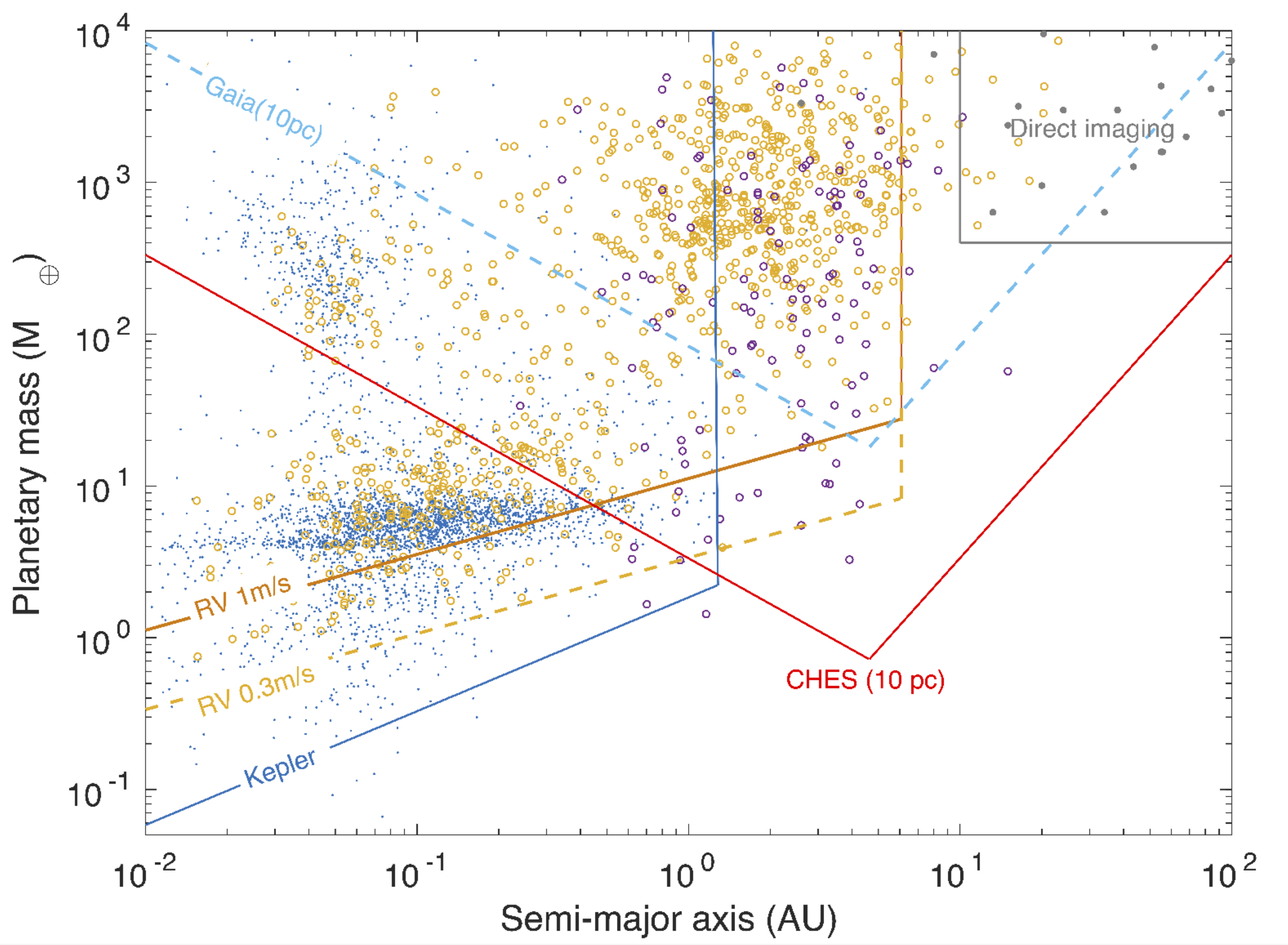}
   \caption{The detection capacity of terrestrial planets for CHES mission.}
   \label{fig:2-3}
   \end{figure}

\subsubsection{Detection of terrestrial planets about nearby stars}
The detection of habitable planets requires high precision (astrometry requires a micro-arcsecond level), and ground-based detection equipment is hard to break through the limit of $0.1$ milliarcsecond accuracy due to atmospheric disturbance and other factors even if it adopts technologies such as active optics. Therefore, space telescopes have turned out to be ideal for high-precision astronomical observations in addition to terrestrial spectroscopy (suitable for detecting planets with very short periods). However, it is rather difficult to detect habitable planets around sun-like stars using the radial velocity due to the influence of stellar activities, which cannot attain the true mass of the planet; the transit requires that the orbital plane of the planet aligns with the line of sight of the observers, which leads to a tiny probability of detection; the direct imaging is inclined to find planets with larger masses and distant orbits from the host stars based on the present best accuracy of observation; the microlensing can only obtain information about individual planets that are relatively far away and cannot be applicable to detecting the nearby planets. As compared to above-mentioned methods, the exoplanet detection of astrometry will have unbiased observations. Figure \ref{fig:2-3} shows the detection capacity of terrestrial planets for CHES mission in comparison with Kepler and high-precision RV (see also \citet{ZhuDong2021}), indicating that the astrometry of CHES will have an advantage in discovering the habitable-zone Earth-like planets revolving the nearby stars.

In the next decade or two, habitable planets in proximate to our solar system are the only places where traces of other forms of life can be found by observing the composition of the atmosphere, e.g., \citep{DesMarais2002, Kaltenegger2017}, as well it is first necessary to detect such terrestrial planets in the habitable zone to explore the atmospheric composition of habitable planets. If many of these "Proxima-type planets" \citep{Anglada-Escude2016} are observed in the future, we can preferentially select observation objects like Earth-mass planets, or those uniform mass planets among these candidates. Therefore, the finding of terrestrial planets in the habitable zone of the nearby stars will greatly contribute to the primary goal of exploration extraterrestrial life.

\subsubsection{Planetary Habitability}
The planetary atmospheres are critical for characterizing planetary features, which may provide potential clues to the history of planetary formation and evolution, and are the basis for assessing planetary habitability \citep{Kaltenegger2017, Madhusudhan2019, Zhang2020}. At present, the observation and characterization of exoplanet atmospheres mainly focus on gaseous planets, most of which are indirectly observed by transit or Doppler spectroscopy, and a small number of hot gas-giants in broad-distance orbits can be directly imaged. The observation limit of the indirect detection is expected to be extended to rocky planets, even including a small number of terrestrial planets located in the habitable zone of M dwarf stars after JWST, ARIEL and other missions are performed. However, the luminosity and spectral line profile changes caused by the stellar activity make it extremely challenging to use indirect means to characterize the atmospheres of terrestrial planets in the habitable zone of solar-type stars. The effects of stellar activity can only be bypassed by direct imaging of planetary signals.

The next-generation flagship ground-based Very Large Aperture Telescope or Space Large Aperture Telescope will change this situation and make the atmospheric characterization of planets in the habitable zone of nearby stars feasible, thereby providing important observational evidence for assessing the habitability of potential Earth twins. According to the previous simulations \citep{Snellen2015, Wang2017, HawkerParry2019}, large aperture ground-based telescopes aided by extreme adaptive optics, equipped with high suppression ratio coronagraph and high dispersion spectrometer, will be able to directly obtain the reflection spectrum and thermal emission spectrum of the atmospheres of terrestrial planets in the habitable zone of nearby stars, and search for the presence of oxygen, water, methane, carbon dioxide and other components. METIS for E-ELT \citep{Brandl2012} and PCS \citep{Kasper2021}, and PSI for TMT \citep{Fitzgerald2019} are expected to make such observations a reality. On the other hand, space-borne telescopes with no influence of the earth's atmosphere, equipped with both high suppression ratio coronagraphs and star shadows, such as HabEx \citep{Gaudi2020} or LUVOIR \citep{TheLUVOIRTeam2019}, would be more effective and direct to characterize the atmospheres of terrestrial planets in the habitable zone of nearby stars, by measuring the chemical abundances of various biomarker molecules, the habitability of the related planets can be obtained, and even the existence of signs of life activities can be assessed \citep{Krissansen-Totton2018}.

\subsection{Science objective 2: a comprehensive census of nearby planetary systems}
To detect planets with orbital periods of 30 days to 10 years around $F, G, K$ stars $\sim 10$ pc, and planets whose masses are greater than that of the Earth, provide the planet's true mass and three-dimensional orbital parameters, and establish a complete exoplanet archive on the nearby planetary systems.

\subsubsection{Completeness census of nearby planets.}
\begin{figure}
   \centering
   \includegraphics[width=8cm]{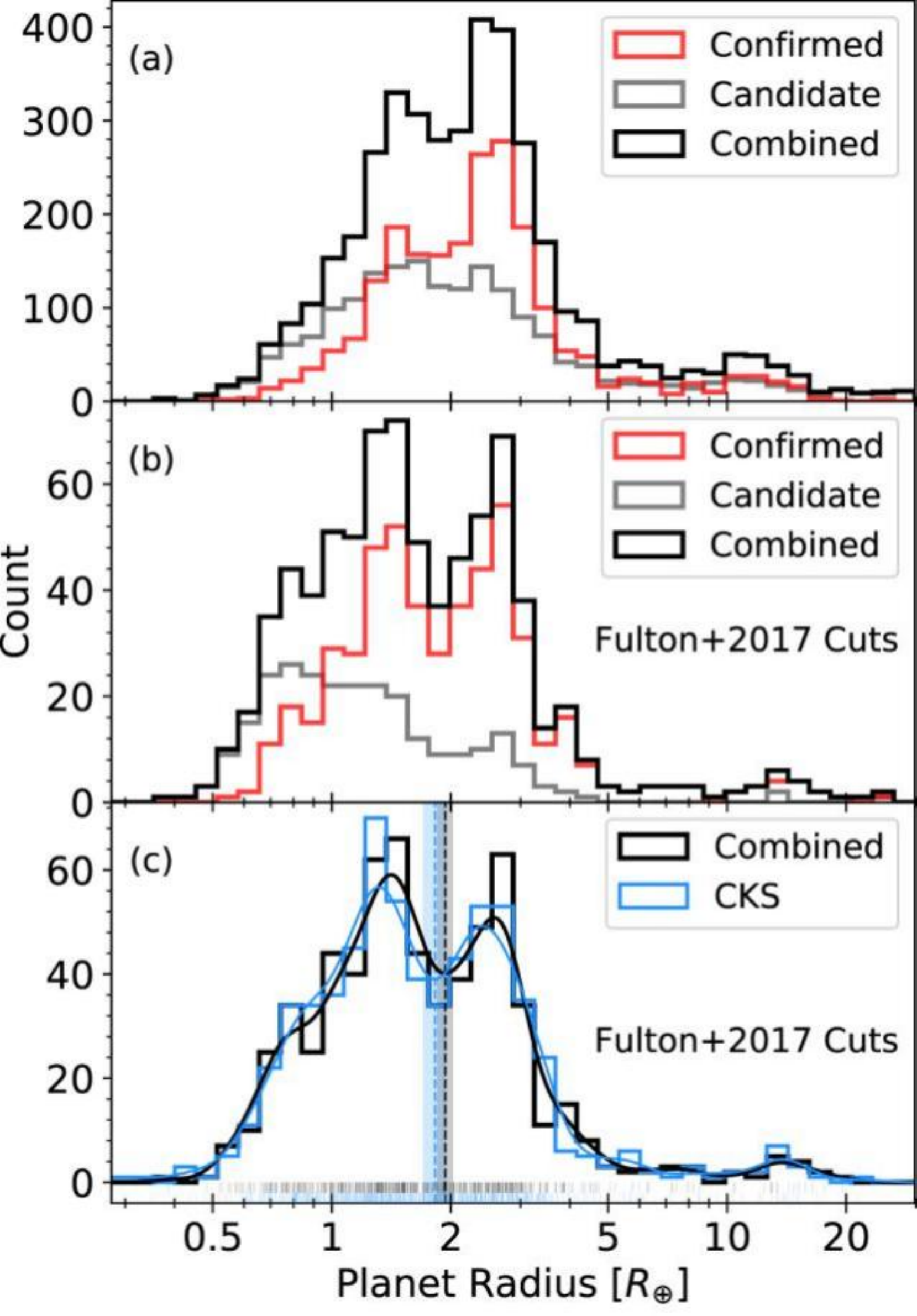}
   \caption{Planetary radius distribution of Kepler's released candidates \citep{Berger2018}.}
   \label{fig:2-4}
   \end{figure}

A more comprehensive exploration of the completeness of exoplanetary systems may cast light on the scenario of planetary formation and orbital evolution characteristics. Currently, various methods have a selective effect on planetary detection, especially radial velocity and transit methods, which are usually simply suitable for the discovery of low-mass planets with orbital periods less than tens of days (Figure \ref{fig:2-4})\citep{Berger2018}, and planetary masses that cannot be decoupled or depend on special orbital configurations. However, CHES can directly give accurate planetary masses and orbital periods of planets, so it will carry out more comprehensive and accurate study on planetary systems of nearby stars, which will play an important role in the completeness of planetary systems. Furthermore, CHES will be helpful to comprehensively and systematically understand the orbital characteristics, formation and evolution of planetary systems, and finally infer whether our solar system is universal or unique.

The planetary formation theory suggests that planetary systems around stars are ubiquitous, and the mass distribution of planets in the system is similar to the distribution of planets in our own solar system; but most of the planets discovered so far are much larger than the mass of the Earth, mainly owing to selection effect induced by the limitations of observational methods: the transit detection is inclined to detect large-size planets with short-period orbits in line of sight of the observer, while the radial velocity method is prone to find massive planets with close-in orbits. The data released by Kepler (Figure \ref{fig:2-4}) shows that the distribution of planetary radii illustrates that there is a deep distribution gap between $1.5$ and $2.0$ Earth radii, and regions with a planetary radius less than $1.14$ Earth radii are less complete. This feature may also imply some undiscovered regularity \citep{Fulton2017, Berger2018}, which may result from the photo-evaporation model \citep{Owen2017}. In addition to the discovery of habitable planets, the ultra-high-precision detection capabilities of CHES will also be able to conduct relatively comprehensive detection and investigation on the nearby planetary systems, to discover the planets from ET2.0 to super Earth in the habitable zones, thereby helping reveal the formation and evolution mechanisms of a wide variety of planets.

\subsubsection{The true mass of a planet}
In addition to its orbital characteristics, the internal structure of the planet also plays a key part in determining its physical properties and evaluates whether it is habitable. According to current theory, a habitable planet must first have a solid or liquid surface; the equilibrium temperature of the planet needs to be able to support the existence of liquid water, so the planet requires an atmospheric envelope. Here CHES can directly measure the planetary mass based on relative astrometry, and then define its final internal structure, which is of great significance for assessing the habitability of a planet.

\begin{figure}
   \centering
   \includegraphics[width=14cm]{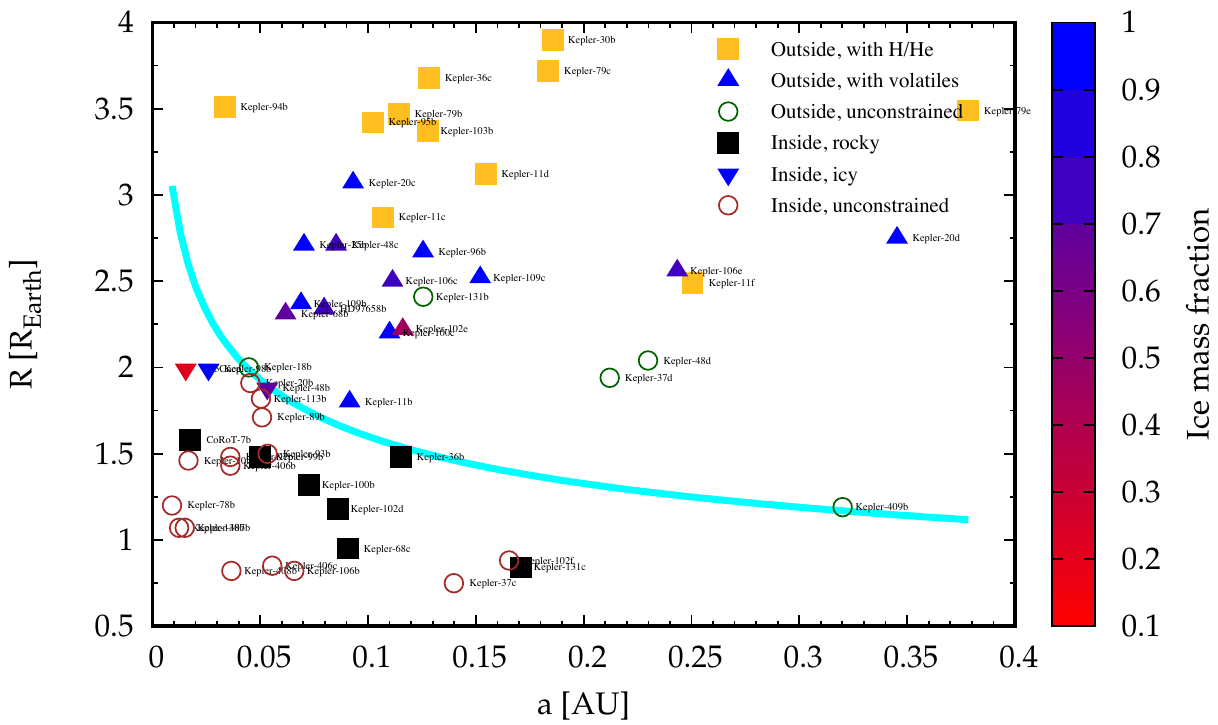}
   \caption{The composition of a planet with its structure and evolution \citep{JinMordasini2018}.}
   \label{fig:2-5}
   \end{figure}

The mass-radius relationship of the discovered exoplanets is a key linkage in defining the model of planet formation and evolution. For low-mass planets, considering the core accretion model of planet formation, it is difficult for them to accrete large amounts of gas; but their mass-radius relationships suggest that some planets' atmosphere-to-mass ratios can vary over a wide range. So, are their larger radius values due to larger initial atmospheric proportions, or implying that they are in the early stages of evolution? Therefore, the measurement of accurate planetary mass can help understand the evolution of short-period low-mass planets, and then compare with the observed mass-radius relationship (Figure \ref{fig:2-5}), which can constrain their formation and evolution \citep{JinMordasini2018}. As afore-mentioned, CHES can accurately measure the mass and radius of planets, investigate the evolution of a large number of planets, and infer their formation scenarios and material compositions.

\subsubsection{Three-dimensional orbits of planets and formation}
The mass and three-dimensional orbit of multiple planetary systems are directly related to the origin and evolution of these systems. The finding of exoplanets in the past two decades has brought us even more cognitive breakthroughs: planetary systems are highly dynamical and constantly changing. The three-dimensional structure of multi-planetary systems increases our knowledge of the origin and evolution of planetary systems, and helps us to statistically understand their migration scenarios and the long-term effects of these dynamical processes on planetary habitability \citep{LiuJi2020}. The accurate mass and three-dimensional orbits of the nearby multi-planetary system that CHES can solely provide will be the most precious observational basis, and will extremely expand human's cognition and theoretical research on planetary systems.

Current theories cannot well shed light on the complex scenarios of planetary formation \citep{Raymond2006, Mordasini2009, Chiang2013, LiuJi2020}. However, CHES mission will offer detailed information on the planetary three-dimensional orbits around nearby bright stars with sufficiently high spatial resolution, and further to provide strong constraints on distinguishing planetary formation theories, thereby being viewed as one of the unique contributions of CHES compared with other missions.

\subsubsection{Planetary diversity and follow-up observations}
A large observational population of exoplanets shows that other worlds around other stars appear to be extremely diverse and quite different from those of our own solar system \citep{Borucki2010,Borucki2011,Batalha2013,Gillon2016}, e.g., hot Jupiters, warm Neptunes, super-Earths and rocky planets. CHES will focus on the detection of habitable-zone planets around nearby solar-type stars, and further can coordinate with follow-up ground-based and space-based telescopes to conduct characterization of habitable planets. For example, CHES can conduct joint measurements with high-precision radial velocity instruments such as future E-ELT \citep{Macintosh2006} and TMT \citep{Skidmore2015}, and can also verify habitable planet candidates discovered by the radial velocity, and can accurately characterize planetary masses and orbital parameters.

Due to the limitation of angular resolution, the most suitable planetary systems for atmospheric characterization by direct imaging to obtain spectra are located around nearby stars, and the target of CHES happens to be close to nearby sun-like stars. CHES will discover unknown planets in the habitable zone of nearby stars. Combined with follow-up radial velocity observations and photometric measurements, the masses and orbital parameters of these planets will be accurately determined, and they will be ideal targets for the next generation of flagship space telescopes via direct imaging for deep spectroscopic observations.

\subsection{Science objective extended: cosmology, dark matter and black holes}
Beside the detection of terrestrial planets in the habitable zones, CHES will find the member stars for satellite galaxies. High-precision measurements of proper motion will provide strong constraints on the mass of the Milky Way and the properties of dark matter particle. Furthermore, CHES may detect the masses of neutron stars through measuring binary orbital motion to constrain the composition and equation of state. Additionally, CHES will measure the proper motion of black hole binaries to understand the orbit and formation scenario. Finally, CHES will carry out cosmological distances measurement.

\section{SCIENTIFIC REQUIREMENTS}
\subsection{Decomposition of detection requirements}
\subsubsection{Positioning Accuracy Requirements}
The major scientific objective of CHES aims at the habitable planets orbiting the nearby solar-type stars, whose requirements of positioning accuracy are well constrained by this detection.

The detection accuracy is required to be $1~\mu$as for a planet with the same mass and orbit as the Earth at a distance of 3 pc (about 10 light years) from the Sun. And it needs to reach $0.3~\mu$as for detecting terrestrial planets (see Section $4.1$ for details) in the habitable zone at 10 $ \mathrm{pc}$. To determine 12 parameters of a planet and its host star (including 7 planetary parameters and 5 stellar parameters), we need to observe at least 12 sets of data in 5 years.

In actual detection, each set of observations is obtained by multiple exposures over 2 hours. Hence, increasing the number of observation groups can obtain a higher signal-to-noise ratio or appropriately reduce the detection accuracy requirements. In order to find terrestrial planets in the habitable zone located at 10 $ \mathrm{pc}$ (about $32.6$ light years), when the number of observation groups reaches 200, thus the required single-group data observation accuracy must be better than $1~\mu$as.

\begin{figure}
   \centering
   \includegraphics[width=12cm]{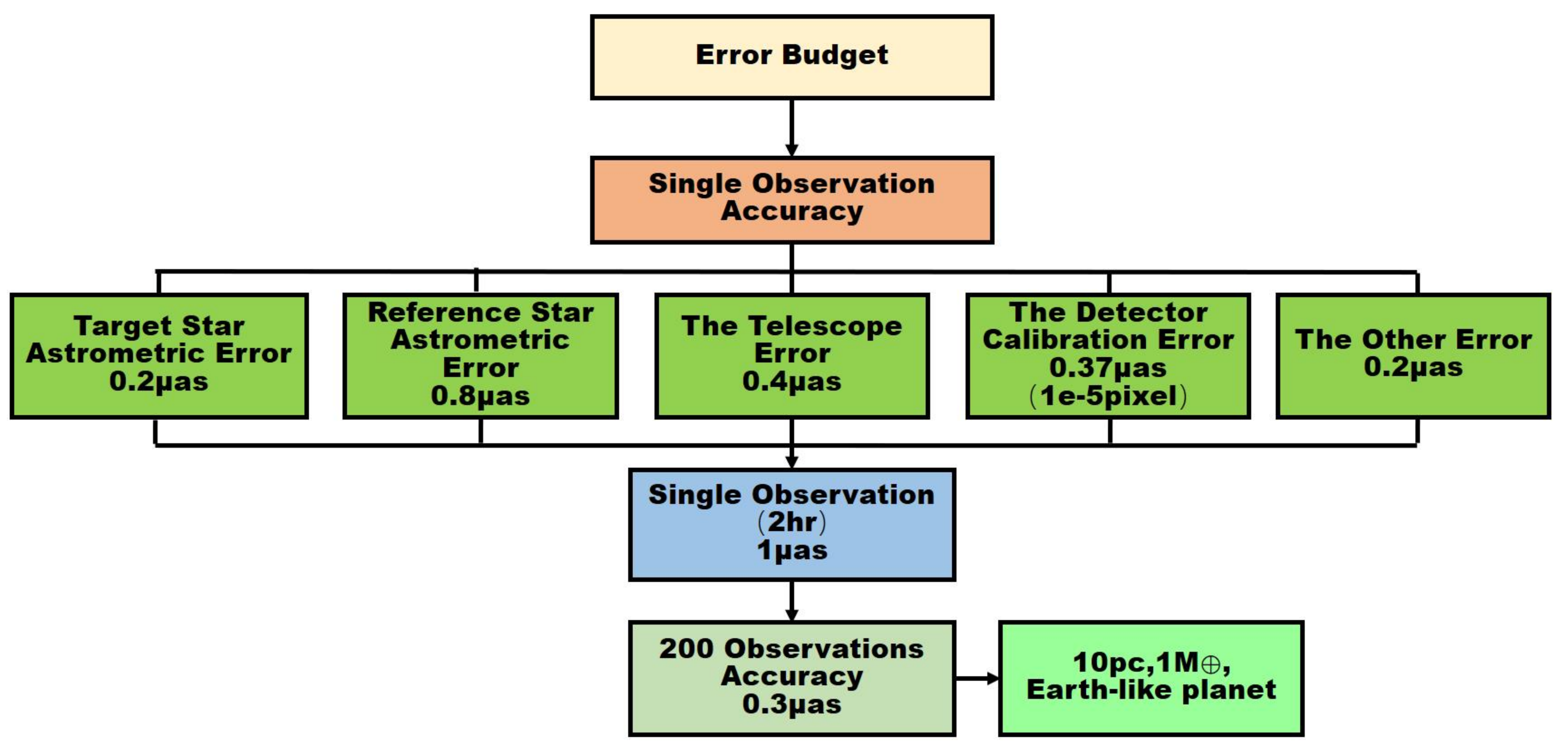}
   \caption{Error budget and analysis of CHES observation accuracy.}
   \label{fig:3-1}
   \end{figure}

It is necessary to analyze the primary errors that may affect the detection accuracy to achieve the detection accuracy of $1~\mu$as with each measurement for CHES. Figure \ref{fig:3-1} shows the block diagram of top-level error budget and the major composition to the overall budget. Preliminary analysis suggests that the dominant factors influencing the detection accuracy with one measurement include the astrometric errors of target star and reference stars, telescope measurement error, detector calibration error and other errors. Taking into account the overall error budget and current engineering implementation capability of the payload, the major terms of each error weighing factor is $0.2$ $\mu$as for the target star, 0.8 $ \mu$as for the reference star, 0.4 $\mu$as for the telescope, 0.37 $\mu$as for detector calibration and other errors 0.2 $ \mu$as, respectively.

The main sources of errors in the payload are telescope wave aberration error, residual error after distortion calibration, and detector calibration error excluding the astrometric error between the target star and the reference star. The simulation analysis shows the requirements for the telescope imaging quality, distortion calibration and detector calibration are as follows to satisfy the corresponding error distribution: the telescope wave aberration is not greater than $\lambda / 12$; the residual error after the maximum field of view relative distortion calibration is not low less than $0.4~\mu$as; detector calibration residual is not more than 0.37 $\mu$as.

\subsubsection{Aperture requirements}
The telescope aperture depends on the effect of photon noise on positioning accuracy. The magnitude of selected reference star must not be less than 12 . Therefore, according to the estimation of the aperture of the CHES primary mirror of $1.2$ meters, a star with a brightness of 12th magnitude will receive about $1.6 \times 10^{9}$ photons after 2-hour exposure. In the case of observing 8 reference stars simultaneously for 2 hours, the error caused by photon noise is about 1 $\mu$as $((\lambda / 2 D) / \sqrt{8 N})$. In the mission, roughly 200 observations for each target star have been accumulated in 5 years, the equivalent precision is 0.05 $\mu \mathrm{as}$ in RA and Dec, corresponding to the signal with an intensity of 0.3  $\mu \mathrm{as}$ that can be measured, where the SNR $\approx $ 6 \citep{Brown2009}. Therefore, it is decided to select the diameter of the CHES telescope as $1.2$ meters after comprehensively weighing the existing domestic mirror processing level, cost, weight, photon noise and other factors.

\subsubsection{Focal length requirements}
The requirement for the focal length of the telescope is based on the diffraction limit of the telescope and the Nyquist sampling theorem. The primary mirror diameter of CHES is $1.2$ meters. At the observation wavelength of 500 nanometers, the corresponding full width at half maximum (FWHM) of the PSF is about $0.086$ arcseconds. If the focal length of the telescope is 36 meters, the FWHM of the target star image on the sCMOS detector is 15 $\mu \mathrm{m}$. The pixel size of the CMOS detector to be selected is 6.5 $\mu \mathrm{m}$, which satisfies Nyquist sampling theorem. The simulations show that a star image size of $32 \times 32$ pixels can be in line with the accuracy requirements.

\subsubsection{Field of View Requirements}
CHES achieves high-precision positioning mainly by measuring the relative position between the target star and the reference star. At least 6 to 8 reference stars are required to enter the CHES field of view. According to the statistics of candidate observation targets, the field of view is greater than $0.44^{\circ} \times 0.44^{\circ}$. It can be achieved that all target stars have at least $6 \sim 8$ reference stars, which is in agreement with the number of reference stars required for positioning.

\subsubsection{Requirements of interstellar distance measurement accuracy}
Since the pixel size of the detector is 6.5 $\mu \mathrm{m}$ and the focal length of the telescope is $36 \mathrm{~m}$, the angular displacement corresponding to one pixel of the detector can be calculated to be $0.037^{\prime \prime}$. If the measurement accuracy of the distance between stars of 1 $\mu$as is to be achieved, the relative positioning accuracy needs to reach $1 \times 10^{-5}$ pixels.

\section{PROPOSED PAYLOAD}
\label{sect:data}
\subsection{Measurement principle}
The major scientific objective of CHES aims at detecting the habitable planets orbiting nearby solar-type stars via unveiling the very tiny periodic signals in the position of the central star induced by the surrounding planets. By measuring the small change and dynamical perturbance of the center position of the target star relative to each reference star (at least 6-8 reference stars for each target star), CHES will detect the planets with different masses and periods around the target stars. High-precision astrometry is much less affected by stellar surface activities than other observation methods \citep{Shao2009}, so high-precision astrometry is the best way to find terrestrial planets in the habitable zone around $\mathrm{F}, \mathrm{G}$, and $\mathrm{K}$-type nearby stars, while ground-based high-precision astrometry cannot be carried out due to the complex Earth's precession-nutation and the influence of atmospheric refraction. Thus, CHES is likely to achieve astrometric precision of micro-arcsecond in space.

The astrometry amplitude $\alpha$ caused by the planet around the target star can be expressed as
$$
\alpha=3\left(\frac{M_{p}}{M_{\oplus}}\right)\left(\frac{a_{p}}{1 \mathrm{AU}}\right)\left(\frac{M_{*}}{M_{\odot}}\right)^{-1}\left(\frac{D}{1 \mathrm{pc}}\right)^{-1} \mu \mathrm{as}
$$
where $M_{p}$ and $a_{p}$ are the mass and semi-major axis of the planet, respectively. $M_{*}$ and $D$ are, respectively, the mass of star and the distance from Sun. For an Earth-mass planet at $1 \mathrm{AU}$ around a solar-type star at 10 $\mathrm{pc}$, the astrometry amplitude $\alpha$ is 0.3 $\mu \mathrm{as}$. Thus, to detect Earth-mass planets, CHES will employ the laser interferometric calibration method, which can improve the relative position accuracy of stars over the images up to micro-pixel level.

\subsubsection{Target stars}
\begin{figure}
   \centering
   \includegraphics[width=8cm]{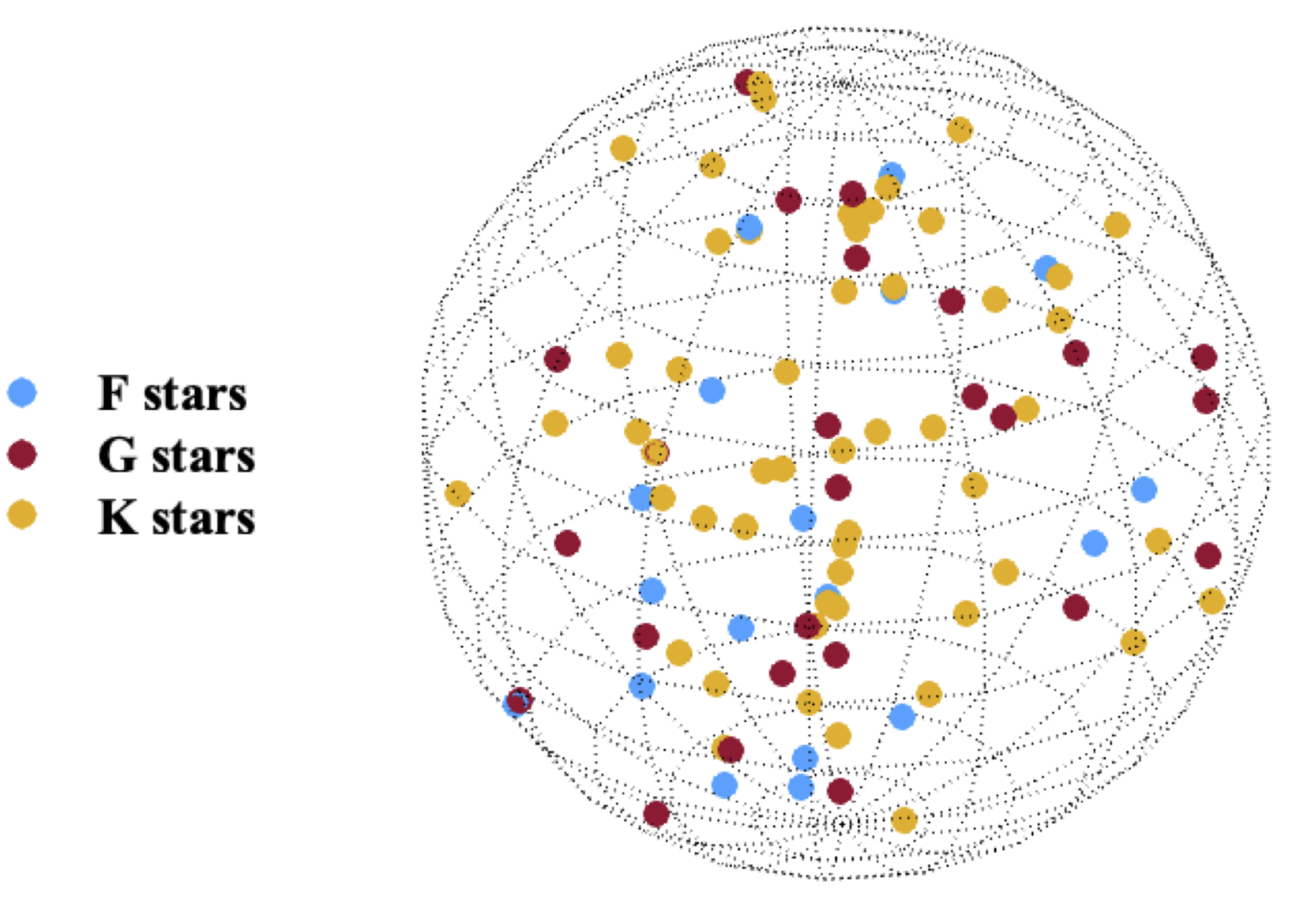}
   \caption{Distribution of CHES target stars within $\sim$ 10 pc.}
   \label{fig:4-1}
   \end{figure}

CHES has preliminarily completed the selection of $100 \mathrm{~F}, \mathrm{G}$ and $\mathrm{K}$ types of nearby stars as primary targets of the space mission. The distribution in the celestial sphere is shown in Figure \ref{fig:4-1}, where the colors represent the types of the target stars (e.g., $\mathrm{F}, \mathrm{G}, \mathrm{K})$ and all stars are within 10 pc away from the Earth. In addition, a portion of the samples of selected target stars are summarized in Table \ref{Tab4.1}, where the stellar name, spectral type, the distance and the magnitude, along with the number of the reference stars with respect to each target are listed.

\begin{table}
\begin{center}
\caption[]{Examples of the $\mathrm{F}, \mathrm{G}$, and $\mathrm{K}$ target stars and number of reference stars.}\label{Tab4.1}
\begin{tabular}{|l|l|l|l|l|l|}
\hline HIP & Name & V mag & Spect. & Dist. & Ref. stars \\
\hline & & mag. & Type & $p c$ & No. \\
\hline 71683 & $\alpha$ Cen A & $0.0$ & G2V & $1.34$ & 26 \\
\hline 71681 & $\alpha$ Cen B & $1.4$ & K1V & $1.34$ & 48 \\
\hline 108870 & $\epsilon$ Ind & $4.7$ & K5V & $3.6$ & 9 \\
\hline 96100 & $\sigma$ Dra & $4.7$ & K0V & $5.8$ & 14 \\
\hline 3821 & $\eta$ Cas & $3.5$ & G0V & 6 & 24 \\
\hline 22449 & $\pi 3$ Ori & $3.2$ & F6V & 8 & 16 \\
\hline 1599 & cTuc & $4.2$ & G0V & $8.6$ & 10 \\
\hline 27072 & $\gamma$ Lep & $3.6$ & F6V & 9 & 20 \\
\hline 105858 & $\gamma$ Pav & $4.2$ & F9V & $9.2$ & 12 \\
\hline 14632 & $\iota$ Per & $4.1$ & G0V & $10.5$ & 20 \\
\hline
\end{tabular}
\end{center}
\end{table}

\subsubsection{Reference stars}
The reference stars used by CHES are mostly K-type stars with a distance of at least 1 kpc. Typical values for their proper motion and parallax are 1 $\mathrm{mas} / \mathrm{yr}$ and 1 millisecond, respectively. CHES is designed to observe the target star with an accuracy of 1 micro-arcsecond, so the motion of the reference star are need to be considered. However, the distance of the reference star is much larger than that of the target star (average 10 $\mathrm{pc}$), the potential planets of the reference star have at least 100 times less influence on the observation than the target star. Only Jupiter-mass planet are possible to affect the measurement of planetary parameters around the target star. Statistically, only $10 \%$ of stars have Jupiter-like planets \citep{Nemati2020}. After obtaining observations, those reference stars which have Jupiter-like planets can be first identified, and then can be removed from the list. In addition, ground-based radial velocity measurements can also be used to find and eliminate reference stars with Jupiter-like planets.

The stellar activity (spots and bright structures) changes its luminosity over time, affecting the detection of planets. At a distance of 10 $\mathrm{pc}$, the position measurement error caused by stellar activity is much smaller than that induced by terrestrial planets in habitable zone. For CHES, the criteria to select a reference star corresponds to its activity index is less than $-4.35$.

   \begin{figure}
   \centering
   \includegraphics[width=8cm]{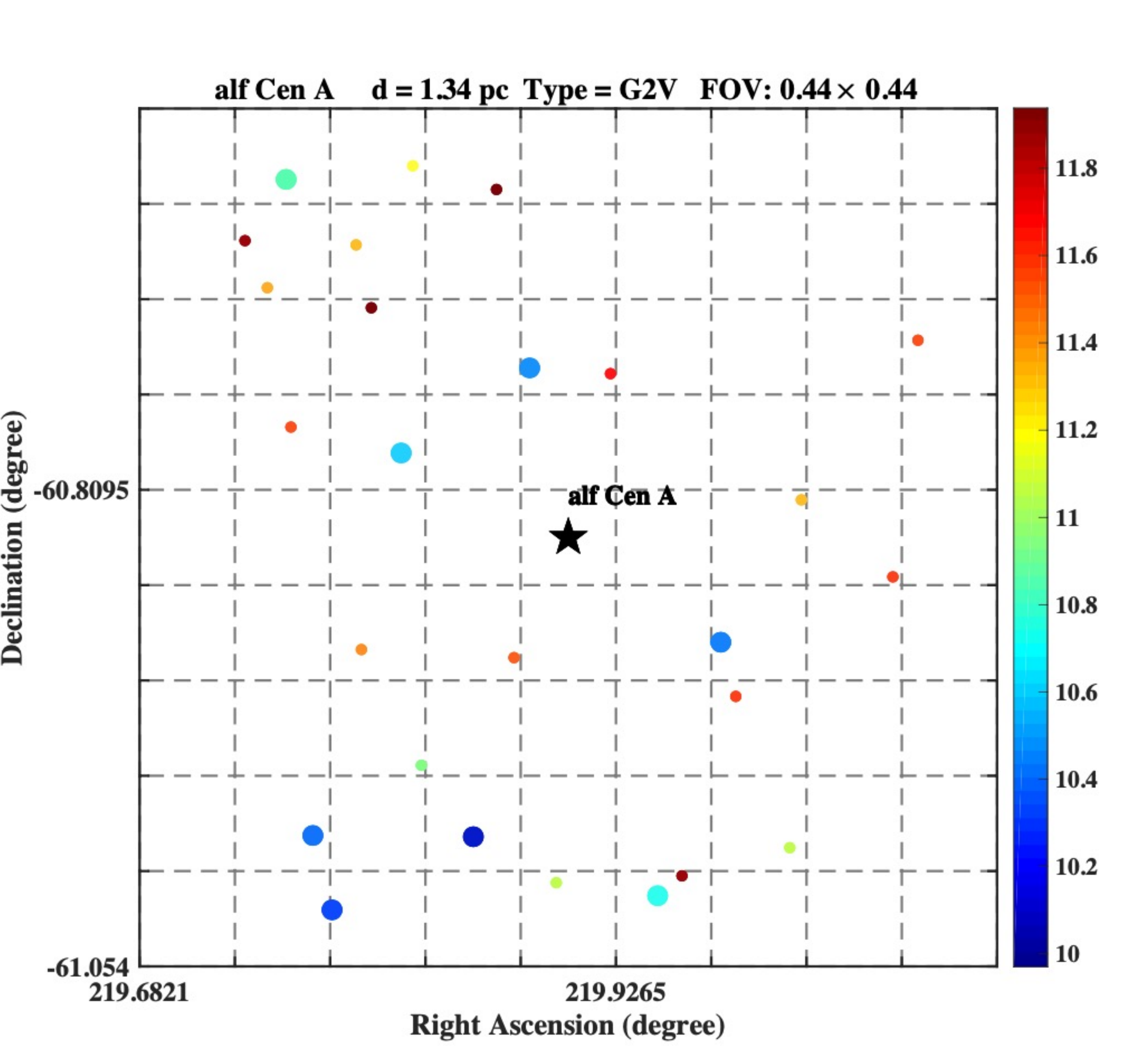}
   \caption{An example of the target star $\alpha$ Cen A with its reference stars.}
   \label{fig:4-2}
   \end{figure}

An example of target star $\alpha$ Cen A, along with 26 possible reference stars within FOV $\left(0.44^{\circ} \times 0.44^{\circ}\right)$, are shown in Figure \ref{fig:4-2}, where the color represents the magnitude of reference stars while the larger eight circles suggest our preferred reference stars.

\subsubsection{Orbital retrieval simulations}
The motion of the star projected on the celestial sphere is described by 5 parameters ( 2 position parameters, 2 proper motion parameters and 1 parallax parameter), and the influence of any planet on the star is determined by 7 parameters (the mass, orbital period, pericentre, orbital eccentricity, orbital inclination, and two other angular parameters], so the projected motion of a star with $p$ planet on the celestial sphere can be determined by $m=5+7 p$ parameters. For a star harboring 3 planets, the number of unknown parameters to be determined is 26 , and 200 measurements are required to determine the parameters of the planetary system.

\begin{figure}
   \centering
   \includegraphics[width=16cm]{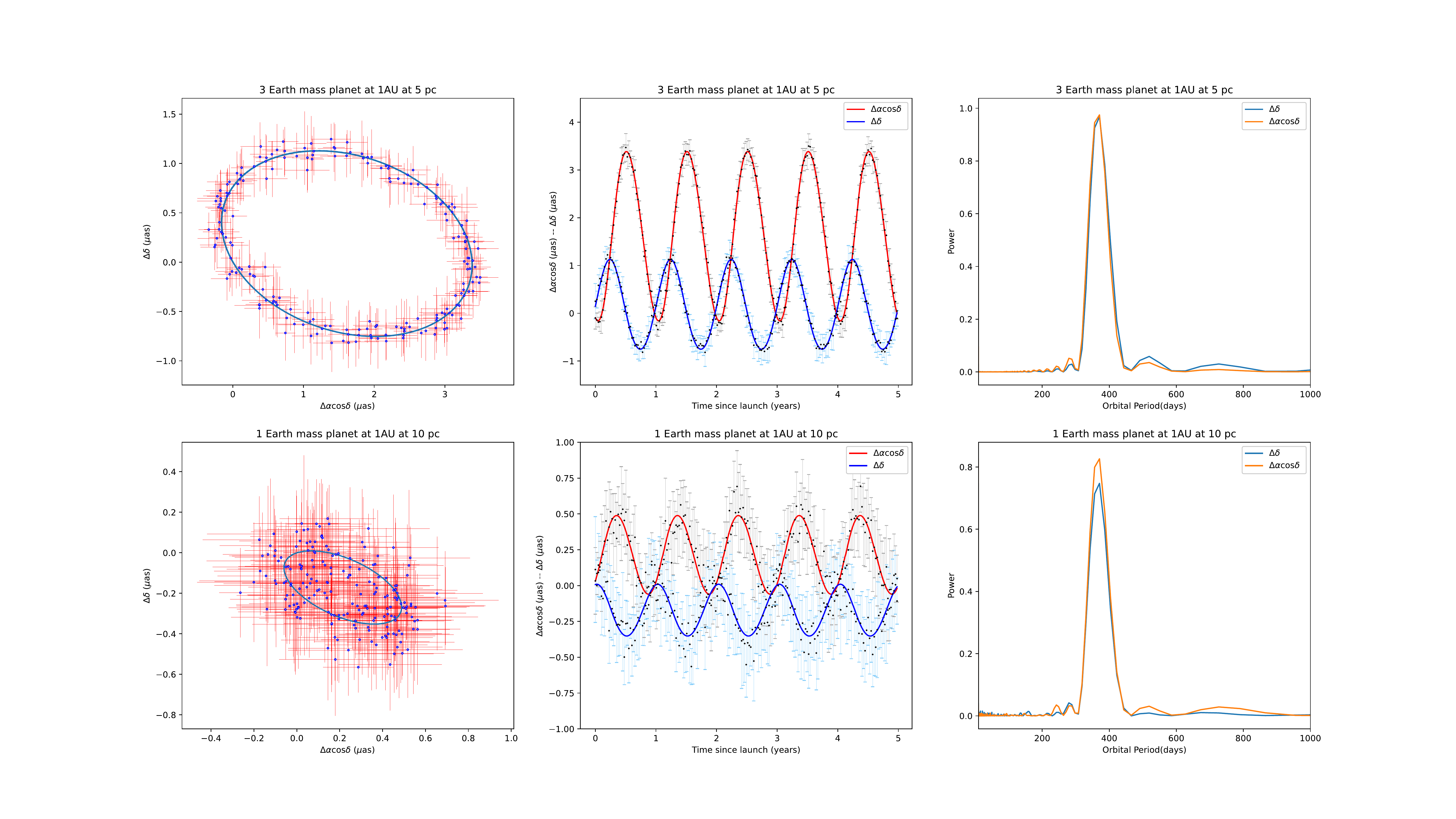}
   \caption{Simulations of astrometric detection of two types of terrestrial planets.}
   \label{fig:4-3}
   \end{figure}

Figure \ref{fig:4-3} shows the simulations of astrometric detection of two types of the terrestrial planets with 200 CHES measurements (each corresponds to 1 $\mu$as astrometric precision for $2 \mathrm{~h})$, where the top panels correspond to a detection of 3 Earth-mass planet at $1 \mathrm{AU}$ around a sun-like star at 5 $\mathrm{pc}$ (easy detection), whereas the bottom panels refer to a detection of 1 Earth-mass planet at 1 $\mathrm{AU}$ around the star at 10 $\mathrm{pc}$ (hard detection). The solid lines in the left and middle columns stand for the projected orbits (right ascensions and declinations) of the terrestrial planets after deriving their parameters, while the peak of the profiles in the right columns represent the orbital periods at one year via Lomb-Scargle periodogram throughout 5-yr mission. The CHES team also developed codes to perform the retrieval of orbital parameters of terrestrial planets via the simulated astrometric data \citep{Jin2022}.

\subsubsection{Planetary occurrence rate}
Based on the exploration of the Kepler released data, an occurrence rate of a rocky planet within the habitable zone is estimated \citep{Traub2016, Kaltenegger2017}. The occurrence rate of a planet with an Earth-radius in the range of $[0.5,1.35]$ in the habitable zone is roughly $0.66 \pm 0.14,1.03 \pm 0.10$ and $0.75 \pm 0.11$, respectively, for $\mathrm{F}, \mathrm{G}$ and $\mathrm{K}$-type stars. In the set of CHES targets, there are 18 F-type, 25 G-type and $57 \mathrm{~K}$-type nearby stars. According to the above estimation of occurrence rate of habitable planets, CHES will detect roughly 69-92 habitable planets by the end of 5-yr mission, while the resultant number of detection of planets can increase if consider a broader range of the planetary radius and orbital periods for solar-type stars (Zhu \& Dong 2021).

\subsection{General description of the payload and challenges}
\begin{figure}
   \centering
   \includegraphics[width=10cm]{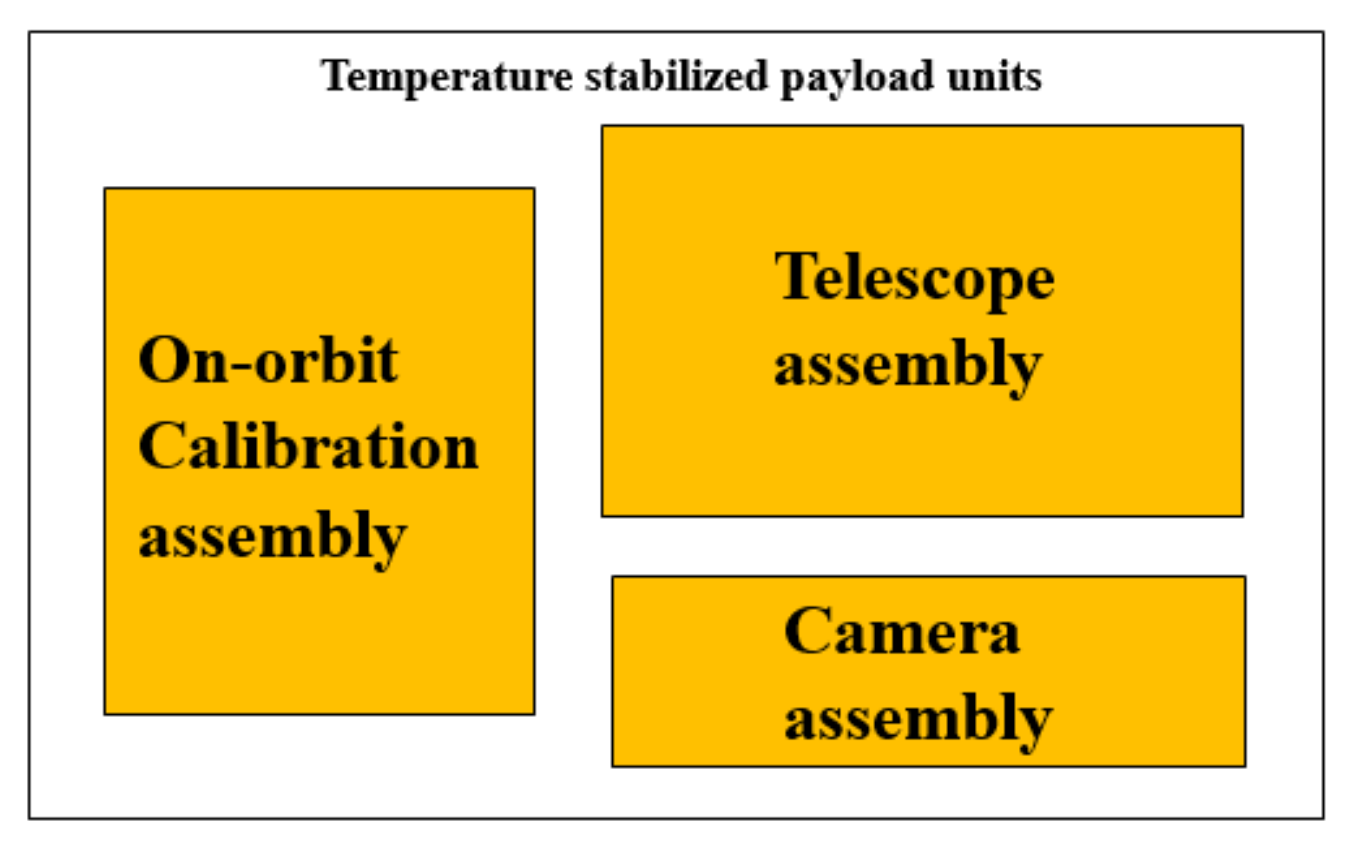}
   \caption{Block diagram of the CHES payload Hardware.}
   \label{fig:4-4}
   \end{figure}

The CHES payload is simple in design and consists of three assemblies: the telescope assembly, the camera assembly, and the on-orbit calibration assembly. The composition of the entire payload configuration is shown in Figure \ref{fig:4-4}.

\subsubsection{Instrumental challenges}
CHES telescope has a $1.2 \mathrm{~m}$ aperture and a $36 \mathrm{~m}$ focal length. For a camera with the pixel size of 6.5 $\mu \mathrm{m}$, the corresponding single pixel FOV is $0.037^{\prime \prime}$ as aforementioned. To obtain the micro-arcsecond relative astrometric accuracy, the measurement accuracy of the stellar image centroid displacement needs to reach $10^{-5}$ pixel, which is the major challenge for the instrument.

In order to achieve the micro-arcsecond relative astrometric precision, three issues should be solved: (1) A telescope with high-quality, low-distortion in imaging. Near diffraction-limited imaging is required to provide perfect PSF over the full FOV of the telescope. To correct the errors induced by the temporal variation of the shape of the mirrors, it is necessary to perform micro-arcsecond accuracy on-orbit calibration of the distortion of the telescope; (2) Micro-pixel accuracy centroid displacement measurement and detector calibration. All effects that may impact the shape and the position of the PSF should be considered, such as the geometric stability of the focal plane array, the photon noise of the stars, and the inhomogeneity of the quantum efficiency between pixels and within pixels; (3) The satellite needs to remain ultra-high attitude stability and telescope thermal control accuracy to reach an ultra-high-precision angle offset between the target star and the reference stars.

\subsubsection{Instrumental concept}
The CHES payload concept consists of an on-axis Three Mirror Anastigmatic (TMA) telescope with high imaging quality, high stability in the visible band. To obtain the PSF of stellar images for Micro-pixel accuracy centroid displacement measurement, the optical system of the telescope should realize near diffraction-limited imaging. A large focal plane covers a $0.44^{\circ} \times 0.44^{\circ}$ $\mathrm{FOV}$ with a mosaic of detectors. Its dynamic range, frame rate, full well charge and other parameters meet the requirements of observing different magnitudes stars in the full FOV. To monitor the mosaic geometry and its quantum efficiency, a focal plane metrology is performed. While to calibrate the telescope distortions with micro-arcsecond precision, an on-orbit calibration algorithm based on micro-pixel accuracy centroid displacement measurement is used.

\subsection{Design of the payload subsystems}
\subsubsection{Telescope assembly}

Considering the demanding image requirements for the PSF and the relative centroid motion due to temporal variations on the optics positions and shapes, an on-axis TMA configuration is adopted for controlling the optical aberrations up to third order. It is possible to correct astrometric displacements on the entire FOV. In order to achieve the scientific goals, the key characteristics of CHES telescope are confirmed as shown in Table \ref{Tab4.2}.

\begin{table}
\begin{center}
\caption[]{Key characteristics of CHES telescope.}\label{Tab4.2}
\begin{tabular}{|c|c|c|}
\hline NO & Specification & Value \\
\hline 1 & Aperture & $\Phi 1.2 \mathrm{~m}$ \\
\hline 2 & Focal length & 36 m\\
\hline 3 & Field of view & $0.44^{\circ} \times 0.44^{\circ}$\\
\hline 4 & Wavelength coverage & $500 \mathrm{~nm}-900 \mathrm{~nm}$ \\
\hline 5 & Detector pixel size & $6.5 \mu \mathrm{m} \times 6.5 \mu \mathrm{m}$ \\
\hline 6 & Focal plane size & $276 \mathrm{~mm} \times 276 \mathrm{~mm}$ \\
\hline 7 & Wave aberration & RMS $\leq \lambda / 12$ \\
\hline 8 & Distortion after calibration & $\sim 1 \mu \mathrm{as}$ \\
\hline
\end{tabular}
\end{center}
\end{table}

\begin{figure}
   \centering
   \includegraphics[width=8cm]{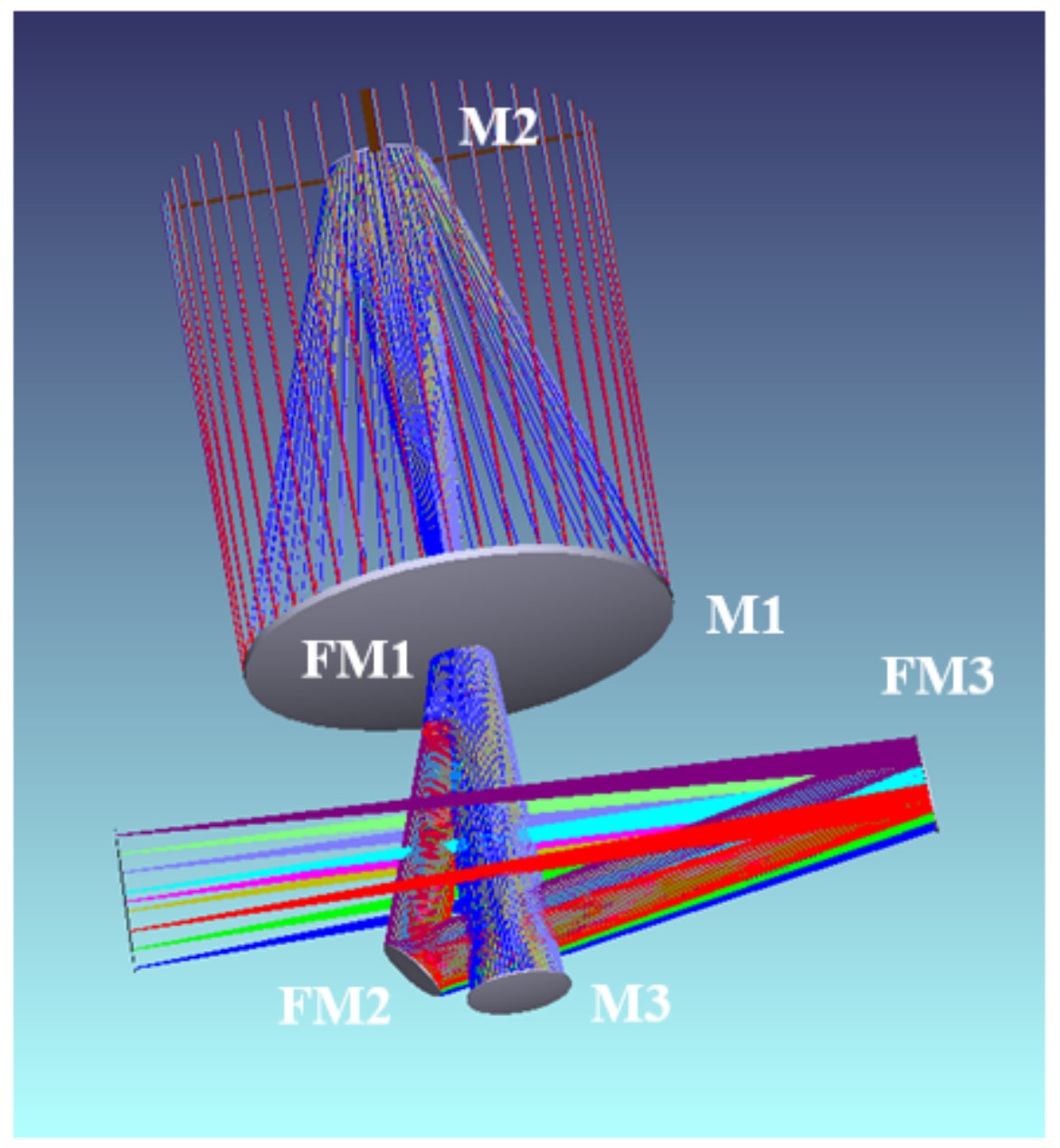}
   \caption{Ray tracing of CHES telescope.}
   \label{fig:4-5}
   \end{figure}

\begin{figure}
   \centering
   \includegraphics[width=12cm]{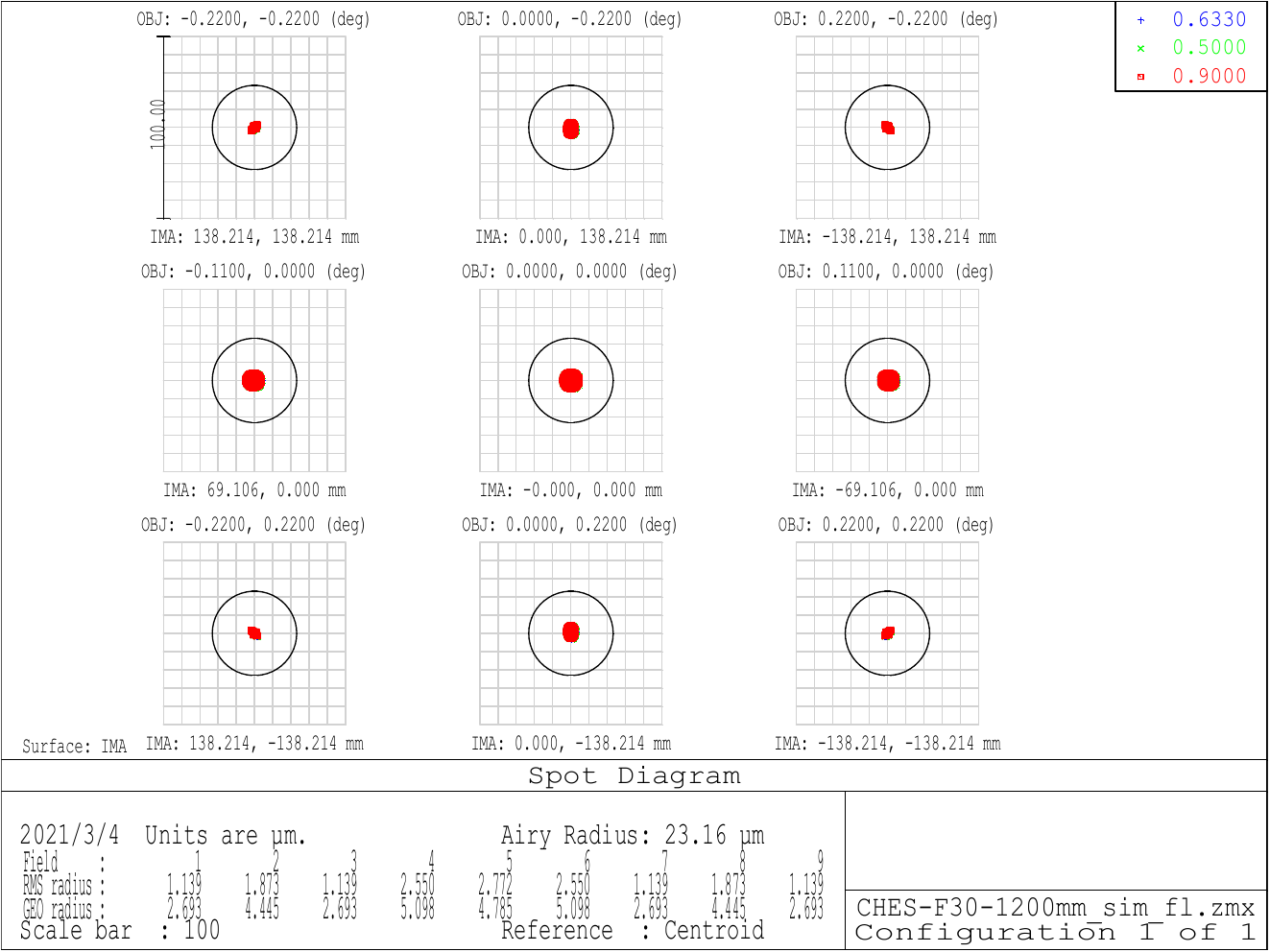}
   \caption{Spot diagrams for the full FOV of CHES telescope.}
   \label{fig:4-6}
   \end{figure}

Figure \ref{fig:4-5} shows the ray tracing of CHES telescope. The stars are imaged on the Cassegrain focal plane after passing through the primary and secondary mirrors. The stellar image at Cassegrain focal plane is re-imaged by M3 mirror and then formed final image on the detector by three fold mirrors. It noted that the first fold mirror FM1 is located near the Cassegrain focal plane. There is a hole in the center of FM1, the size of the hole is the same as the size of the intermediate image plane. The purpose of the hole is to allow the ray of the intermediate image plane to pass. Due to the hole on FM1, a fraction of the light energy is lost, but no loss of FOV. The imaging quality of the designed optical system can be evaluated by the spot diagrams, as shown in Figure \ref{fig:4-6}. The results indicate the telescope achieves near-diffraction-limited imaging in the entire field of view.

\begin{figure}
   \centering
   \includegraphics[width=8cm]{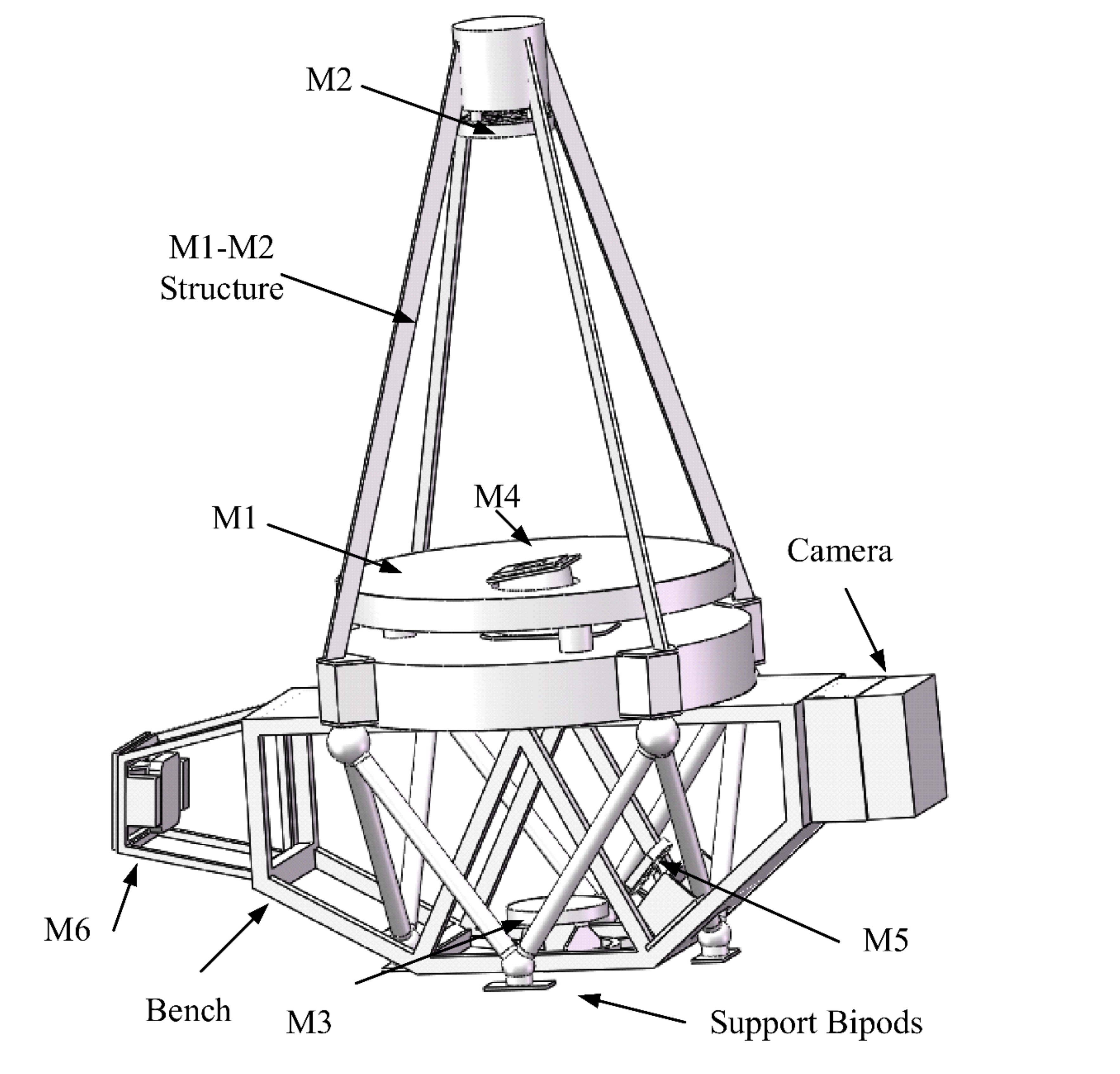}
   \caption{Overall layout of the CHES telescope.}
   \label{fig:4-7}
   \end{figure}

According to the optical design and thermos-elastic effects, the structural design of the CHES telescope is performed as shown in Figure \ref{fig:4-7}.

The mirrors adopt low temperature optimized Zerodur or ULE \citep{Westerhoff2014} and the light weighting techniques is used to reduce the weight of the prime mirror \citep{Krodel2014}. To minimize thermo-elastic impacts in the position of the mirrors, SiC and TC4 could be adopted for the telescope structure \citep{Verlaan2012}. Aluminum coatings can be used as optical mirror surface. The total mass of telescope under consideration of ULE and SiC is estimated to be about 500 kg.

\subsubsection{Focal Plane Array assembly}
With the focal length of the telescope of $36 \mathrm{~m}$ and the $\mathrm{FOV}$ of $0.44^{\circ} \times 0.44^{\circ}$, the geometric size of the focal plane is about $276 \mathrm{~mm} \times 276 \mathrm{~mm}$. To satisfy the requirements of such a large-format focal plane, a mosaic of detectors will be assembled on the focal plane.

CMOS detectors present a high QE over a larger visible spectral band, faster readout, lower readout noise, and better radiation hardness \citep{Marco-Hernandez2020}. While CCD detectors have high readout noise in high frame rate readout mode. Due to the existence of a mechanical shutter for a large size CCD, it will cause the vibration of the focal plane, which has an impact on the measurement accuracy of the centroid displacement of stellar images. Here we select CMOS detector, which can be produced by \emph{Gpixel Changchun Optotech, Inc}. The CMOS chip parameters can be customized to further reach the requirement of CHES mission. The typical characteristics of detectors required for CHES is summarized in Table \ref{Tab4.3}.

\begin{table}
\begin{center}
\caption[]{The typical characteristics of detectors.}\label{Tab4.3}
\begin{tabular}{|c|c|}
\hline Specification & Value \\
\hline Detector size (pixels) & $4096 \times 4096$ pixels \\
\hline Detector pixel size & $6.5~\mu \mathrm{m} \times 6.5~\mu \mathrm{m}$ \\
\hline Focal plane size & $276 \times 276 \mathrm{~mm}$ \\
\hline Pixel scale & 37 mas $/$ pixel \\
\hline Pixel full well capacity $>$ & $170~\mathrm{ke}-$ \\
\hline Wavelength coverage & $500 \mathrm{~nm}-900 \mathrm{~nm}$ \\
\hline Quantum efficiency & $>90 \%(600 \mathrm{~nm})$ \\
\hline Dark signal $\left(@-50^{\circ} \mathrm{C}\right)$ & $<0.05 \mathrm{e}-/ \mathrm{pix} / \mathrm{s}$ \\
\hline Readout noise $(@ 10 \mathrm{kHz})$ & $<2 \mathrm{e}-/ \mathrm{pix}$ \\
\hline
\end{tabular}
\end{center}
\end{table}

\begin{figure}
   \centering
   \includegraphics[width=8cm]{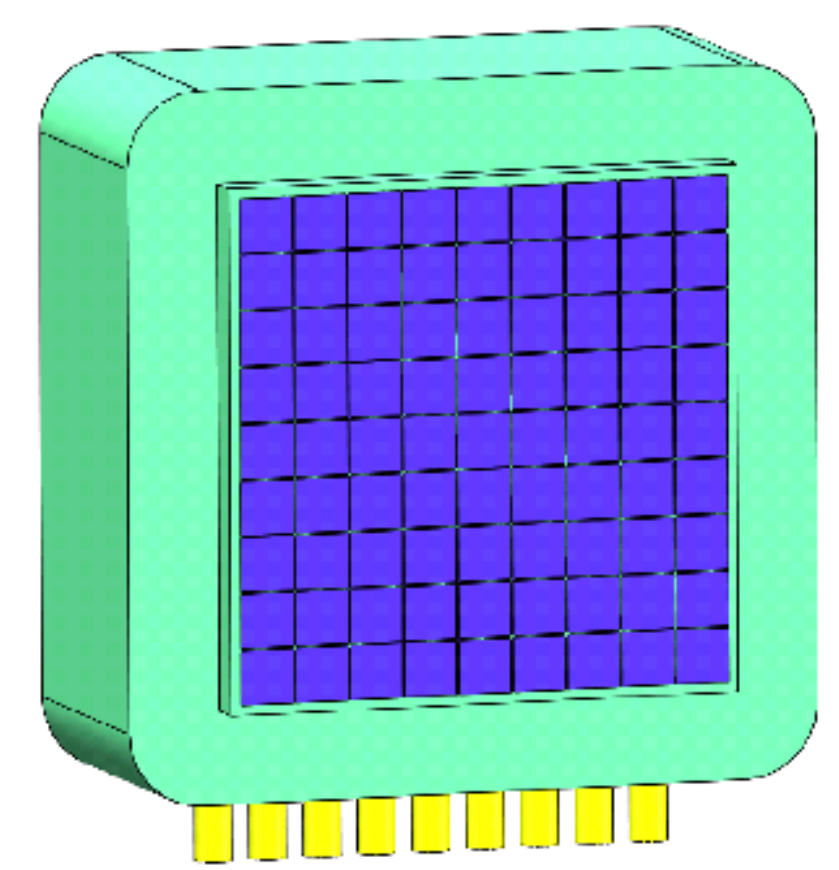}
   \caption{Concept for CHES Camera.}
   \label{fig:4-8}
   \end{figure}

The pixel size of each CMOS chip is 6.5 $\mu \mathrm{m}$, the number of pixels is $4 \mathrm{~K} \times 4 \mathrm{~K}$, and each size is about $32 \mathrm{~mm} \times 32 \mathrm{~mm}$ (including package size). Thus 81 mosaic CMOS detectors are required to achieve full focal plane coverage. The fundamental architecture of focal plane is shown in Figure \ref{fig:4-8}.

Considering the consistency with the linear expansion coefficient of the detector packaging material, INVAR steel or SiC material is selected as the substrate material. 81 mosaic CMOS detectors are fixed to the substrate. To ensure the position accuracy, geometrical tolerances of the substrate are strictly specified and the processing technology is strictly controlled when the substrate is structurally designed. The position of the detectors is independently detected when assembling. After the assembling is completed, a high-precision optical measurement is adopted for quality inspection of the mosaic detectors.

\subsubsection{On-orbit calibration assembly}
The on-orbit calibration assembly consists of a focal plane array calibration module and a telescope optical distortion calibration module. The calibration mode lasts roughly 2.5 hours and includes 2 hours optical distortion calibration mode and 0.5 hours focal plane metrology mode after each 2 hours scientific exposures. The average scientific exposure is about 2 hours for each target through scientific mode. The CHES satellite enters into calibration mode after 2 hours scientific exposures. Firstly, a $~$20 arcsecond dithering is used in the optical distortion calibration mode. The optical distortion calibration mode takes 2 hours star field integration. The scientific exposures and the dither pattern are used to calibrate the optical distortion. The optical distortion calibration process is described in Section 4.3.3.2. The 0.5 hours focal plane calibration frames will be taken after the optical distortion calibration mode. The focal plane calibration frames consist of 10 minutes one single mode fiber Gaussian pattern exposures for the flat fielding and 20 minutes interference fringe exposures for the pixel position calibration. The focal plane calibration is described in Section 4.3.3.1.

4.3.3.1 Calibration of focal plane array

\begin{figure}
   \centering
   \includegraphics[width=8cm]{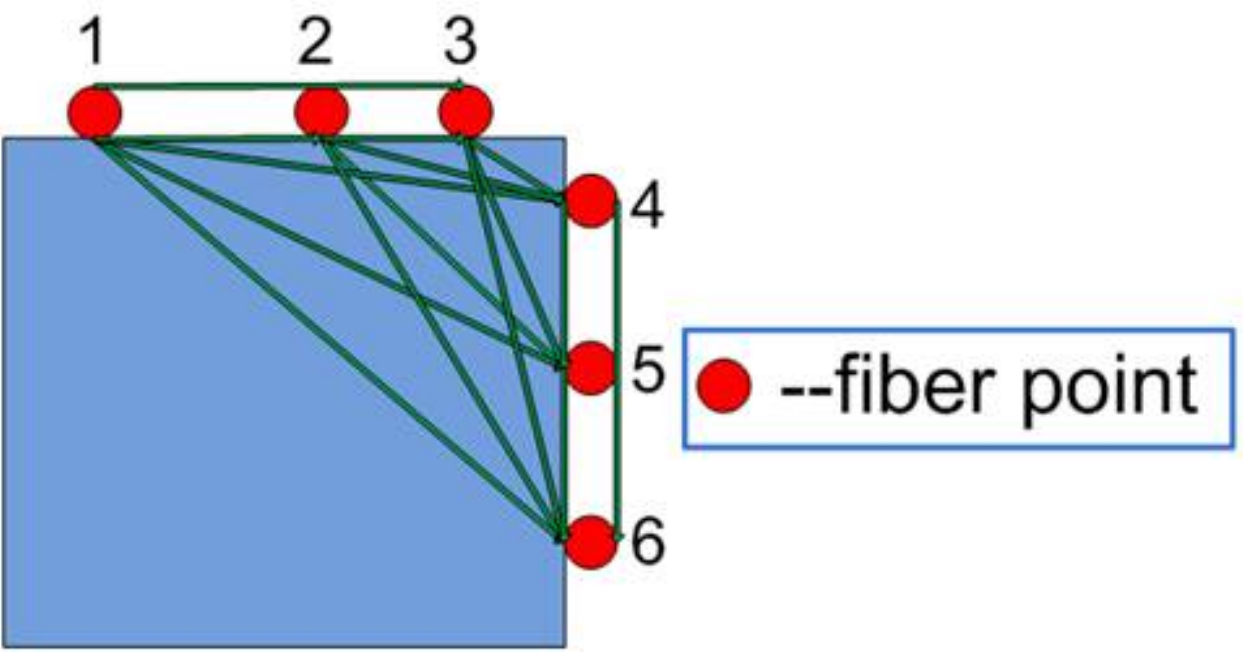}
   \caption{The baselines of the FPA calibration module.}
   \label{fig:4-9}
   \end{figure}

The focal plane array calibration subsystem in orbit consists a module of heterodyne laser interferometry. This module is mainly used to calibrate the non-uniformity of the pixel positions and inter-/intra-pixel response of the scientific CMOS detector. With the Fourier re-sampling algorithms and the calibrated characteristics of the scientific CMOS detector, the micro-pixel level relative centroiding can be reached on the FPA (Figure \ref{fig:4-9}). It meets the requirements of CHES mission for the $1~\mu$as measurement accuracy of the angular distance between the target star and the reference star.

   \begin{figure}
   \centering
   \includegraphics[width=12cm]{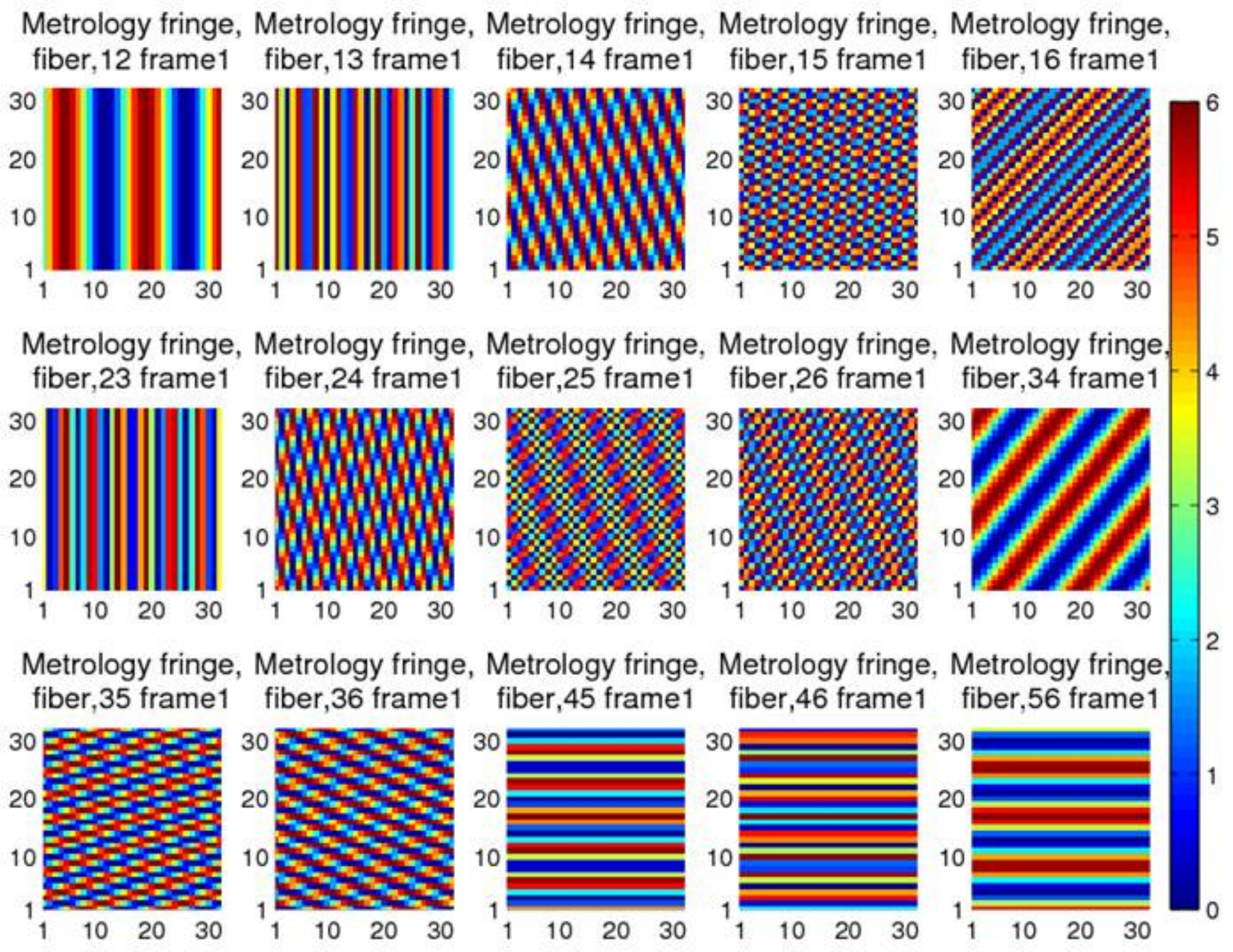}
   \caption{15 sets of interference fringes with various orientations and spaces.}
   \label{fig:4-10}
   \end{figure}

The CMOS detector calibration module is mainly composed of six optical fibers. A total of fifteen sets of baselines can be obtained by combining them in pairs, so that fifteen sets of interference fringes with different orientations and spaces can be formed. These fringes are utilized to calibrate CMOS detectors. The interference baselines and interference fringes are shown in Figure \ref{fig:4-10}.

\begin{figure}
		\begin{subfigure}{.5\textwidth}
			\centering
			\includegraphics[width=\textwidth]{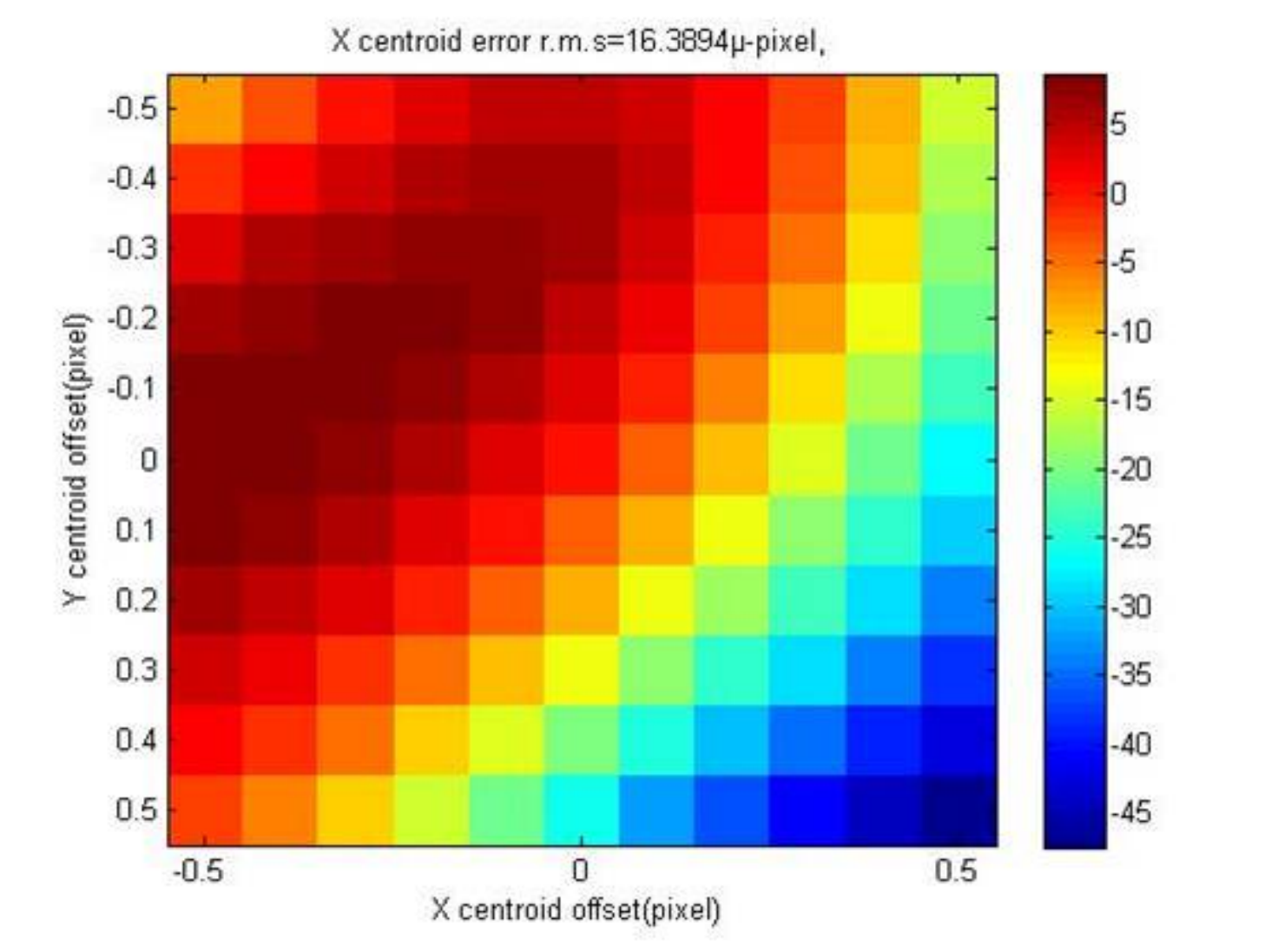}
		\end{subfigure}
		\begin{subfigure}{.5\textwidth}
			\centering
			\includegraphics[width=\textwidth]{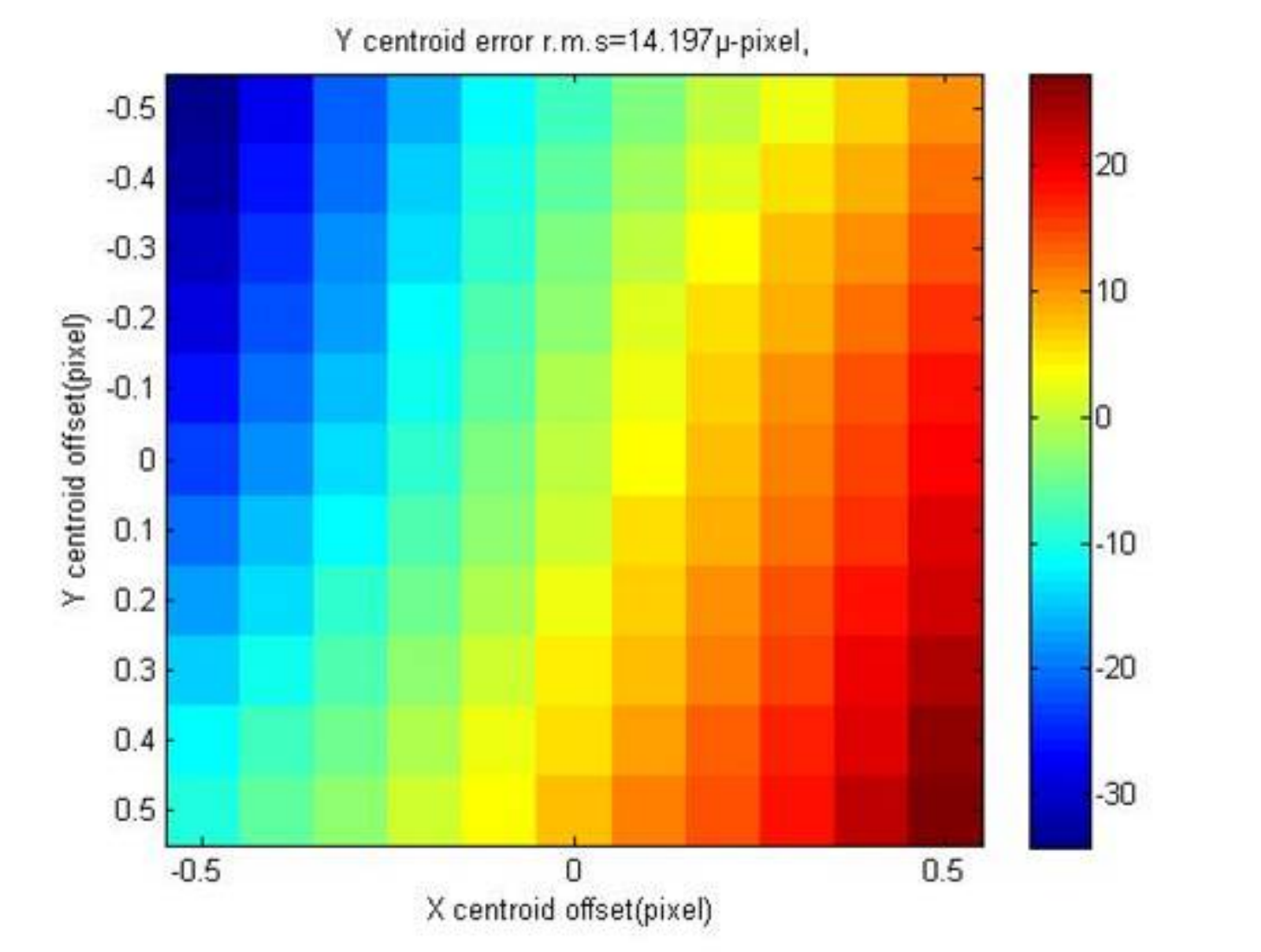}
		\end{subfigure}
				\caption{The centroid displacement measurements accuracy between the target star and reference star for X and Y directions.}
		\label{fig:4-11}
	\end{figure}

For an actual detector, the numerical simulation results exhibit that the centroid displacement measurements accuracy between the target star and reference star can amount to 20 micro-pixel if we calibrate the higher order terms of the Taylor series of the Fourier transform of the pixel response function, as shown in Figure \ref{fig:4-11}.

\begin{figure}
   \centering
   \includegraphics[width=12cm]{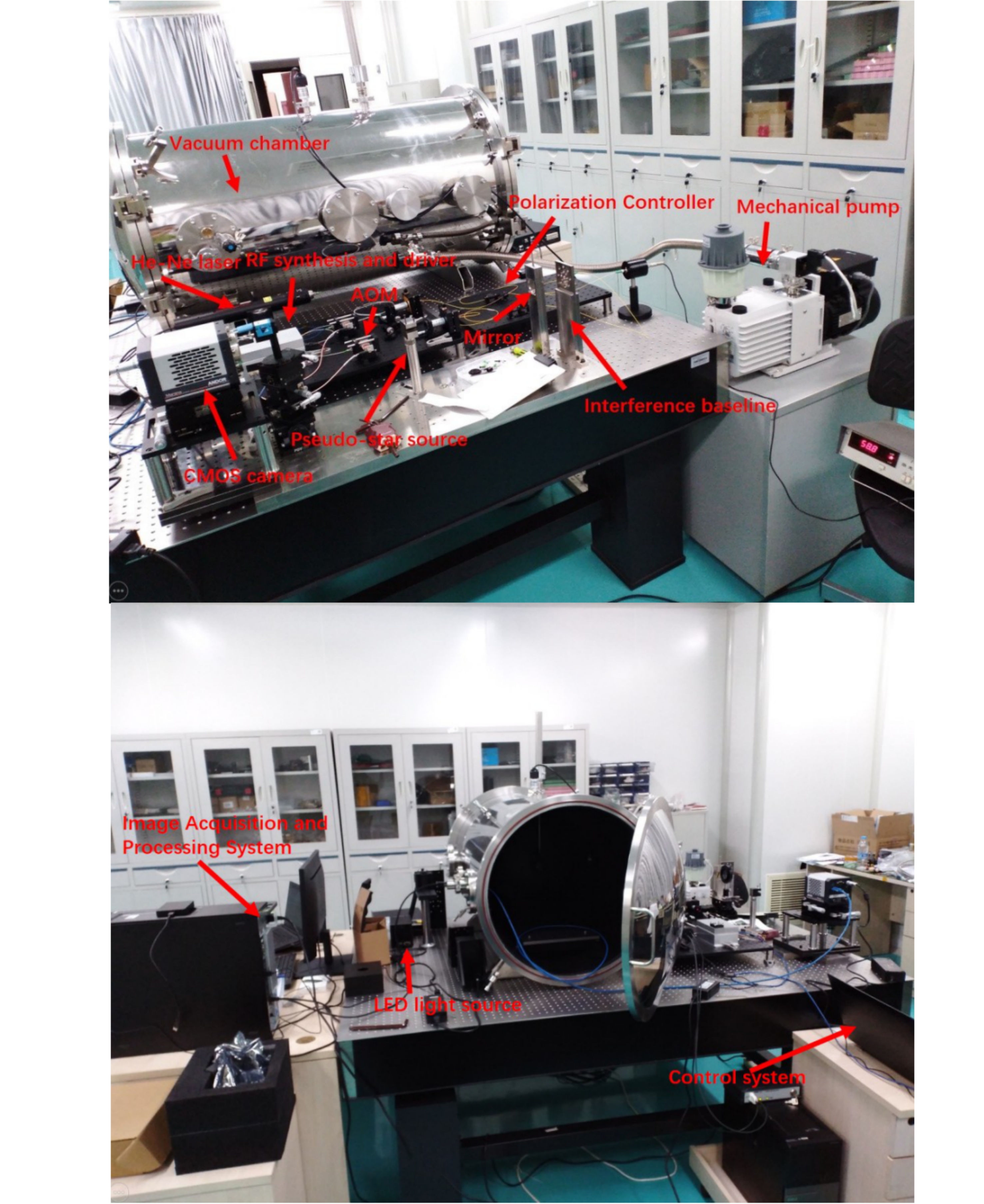}
   \caption{Testbed for heterodyne laser interferometry and micro-pixel centroid displacement measurement.}
   \label{fig:4-12}
   \end{figure}

\begin{figure}
   \centering
   \includegraphics[width=12cm]{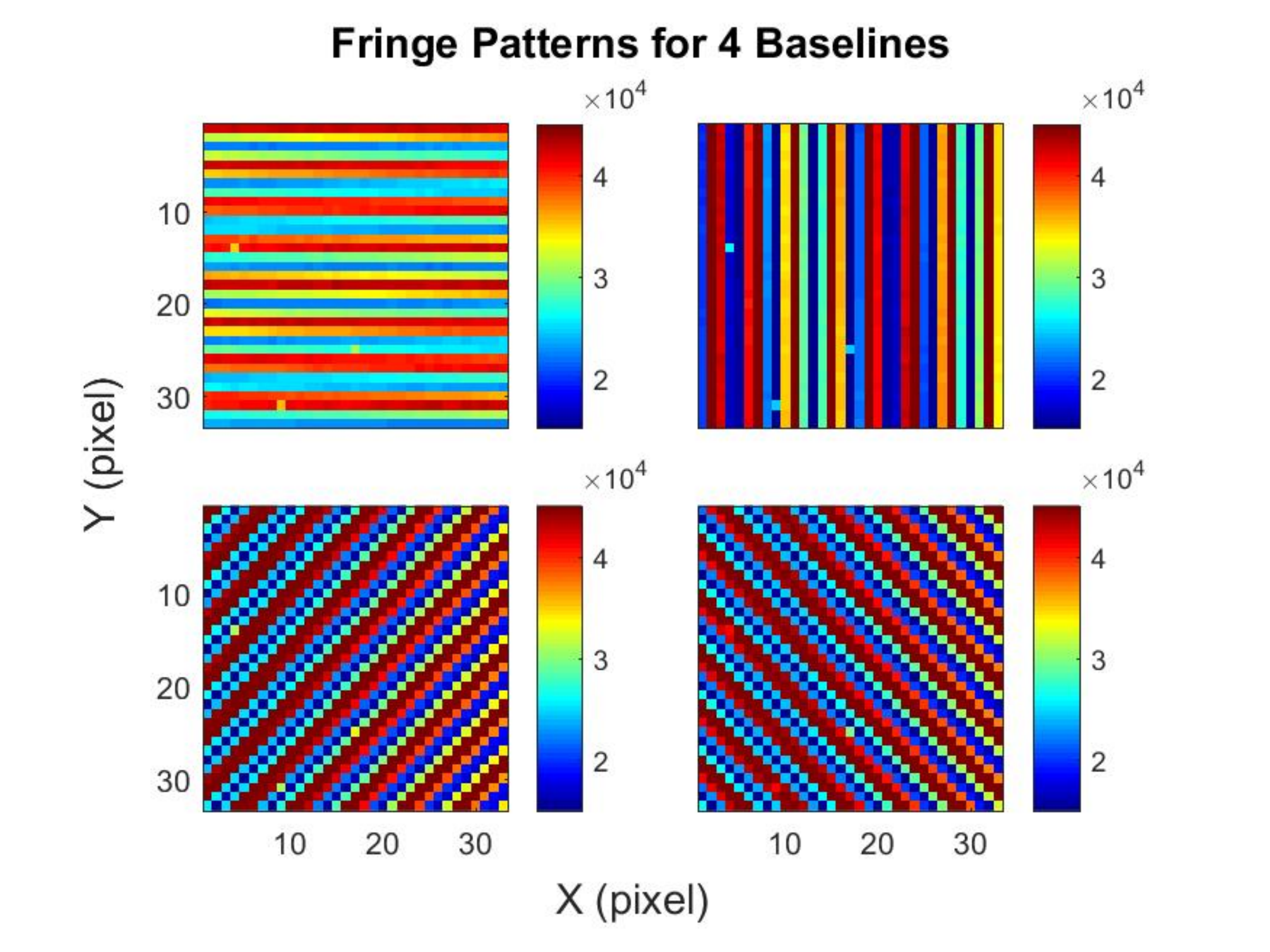}
   \caption{The interference fringe images from the testbed for four different baselines.}
   \label{fig:4-13}
   \end{figure}

\begin{figure}
   \centering
   \includegraphics[width=12cm]{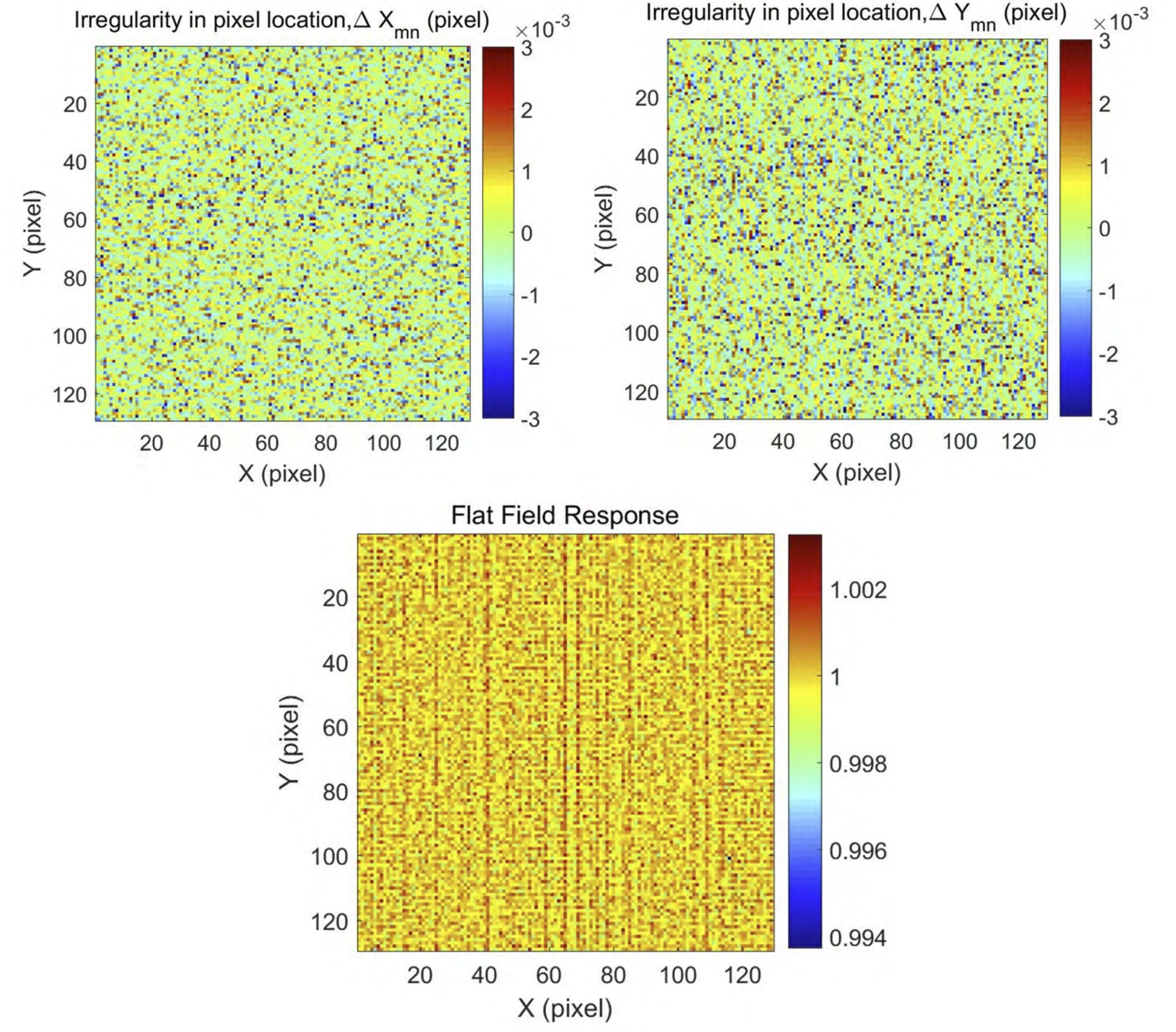}
   \caption{Results of flat field calibration and the pixel position measurements.}
   \label{fig:4-14}
   \end{figure}

\begin{figure}
   \centering
   \includegraphics[width=8cm]{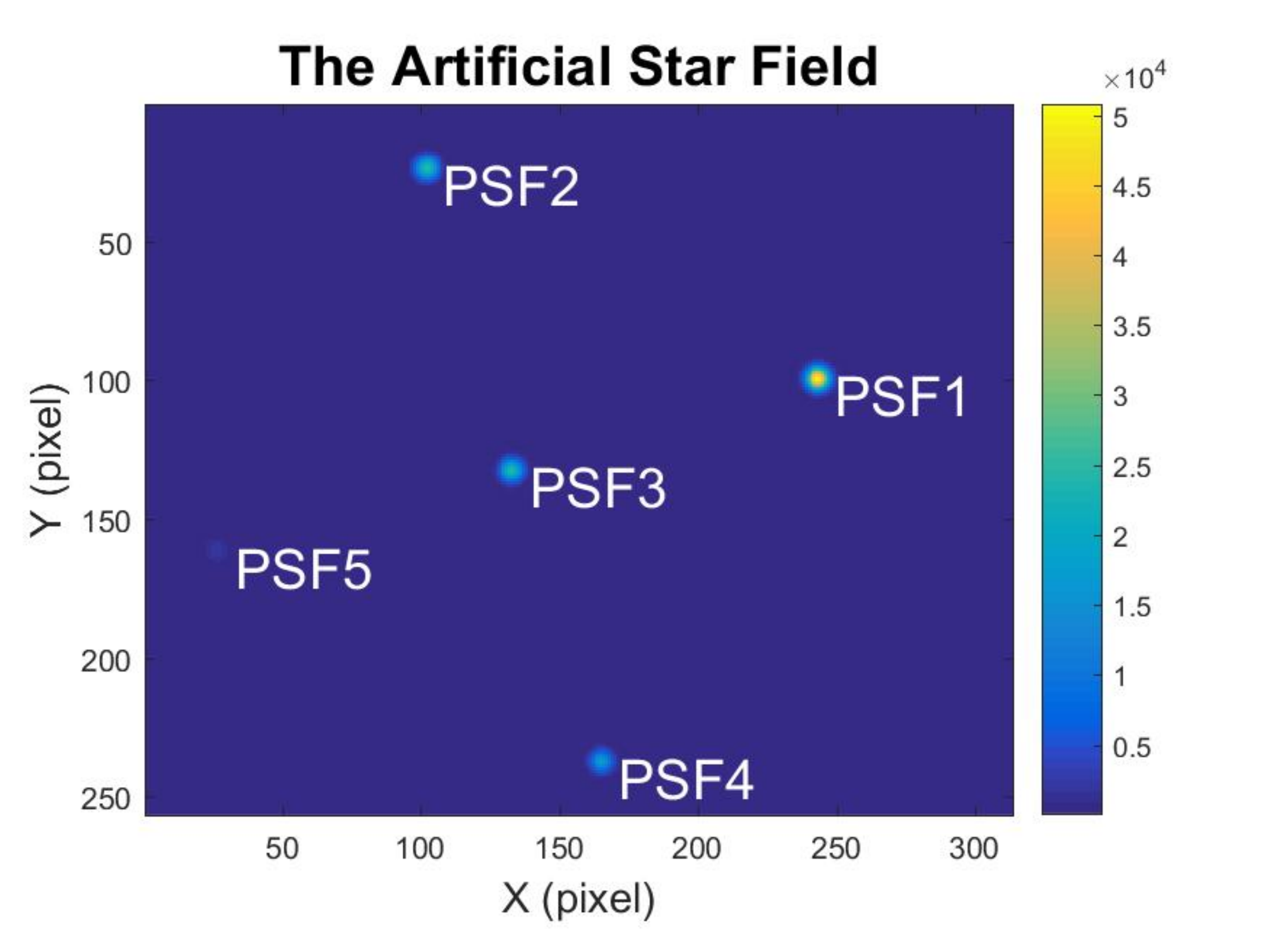}
   \caption{The artificial star field from testbed.}
   \label{fig:4-15}
   \end{figure}

\begin{figure}
   \centering
   \includegraphics[width=8cm]{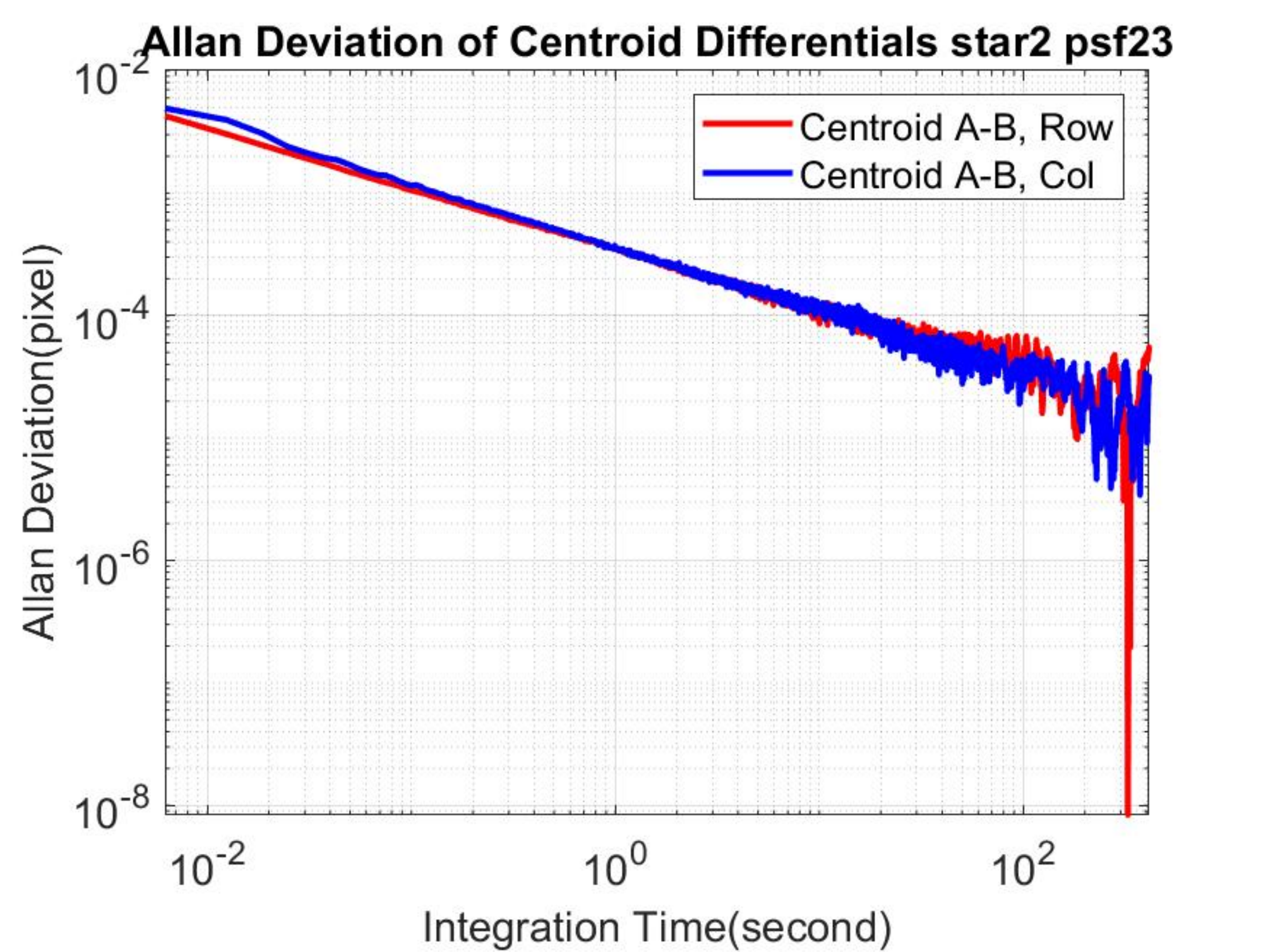}
   \caption{The results for centroid displacement measurements.}
   \label{fig:4-16}
   \end{figure}

We built an heterodyne laser interferometry and micro-pixel accuracy centroid displacement measurement testbed (Figure \ref{fig:4-12}), called LIMT (Laser Interferometric Metrology Testbed). We have conducted the flat-field response calibration, pixel position calibration, and micro-pixel centroid displacement estimation with laboratory demonstration based on scientific-grade CMOS image sensors (Figs \ref{fig:4-13}-\ref{fig:4-15}). At present, the centroid displacement measurement accuracy between the target star and reference star (in the artificial star field) can achieve 2 $\times~10^{-5}$ pixel (Figure \ref{fig:4-16}) with flat-field response calibration and pixel position calibration.

4.3.3.2 Calibration of telescope optical distortion

Static and temporal varying shape errors of the primary mirror will cause a shift on the centroid of the star image, but this shift is a common mode over the full FOV. For relative astrometry, its impact on measurement accuracy is small and negligible \citep{Tan2022}. However, temporal variation of the shape of mirrors after the primary mirror would produce non-common centroid shifts. Such error will dramatically affect the estimation accuracy. Therefore, the measurement error rising from the variation of the mirror surface shape needs to be calibrated through on-orbit calibration subsystem.

Multiple independent interferometers can be used to monitor the relative position shift between two mirrors in the telescope, but the monitoring system is extremely complicated and cannot detect the variation of the mirror surface shape. Therefore, there is no telescope metrology in CHES payload, but a method to calibrate the star image position affected by the distortion, imaging quality and other relevant factors is proposed and employed. The calibration accuracy arrives at the order of micro-arcsecond.

The calibration method does not depend on the absolute position coordinates of the stars in the current stellar catalogue, instead of using micro-pixel accuracy centroid displacement measurement technology and the characteristic of zero distortion of the center FOV of the optical system. Based on the principle of wavefront reconstruction in adaptive optics, the on-orbit calibration of the telescope optical distortion with micro-arcsecond accuracy is realized.

The calibration process is as follows: First, make the pointing direction of telescope change only in yaw and pitch, and use the micro-pixel accuracy centroid displacement measurement technology to obtain the displacement of each star image at the center FOV and edge FOV. Then, the displacements of the star images in the edge FOV and the center FOV in the $\mathrm{x}$ and $\mathrm{y}$ directions are subtracted to obtain the distortion difference and distortion gradient at each star image. The gradient is fitted to obtain the coefficients of the terms in the expression of the distortion gradient function. Finally, the expression of the distortion gradient function is integrated. According to the fact that the distortion at the central FOV of the telescope is zero, the constant term of the expression can finally obtain and the distortion function is determined, so as to realize the high-precision calibration of the optical distortion of the telescope.

In order to verify the feasibility of the proposed calibration method, a calibration simulation experiment was carried out on the CHES optical system derived from the optical design software. Preliminary results show that, assuming that the measurement accuracy of the centroid displacement measurement is 1 $\mu$as, the average value of the distortion residual after calibration is about 1.4 $\mu$as.

\begin{figure}
   \centering
   \includegraphics[width=8cm]{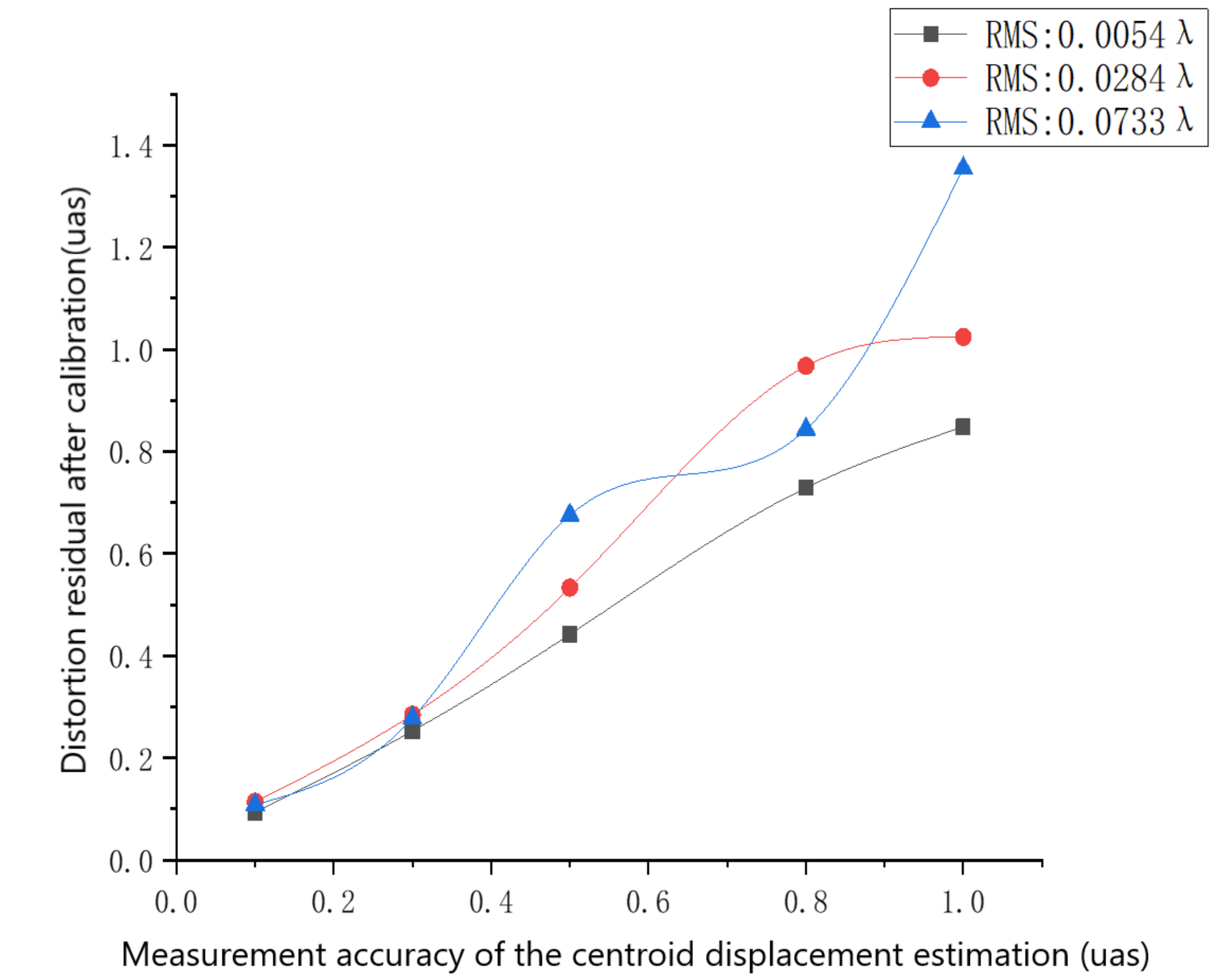}
   \caption{The variation of distortion residual after calibration.}
   \label{fig:4-17}
   \end{figure}

Furthermore, we give a graph of the variation of the distortion calibration accuracy with the centroid displacement measurement accuracy under different telescope wave aberration conditions (Figure \ref{fig:4-17}). It can be seen that the calibration accuracy is comparable to the centroid displacement measurement accuracy. If the centroid displacement measurement accuracy reaches sub-micron arcseconds, the calibration accuracy can also arrive at sub-micron arcseconds, which meets the requirements of calibration accuracy for CHES mission.

\subsection{TRL assessment}
\begin{figure}
   \centering
   \includegraphics[width=8cm]{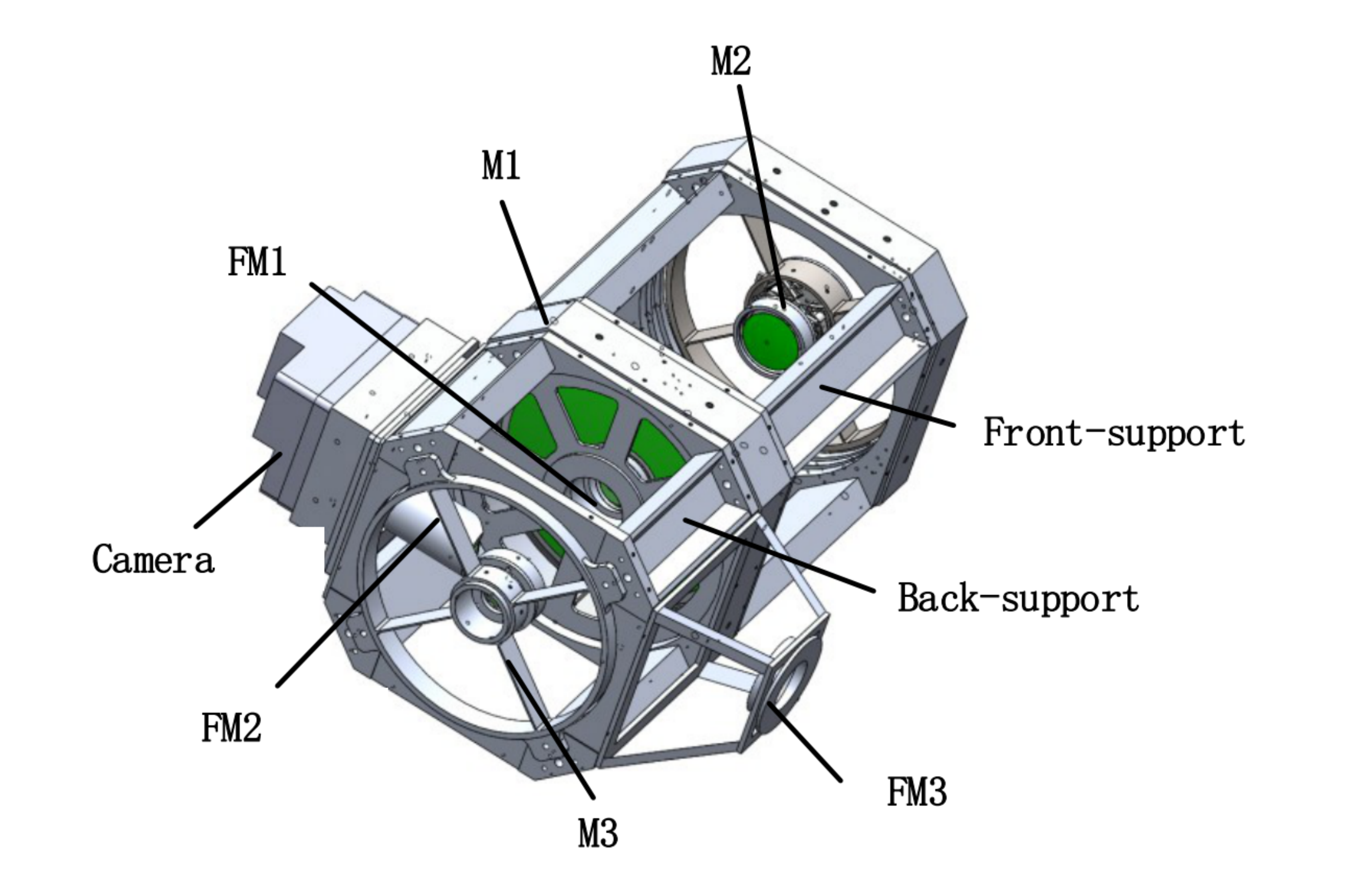}
   \caption{Structure diagram of scale prototype of the telescope.}
   \label{fig:4-18}
   \end{figure}

\begin{figure}
   \centering
   \includegraphics[width=8cm]{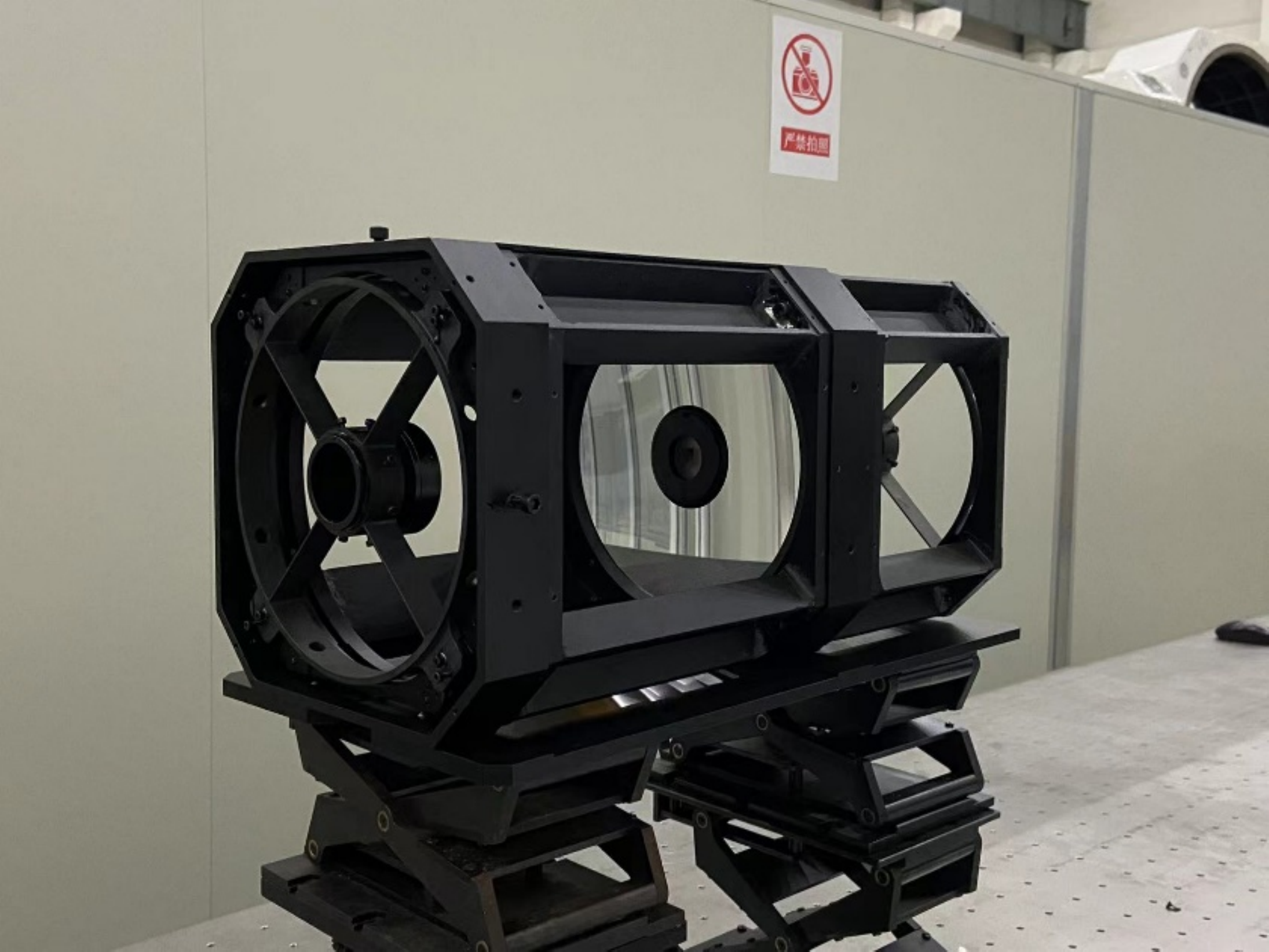}
   \caption{Completed 1/6 scale prototype of the telescope.}
   \label{fig:4-19}
   \end{figure}

CHES telescope is based on traditional optical telescope technology, and the key technologies of large-diameter high-image quality telescopes and focal plane splicing can be inherited in the previous programs. To verify the optical performance of the telescope and perform the experimental verification of the micro-pixel accuracy centroid displacement measurement  on the telescope optical system, a $1 / 6$ scale prototype of the telescope has been developed. The scaled prototype bears the same optical structure as the full-aperture of CHES telescope, and has the identical optical parameters such as F$\#$, FOV, imaging quality, and distortion. The scaled prototype of telescope includes the main mirror FM1, the secondary mirror FM2, the third mirror FM3, the front-support, the back-support, and the camera (Figure \ref{fig:4-18}). The overall structure adopts a cuboid structure, and the outer shell is employed to avoid the influence of external stray light. Figure \ref{fig:4-19} displays the completed $1 / 6$ scale prototype of the telescope.
\begin{table}
\begin{center}
\caption[]{Test results of wave aberration in different FOV of the scale prototype.}\label{Tab4.4}
\begin{tabular}{|c|c|c|c|c|c|}
\hline FOV & $\left(0^{\circ},+0.22^{\circ}\right)$ & $\left(-0.22^{\circ}, 0^{\circ}\right)$ & $\left(0^{\circ}, 0^{\circ}\right)$ & $\left(+0.22^{\circ}, 0^{\circ}\right)$ & $\left(0^{\circ},-0.22^{\circ}\right)$ \\
\hline RMS & $\sim \lambda / 14$ & $\sim \lambda / 19$ & $\sim \lambda / 26$ & $\sim \lambda / 14$ & $\sim \lambda / 13$ \\
\hline
\end{tabular}
\end{center}
\end{table}

The wave aberration of the full FOV of the scaled telescope prototype is tested, and the test results are shown in Table \ref{Tab4.4}. It can be seen that the wave aberration RMS of each field of view is better than $\lambda / 13$. Although only a scaled prototype is completed, our payload research team is very experienced in the micro-stress fixation of large-aperture telescope mirrors and the control of surface shape changes. We have successfully developed a meter-scale large-aperture space-based telescope and on-orbit performance is good. According to the standards for classification and definition of Technology Readiness Level (TRL), the TRL of CHES telescope technology can attain level 7 (Table \ref{Tab4.5}).

\begin{table}
\begin{center}
\caption[]{Payload TRL assessment.}\label{Tab4.5}
\begin{tabular}{|c|c|l|}
\hline \multicolumn{1}{|c|}{ Item or function } & TRL & \multicolumn{1}{|c|}{ Heritage or comment } \\
\hline \multirow{2}{*}{Telescope structure} & \multirow{2}{*}{7} & Rich experience in the development of\\
& & large-aperture space-based telescopes; \\
\hline \multirow{2}{*}{Telescope optics} & \multirow{2}{*}{5} & Mature TMA optical structure;\\
& & Scaled prototype has been completed \\
\hline MOSAIC detectors & 6 & Chinese space station survey telescope \\
\hline \multirow{2}{*}{On-orbit calibration} & \multirow{2}{*}{5} & Laboratory validated, but not to CHES FPA\\
& & scale. \\
\hline
\end{tabular}
\end{center}
\end{table}

In terms of MOSAIC detectors technology, our payload research team is participating in the development of the focal plane array of the Chinese space station survey telescope. The effective physical size of the focal plane is about $560 \mathrm{~mm} \times 490 \mathrm{~mm}$, and the focal plane area is about 4 times that of the CHES telescope. At present, the key technologies such as low-temperature flatness detection, self-gravity surface shape variation detection, and focal plane surface shape control have been completed, and the surface shape accuracy of the mosaic focal plane meets the requirements of the project. The related technology can be inherited and applied to the focal plane assembly of the CHES telescope.
The TRL for MOSAIC detectors can reach level 6.

An on-orbit calibration testbed was built under vacuum environment and the experiment of the focal plane array calibration was carried out. The calibration of micro-pixel accuracy was laboratory validated. Here the TRL can attain level 5.

\section{PROPOSED MISSION CONFIGURATION AND PROFILE}
\subsection{Mission Requirements and analysis}
\subsubsection{Mission requirements}

The CHES mission will detect the habitable-zone terrestrial planets orbiting the nearby solar-type stars with the aid of relative astrometric method. Hence, here we design the satellite system with ultra-high attitude stability and thermal control accuracy to meet the requirements of ultra-high-precision centroid displace measurement between the target star and reference stars. The satellite adopts a 3-axis stabilization zero-momentum attitude control mode, combining telescope payload, star sensor, gyroscopic to determine the attitude, and with super-stable structure and thermo-elastic control to achieve high accurate $(0.07$ arcsecond) and high stability attitude control (0.0036 arcsecond / 0.02sec); the satellite adopts a compartmentalized thermal design and a combination of active and passive thermal control mode. It also adopts $X$-band measurement, control and data transmission integrated design, equipped with $512 \mathrm{~Gb}$ large-capacity storage and phased array antenna, to realize the daily $84 \mathrm{~Gb}$ of scientific data transmitted at $20 \mathrm{Mbps}$ information rate. The total weight of the satellite when launch is about $2.93~\mathrm{t}$, with a peak power consumption of $1493 \mathrm{~W}$, and is planned to be launched at Xichang using CZ-3C rocket.

\subsubsection{Mission Orbit}
As a satellite for the nearby habitable planet detection using micro-acrsecond relative astrometric measurement, the CHES satellite will have extremely high requirements on attitude stability and thermal control accuracy. By analyzing the solar-terrestrial Lagrange point orbit, geocentric HEO orbit, Earth-Moon resonance orbit, and heliocentric Earth trailing orbit, we find that the solar-terrestrial L2 point orbit has many advantages including free of the Earth's gravity gradient, which enables the all-day scientific observations, and relatively stable thermal radiation environment. Moreover, China has the capability of delivery, measurement and control, and data transmission to carry out the L2 mission. Thus, the solar-terrestrial L2 orbit is designed as the mission orbit.

\subsubsection{Space environment analysis}
The CHES satellite operates at the solar-terrestrial L2 point where the space environment is mainly the interplanetary space environment, which mainly consists of high-energy radiation, solar wind plasma and ultraviolet radiation, etc. The main effects of these on the satellite are the following four aspects.

(1) Single-particle effect analysis: the single event upset rate (SEU) rate is slightly higher than that of geosynchronous orbit, and the SEU rate rises by more than 1 order of magnitude during proton events, requiring attention to the impact of single-particle effect occurring in the device.

(2) Total dose analysis: The dose received under the $3 \mathrm{~mm}$ aluminum shield is between $5.5 \mathrm{Krad} \sim 9.7 \mathrm{Krad}$ in 5 years.

(3) Displacement damage effect analysis: the impact on CMOS detectors should be considered, specific devices need to be shielded and reinforced according to the ability to withstand.

(4) Surface charging effect analysis: The test results show that the charging effect has little impact on the safety of CHES satellites.

\subsubsection{Thermal environment analysis}
The CHES satellite, at the solar-terrestrial L2 point, is mainly influenced by solar radiation, as well as thermal radiation from the Earth and the Moon. (1) Solar radiation: the solar constant (the value of the energy flow density in the visible band received per unit area perpendicular to the direction of sunlight per unit time) at point L2 ranges from $1296 \mathrm{~W} / \mathrm{m}^{2}$ to $1389 \mathrm{~W} / \mathrm{m}^{2}$. (2) Earth and lunar thermal radiation: the sunlight albedo energy flow of the Earth is around $0.0004 \mathrm{~W} / \mathrm{m}^{2}$, and the sunlight albedo energy flow of the Moon is not higher than $0.0002 \mathrm{~W} / \mathrm{m}^{2}$. The infrared radiation energy flow of the Earth and the Moon does not exceed $0.005 \mathrm{~W} / \mathrm{m}^{2}$, which has little effect on the satellite.

\subsubsection{Access analysis}
CHES operates at the solar-terrestrial L2 point. Due to the ultra-high requirement of attitude stability, the satellite data transmission adopts X-band phased array antenna. We can use Kashgar $35 \mathrm{~m}$ deep-space station, Argentina $35 \mathrm{~m}$ deep-space station, Jiamusi $66 \mathrm{~m}$ deep-space station, and $40 / 50 \mathrm{~m}$ equipment of Miyun station of Chinese Academy of Sciences for the data reception.

(1) The budget of measurement and control link

CHES locates at the solar-terrestrial L2 point $1.7$ million km from Earth. The satellite uses a wide-beam antenna, and the ground uses a $35 \mathrm{~m}$ deep-space measurement and control station. The link margin is $3.05 \mathrm{~dB}$ at $2048 \mathrm{bps}$ telemetry rate and $13.52 \mathrm{~dB}$ at $2000 \mathrm{bps}$ remote control rate.

(2) The budget of digital transmission link

Take the $35 \mathrm{~m}$ Kashgar deep space station with the most demanding reception index as an example, the farthest distance between the star and the ground is $1.7$ million $\mathrm{km}$, the data transmission rate is $20 \mathrm{Mbps}$, and the link margin is $3.34 \mathrm{~dB}$ when the phased array antenna is used on the star for downlink data transmission.

(3) Visibility analysis

By setting the minimum elevation angle of the ground station antenna at $5^{\circ}$, the coverage arc of CHES satellite passing through the ground station is simulated and analyzed. The visibility time between the satellite and the three ground stations of Jiamusi, Kashgar and Miyun is between $7.9 \sim 15.3$ hours. The scientific data is about $84 \mathrm{~Gb}$, the information rate is $20 \mathrm{Mbps}$, and the daily data transmission takes about $1.2$ hours, which meets the satellite data downlink demand.

\subsection{Mission Profile}
\subsubsection{Launcher and orbit injection}
5.2.1.1 Launcher

This mission can be launched to the initial orbit using CZ-3C at Xichang Satellite Launch Center. (1) initial orbit requirements: the launch vehicle sends the satellite to a parking orbit with an orbital inclination of $28.5^{\circ}$, a perigee altitude of $200 \mathrm{~km}$ and an apogee altitude of $35958 \mathrm{~km}$; (2) launch capacity requirements: $\geq 2930 \mathrm{~kg}$ (CZ-3C@GTO has a carrying capacity of about 3.8t); (3) launch envelope size: $\geq 3761 \mathrm{~mm}$ x $6487 \mathrm{~mm}$ (including adapters).

5.2.1.2 Orbit injection

According to the mission analysis, the Halo orbit of the solar-terrestrial L2 point
with Az of 110,000 km is designed. The mission orbit entry parameters are shown in
Table \ref{Tab5.1}.

\begin{table}
\begin{center}
\caption[]{Mission orbit entry parameters.}\label{Tab5.1}
\begin{tabular}{|l|l|l|l|}
\hline \text{Orbit injection time} & \multicolumn{3}{|c|}{2025/6/1 00: 00: 00.000  \text{UTCG}} \\
\hline \text{Position-X} & -408023 $\mathrm{~km}$ & \text{Velocity X}  & 0.515501 $\mathrm{~km/s}$ \\
\hline \text{Position-Y} & -1132040 $\mathrm{~km}$ & \text{Velocity Y} & -0.174719 $\mathrm{~km/s}$\\
\hline \text{Position-Z} & -371616 $\mathrm{~km}$ & \text{Velocity Z} & -0.0768254 $\mathrm{~km/s}$\\
\hline
\end{tabular}
\end{center}
\end{table}

\begin{figure}
   \centering
   \includegraphics[width=8cm]{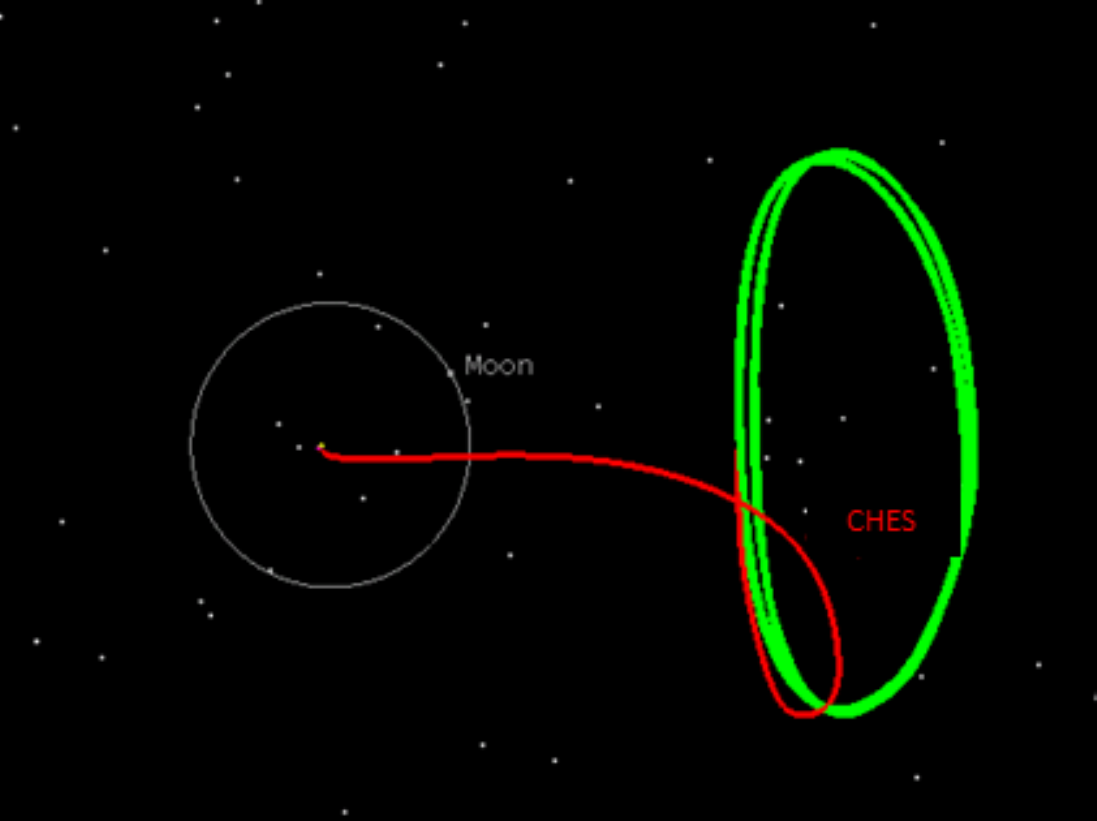}
   \caption{CHES mission orbit.}
   \label{fig:5-1}
   \end{figure}

The satellite is put into a parking orbit with an orbital inclination of $28.5^{\circ}(200 \mathrm{~km} \times$ $35958 \mathrm{~km}$ ) by the launch vehicle. The satellite maneuvers in the perigee of the parking orbit and transfer to the mission orbit. The satellite transfers to the Halo orbit along the invariant flow pattern, which take about 117 days. When the satellite approaches to the Halo orbit along the stable invariant manifold, a smaller velocity pulse needs to be applied to bring it into Halo orbit, as shown in Figure \ref{fig:5-1}. To summarize, the maneuvers during the transfer process include the transfer orbit entry maneuver and the Halo orbit insert maneuver.

Due to the long orbit transfer time of the satellite, it is also necessary to consider the midway correction maneuver to correct the orbit error and ensure the orbit accuracy in the actual flight. The correction required for this process is no more than $150 \mathrm{~m} / \mathrm{s}$. The maneuver required for each orbit maintenance in Halo orbit is about $2 \mathrm{~m} / \mathrm{s}$, and the orbit maintenance period of 90 days can meet the mission requirements. In addition, it is also necessary to consider the attitude control fuel consumption and orbital control design margin. The orbit control engine specific impulse is $315 \mathrm{~s}$, and the attitude control engine specific impulse is $215 \mathrm{~s}$. The satellite fuel is required to be $840 \mathrm{~kg}$.

   \begin{figure}
   \centering
   \includegraphics[width=8cm]{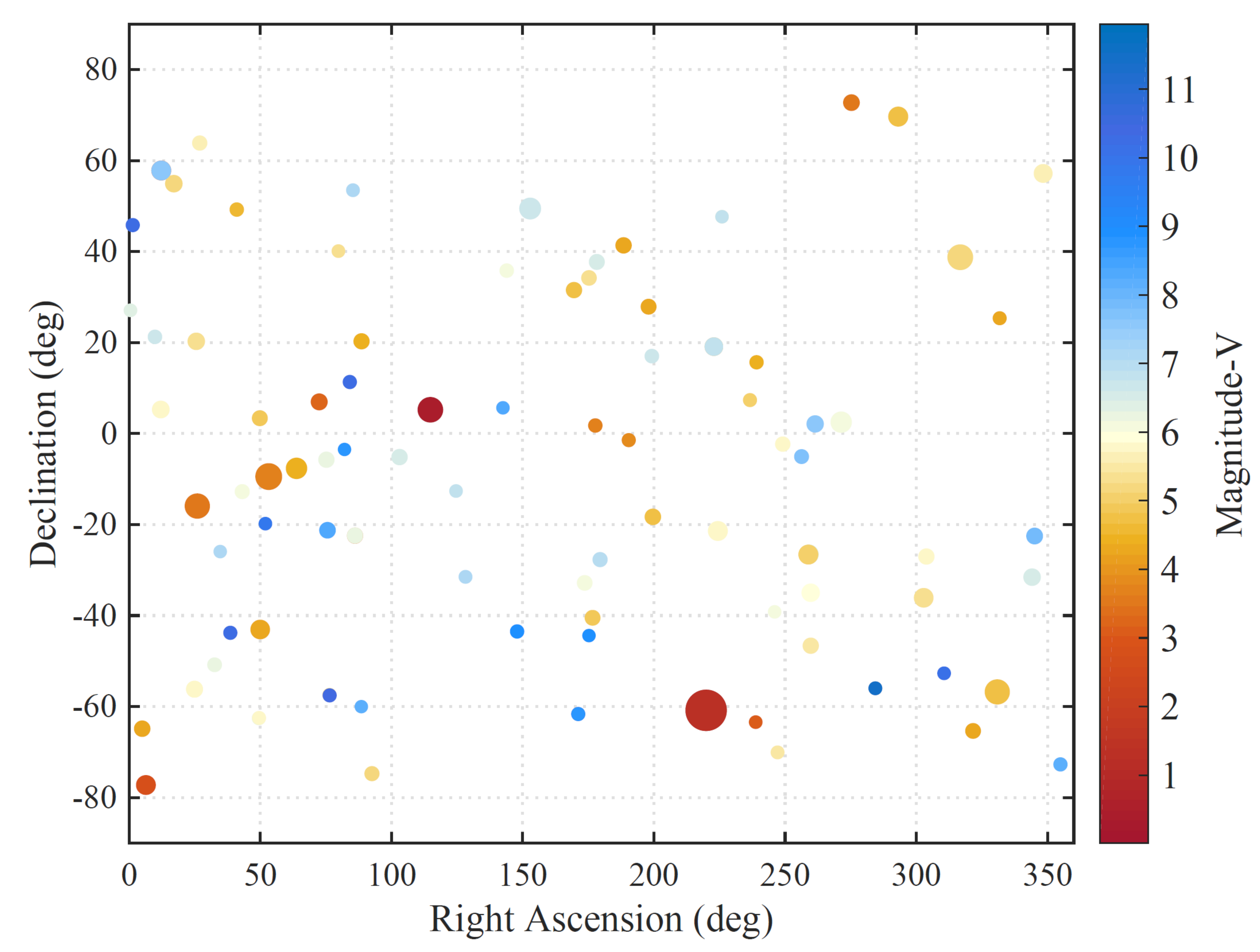}
   \caption{Target stars distribution.}
   \label{fig:5-2}
   \end{figure}

\subsubsection{Observing strategy}
5.2.2.1 Observing strategy design

As previously mentioned, the primary scientific objective of CHES satellite will focus on the discovery of Earth-like planets in the habitable zone around the stars within 10 pc, where Figure \ref{fig:5-2} exhibits the target stars under selection.

\begin{figure}
   \centering
   \includegraphics[width=12cm]{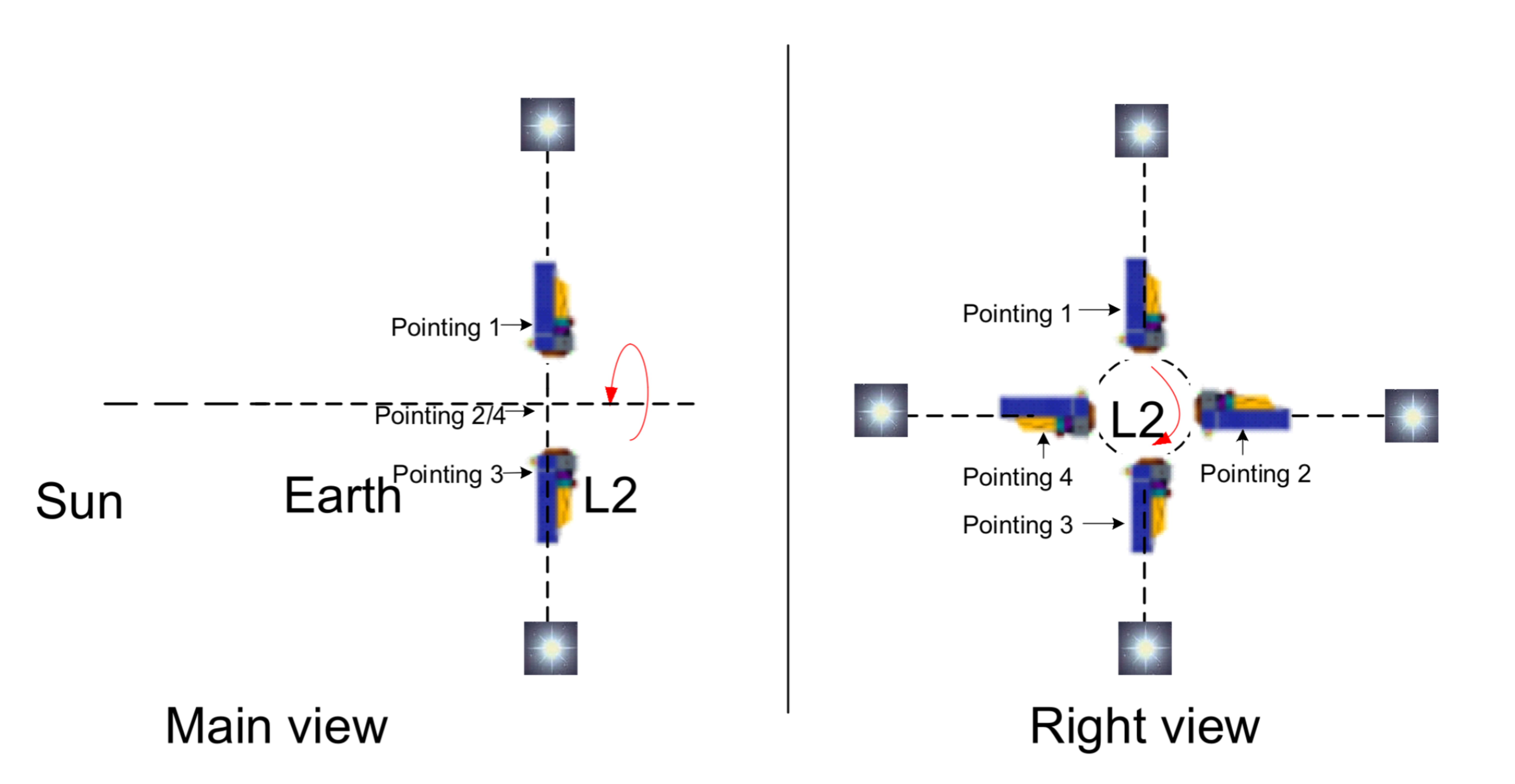}
   \caption{CHES conventional mode pointing diagram.}
   \label{fig:5-3}
   \end{figure}

\begin{figure}
   \centering
   \includegraphics[width=10cm]{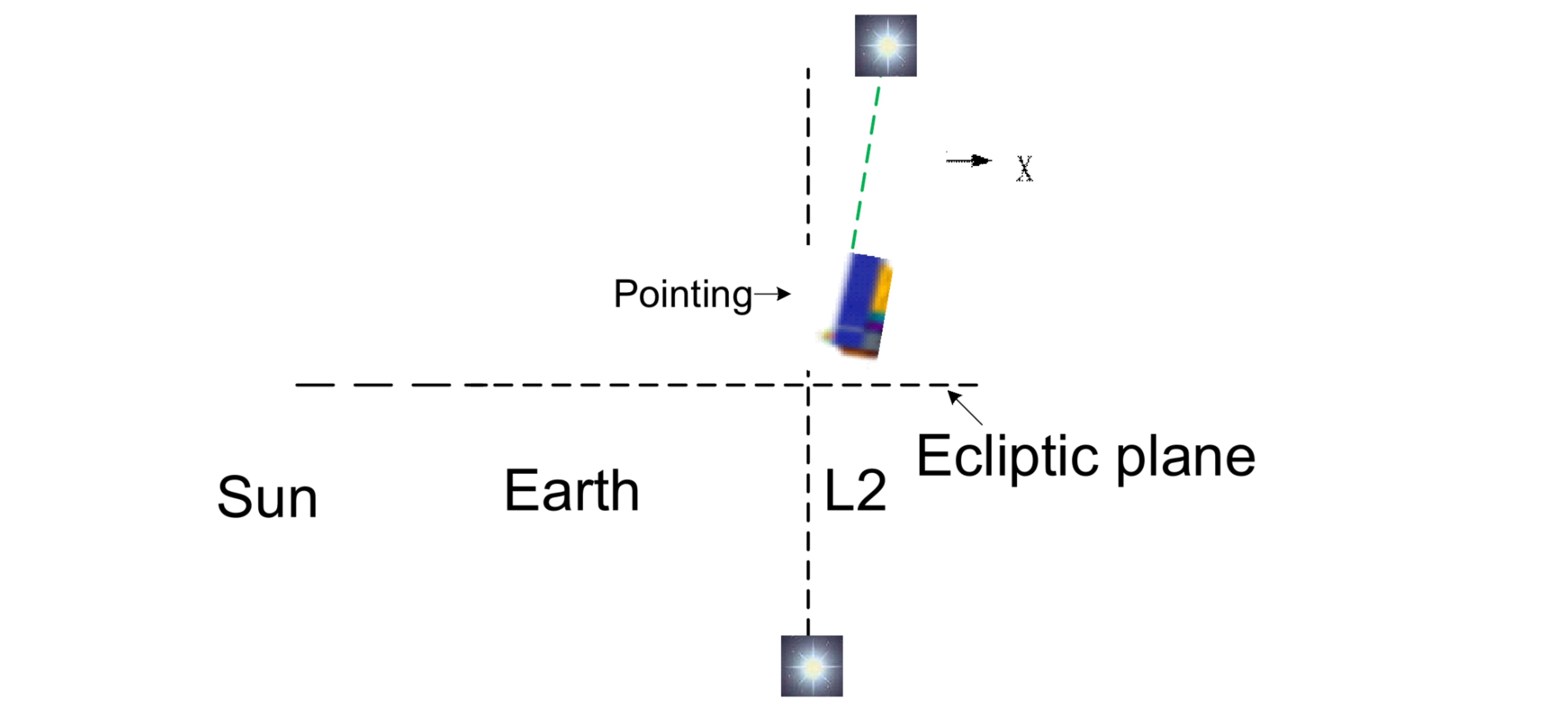}
   \caption{CHES revisited mode pointing diagram.}
   \label{fig:5-4}
   \end{figure}

According to the distribution of the target stars to be observed by CHES satellite and the requirement that the solar suppression angle of the telescope is not less than $60^{\circ}$, the observation modes include conventional pointing and revisit pointing. a. Conventional mode pointing (Figure \ref{fig:5-3}) : the telescope optical axis is perpendicular to the Sun-Earth direction and rotates around the Sun-Earth vector to observe the target stars on the coverage sky area in turn, which can reduce the number of satellite attitude maneuvers and improve the observation efficiency. b. Revisited mode pointing (Figure \ref{fig:5-4}): side-swing observations of the target stars on the left and right sides of the coverage sky area perpendicular to the Sun-Earth. When the Earth revolves around the Sun, all target stars can be covered in six months.

\begin{figure}
   \centering
   \includegraphics[width=12cm]{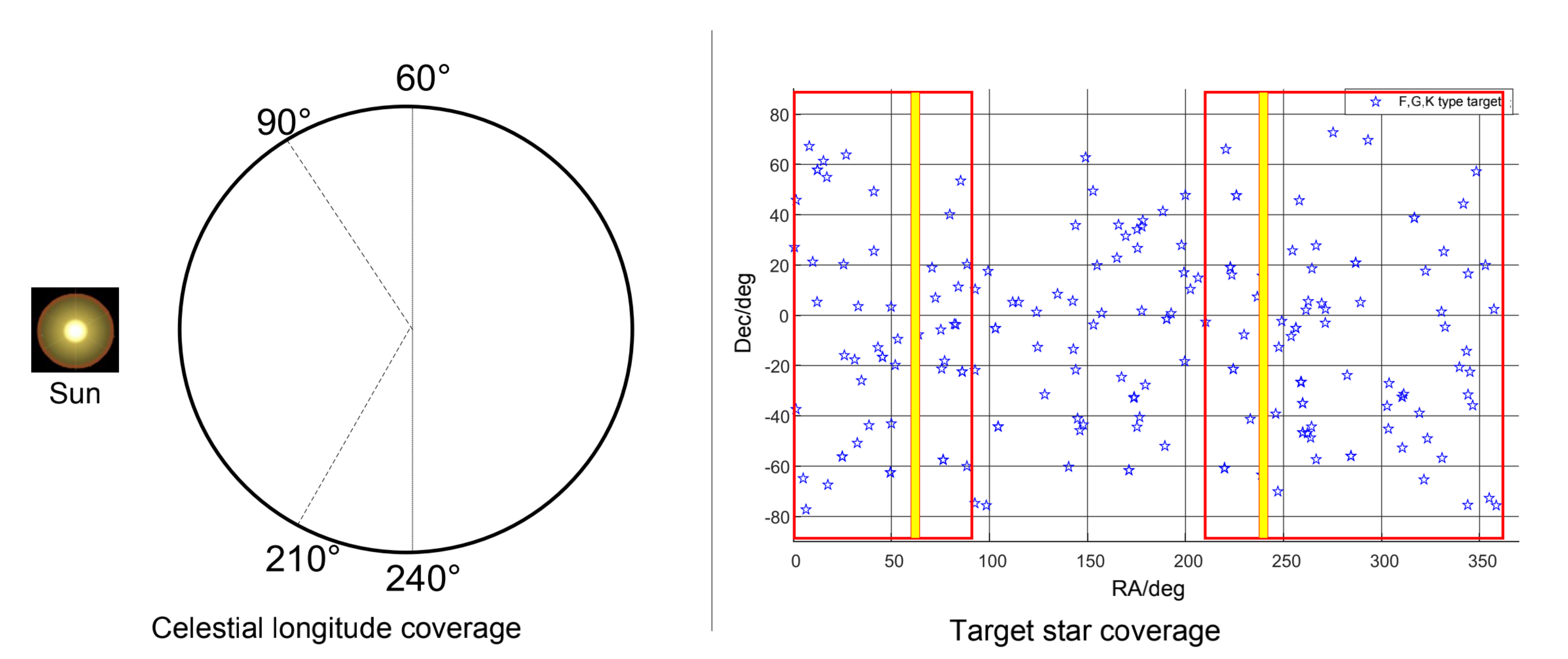}
   \caption{CHES observation mode diagram.}
   \label{fig:5-5}
   \end{figure}

\begin{figure}
   \centering
   \includegraphics[width=10cm]{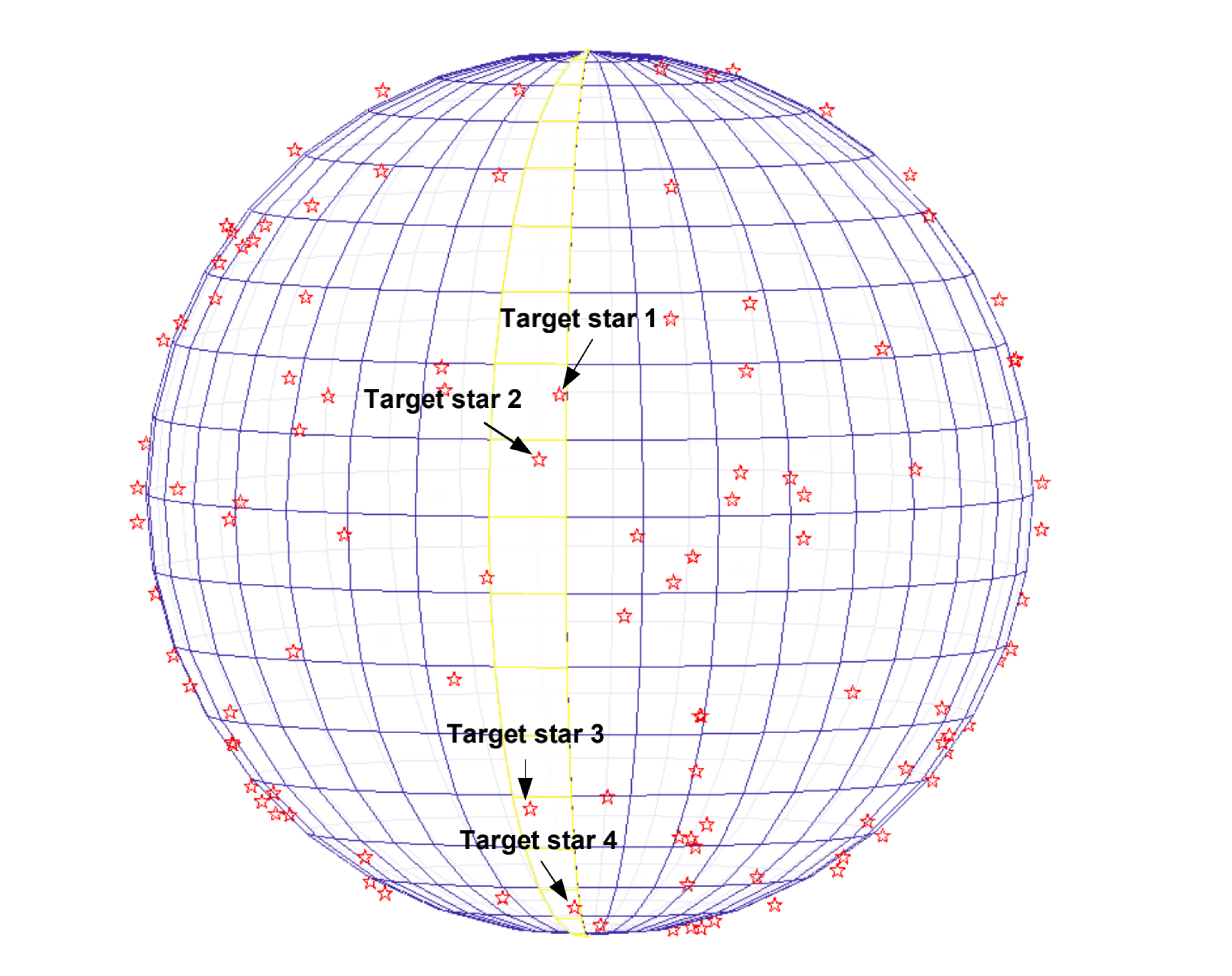}
   \caption{CHES observation sequence diagram (sequence for target stars is
$1->2->3->4$).}
   \label{fig:5-6}
   \end{figure}

Based on the pointing design, the observation mode analysis of CHES satellite in mission orbit to detect the target star is carried out. The observation mode of the satellite is described as an example of satellite observation of a target near $60^{\circ}$ longitude (Figure \ref{fig:5-5}). The satellite can observe the target near $60^{\circ}$ and $240^{\circ}$ longitude at the same time with the conventional pointing around the Sun-Earth vector, as shown in the yellow area on the right side of the observation mode schematic. The observation order is from top to bottom or from bottom to top, as shown in Figure \ref{fig:5-5}. In this case, considering that the solar rejection angle of the telescope is 60 degrees, the satellite can also cover the area of longitude from about 0 to $90^{\circ}$ and 210 to $360^{\circ}$ by the pendulum measurement, as shown in the red box on the right side of the observation mode schematic (Figure \ref{fig:5-6}).

5.2.2.2 Observing Efficiency Analysis

The main events of the CHES satellite in mission orbit include: scientific
observations, communication between the satellite and the ground station, charging of the battery, attitude maneuvering and high stability pointing attitude control for the satellite to observe different targets. The preliminary analysis allows for about 16 hours of scientific observation per day, so the total scientific observation time is about 29,200 hours during the 5-year lifetime.

\subsection{Spacecraft Preliminary Design}
\subsubsection{Spacecraft subsystems overview}
\begin{figure}
   \centering
   \includegraphics[width=16cm]{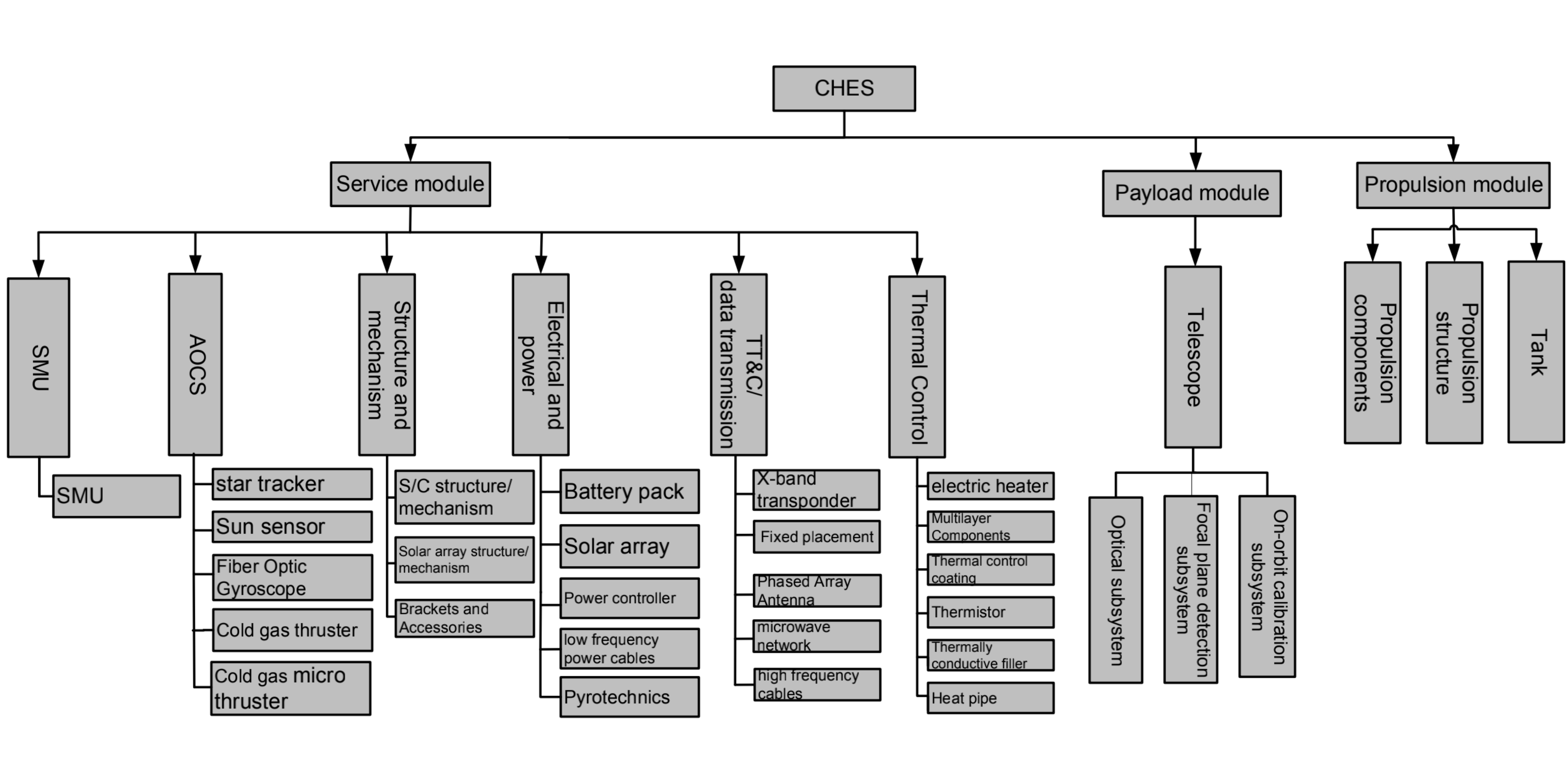}
   \caption{CHES composition.}
   \label{fig:5-7}
\end{figure}

CHES spacecraft consists of satellite platform and payload. The satellite platform mainly consists of structure, thermal control, electrical and power supply, TT$\&$C/ data transmission, attitude and orbit control (AOCS), satellite management unit (SMU), providing energy, communication, thermal control and another operational environment for the payload. The payload is an optical telescope. The composition of the CHES spacecraft is shown in Figure \ref{fig:5-7}. The main technical indicators of the satellite are given in Table \ref{Tab5.2}.

\begin{table}
\begin{center}
\caption[]{CHES main technical indicators.}\label{Tab5.2}
\begin{tabular}{|l|l|l|}
\hline \multicolumn{2}{|l|}{ Item } & Technical indicators \\
\hline \multirow{2}{*}{Mass} & \multirow{2}{*}{Launch} & $2930 \mathrm{~kg}$ ( CZ-3C@GT0 launch capability\\
& & about 3.8t ) \\
\hline \multicolumn{2}{|l|}{Size} & $\phi 3761 \mathrm{~mm} \times 6487 \mathrm{~mm}$ \\
\hline \multirow{6}{*}{Thermal control } & \multirow{2}{*}{Method} & Combination of active thermal control\\
& & mode and passive thermal control mode\\
\cline { 2 - 3 } & \multirow{2}{*}{Telescope optical system} & Working temperature $20 \pm 5^{\circ} \mathrm{C}$,\\
& & Temperature stability $45 \mathrm{mK}$ \\
\cline { 2 - 3 } & S/C platform,$\quad$ Other & \multirow{2}{*}{$-15 \sim+45^{\circ} \mathrm{C}$}\\
& parts of the telescope &  \\
\hline \multirow{4}{*}{ Power } & Solar cell array & $11.8 \mathrm{~m}^{2}$ triple junction GaAs battery array \\
\cline { 2 - 3 } & Battery & $120 \mathrm{Ah}$ Lithium ion battery \\
\cline { 2 - 3 } & Bus voltage & $30 \pm 0.5 \mathrm{~V}$ \\
\hline \multirow{3}{*}{ Attitude control } & Method & Three-axis stabilization \\
\cline { 2 - 3 } & Pointing accuracy & $0.07^{\prime \prime}$ \\
\cline { 2 - 3 } & Pointing stability & $0.0036^{\prime \prime} / 0.02 \mathrm{sec}$ \\
\hline \multirow{7}{*}{ Propulsion } & \multirow{5}{*}{ Thruster } & attitude control thruster:\\ & & $12 * (1 \mu \mathrm{N} \sim 50 \mu \mathrm{N}) ;$\\
& & Attitude control micro thruster: \\
& & $12 * 20 \mathrm{mN} ;$\\
& & orbit control engine: $490 \mathrm{~N}+12^{*} 10 \mathrm{~N} ;$ \\
\cline { 2 - 3 } & \multirow{2}{*}{Propellant} & Propulsion module: $840 \mathrm{~kg} ;$\\
& & Service module: $150 \mathrm{~kg} ;$ \\
\hline \multirow{2}{*}{ Telemetry } & Telemetry code rate & $512 \mathrm{bps}, 2048 \mathrm{bps}, 4096 \mathrm{bps}, 8192 \mathrm{bps}$ \\
\cline { 2 - 3 } & Telecontrol code rate & $500 \mathrm{bps}, 1000 \mathrm{bps} , 2000 \mathrm{bps}$ \\
\hline \multirow{5}{*}{ Data transmission} & Working frequency band & $\mathrm{X}$ \\
\cline { 2 - 3 } & Modulation mode & QPSK \\
\cline { 2 - 3 } & Information rate & $20 \mathrm{Mbps}$ \\
\cline { 2 - 3 } & Storage capacity & $512 \mathrm{Gbits}$ \\
\cline { 2 - 3 } & Reading and writing ways & Write in file order and read randomly \\
\hline \multirow{3}{*}{ SMU } & CPU & AT697 \\
\cline { 2 - 3 } & Basic frequency & $80 \mathrm{MHz}$ \\
\cline { 2 - 3 } & PROM & $128 \mathrm{KBytes}$ \\
\hline \multirow{4}{*}{Launch  vehicle and} & Connection and  & \multirow{2}{*}{Wrapping belt}\\
& separation mode & \\
\cline { 2 - 3 } spacecraft interface & LV and S/C connecting & \multirow{2}{*}{$\Phi 1194 \mathrm{~mm}$} \\
& ring &\\
\hline \multirow{2}{*}{Longevity and Reliability} & Longevity & 5 year \\
\cline { 2 - 3 } & Reliability & End of life better than 0.65 ( TBC ) \\
\hline
\end{tabular}
\end{center}
\end{table}

\subsubsection{System preliminary scheme overview}
5.3.2.1 Structure and accommodation

The design of CHES satellite adopts the idea of compartmentalized design. The whole satellite consists of three parts: the service module, the propulsion module and the payload module. The telescope payload is mounted on the optical reference plate and supported by carbon fiber support rods to form the payload module. The optical reference plate and the support rods are fastened, and the carbon fiber layup design is used to improve the force thermal stability of the payload module. The service module has a framed panel configuration with hexagonal columns for mounting most of the on-board units. The propulsion module is laid out below the service module. The load bay is connected to the service bay main frame by six titanium bearings to shorten the force transfer path. The two modules are insulated to improve the thermal stability of the payload during observation. The satellite contains one - $X$-plane, $\pm Y$-direction fixed solar wing, both insulated.

      \begin{figure}
		\begin{subfigure}{.5\textwidth}
			\centering
			\includegraphics[width=\textwidth]{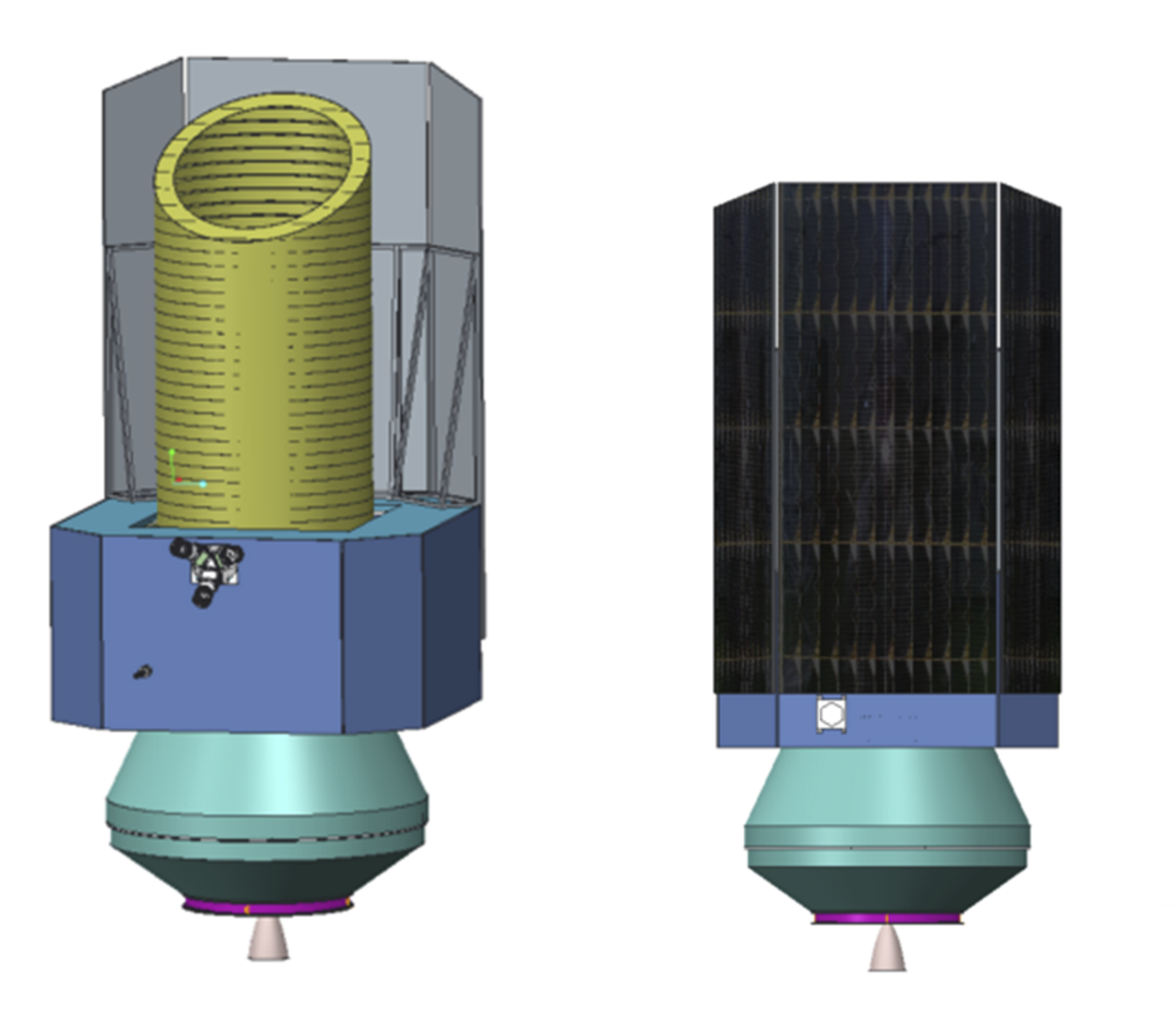}
		\end{subfigure}
		\begin{subfigure}{.5\textwidth}
			\centering
			\includegraphics[width=\textwidth]{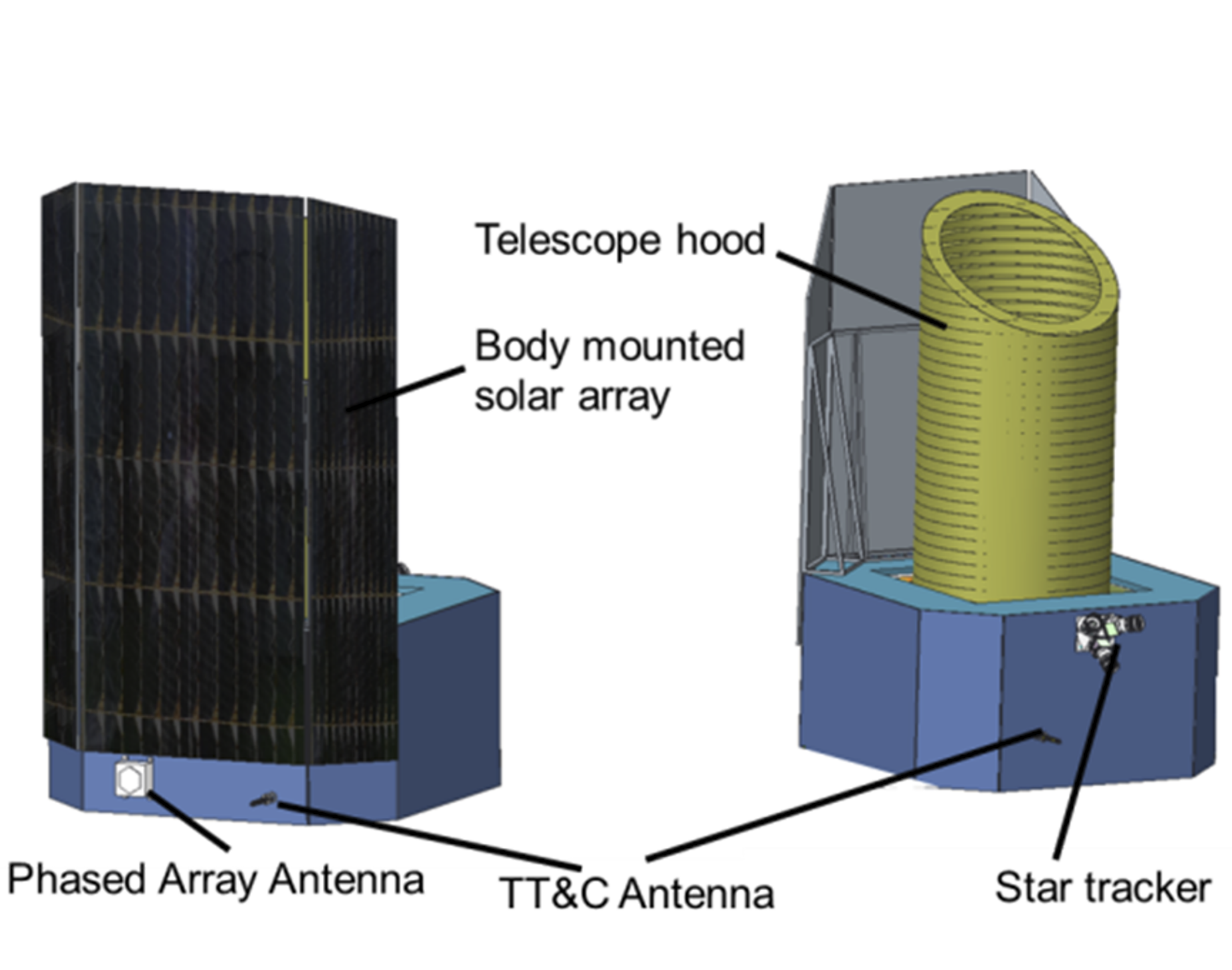}
				\end{subfigure}
				\caption{(a) Satellite retraction status  (b) Extra-satellite equipment layout.}
		\label{fig:5-8}
	\end{figure}
	
5.3.2.2 AOCS

\begin{figure}
   \centering
   \includegraphics[width=12cm]{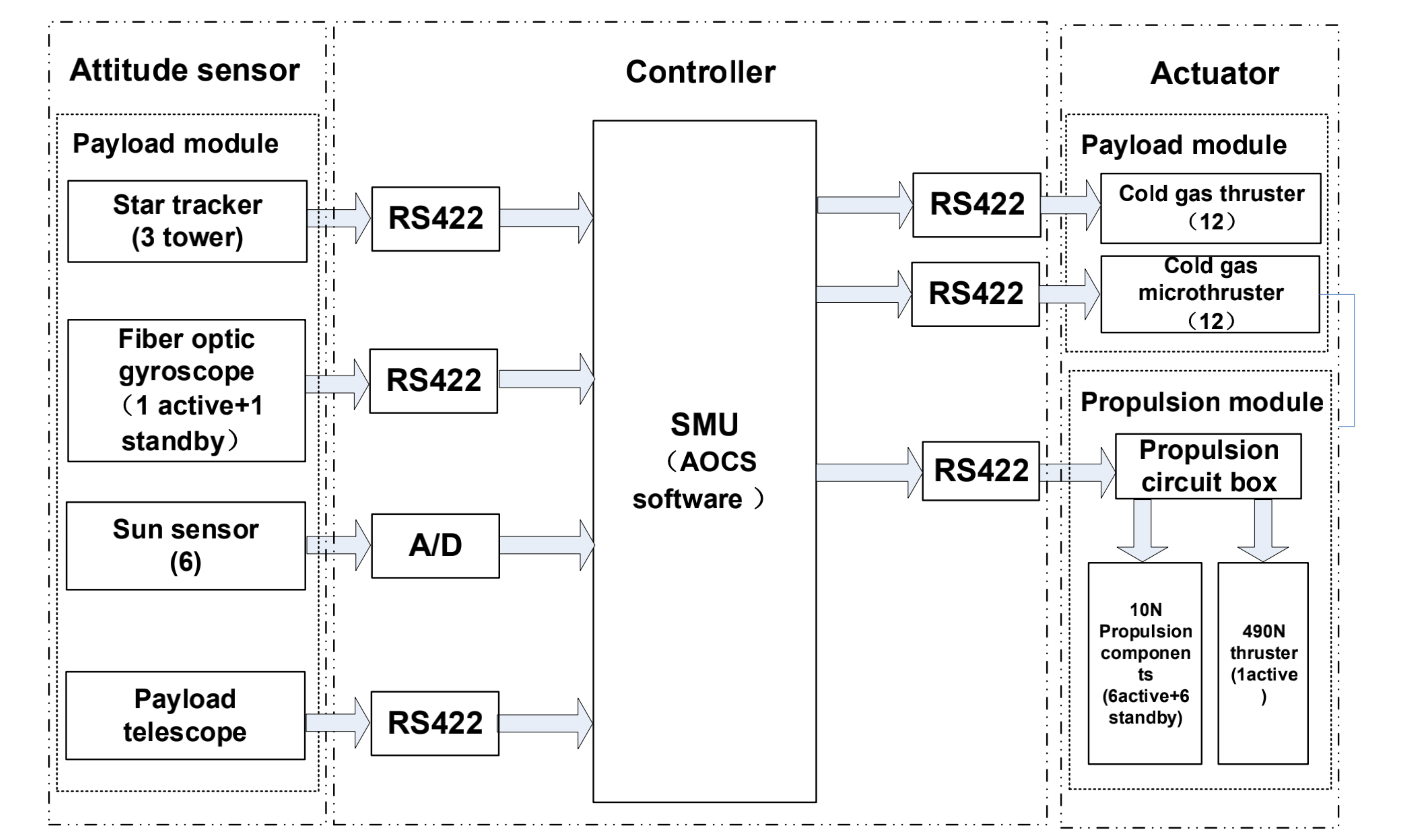}
   \caption{AOCS composition.}
   \label{fig:5-9}
\end{figure}

The satellite uses a three-axis stable attitude control method. The composition of  the attitude orbit control system is as shown in Figure \ref{fig:5-9}.

\begin{figure}
   \centering
   \includegraphics[width=12cm]{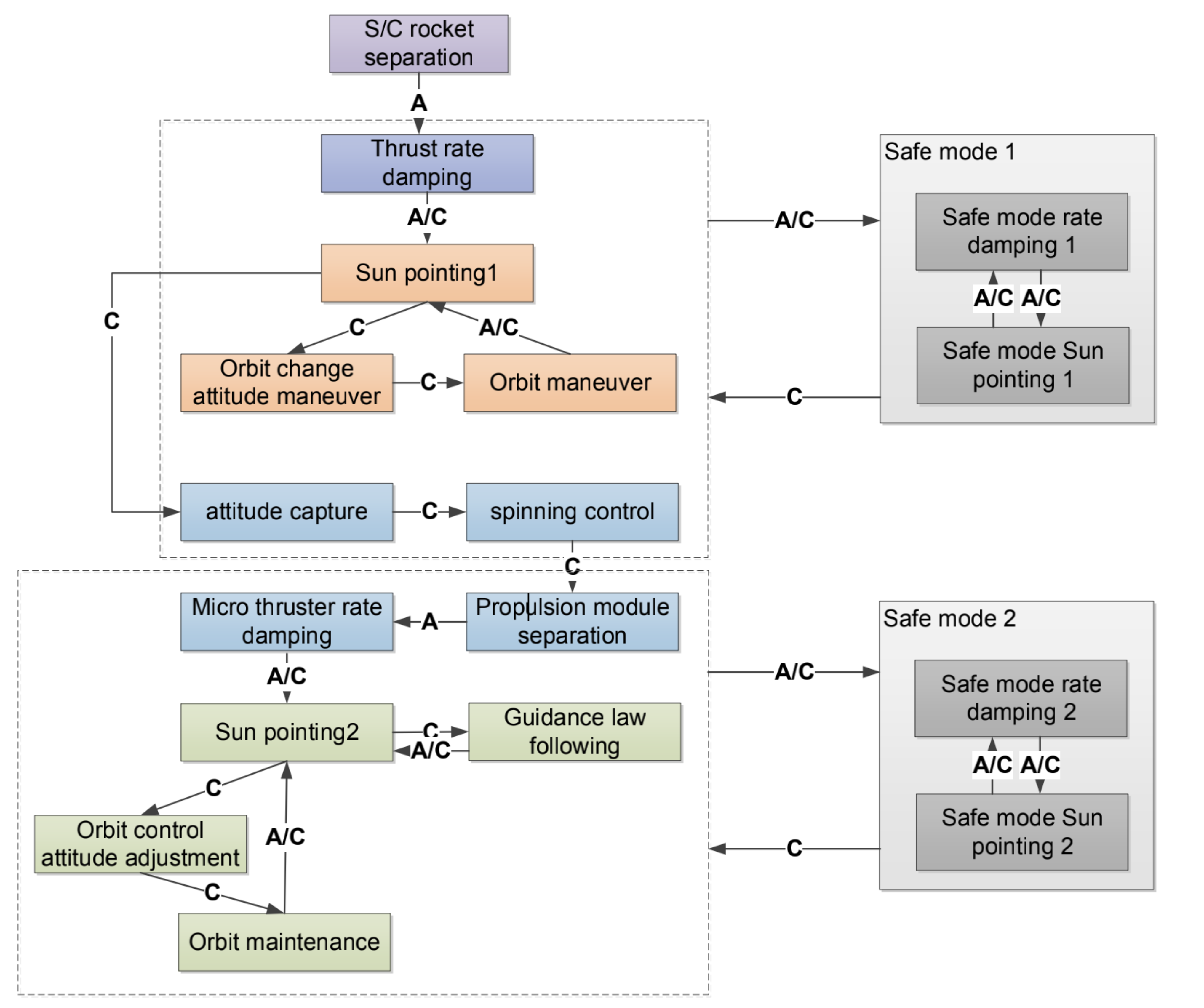}
   \caption{AOCS working modes.}
   \label{fig:5-10}
\end{figure}

The working modes of the attitude and orbit control subsystem are divided into the entry phase (after separation of satellite and rocket), the orbit transfer phase, the mission orbit entry phase, the mission phase, and the safety mode. The transition flow between modes is illustrated in Figure \ref{fig:5-10}.

\begin{figure}
   \centering
   \includegraphics[width=12cm]{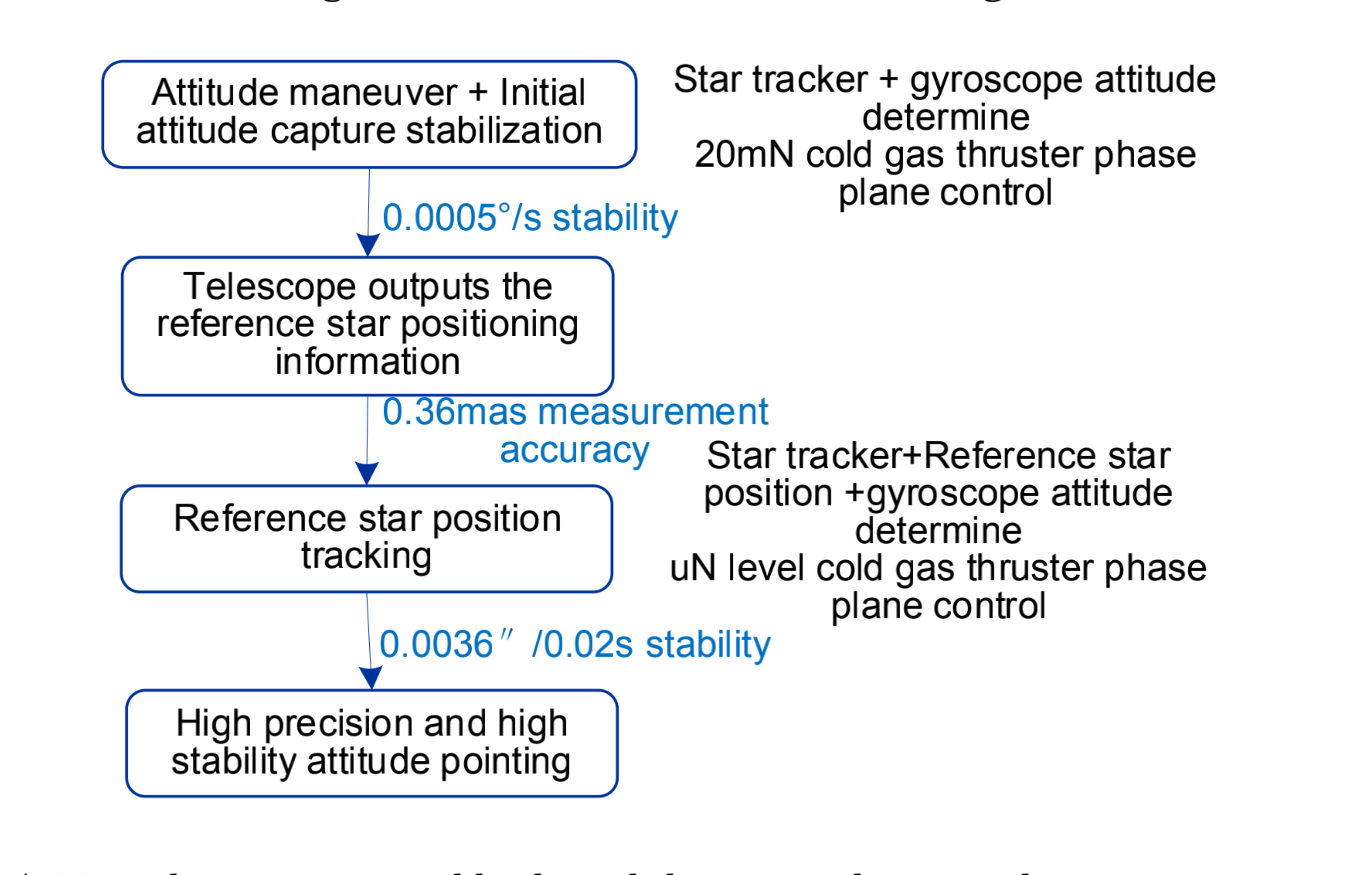}
   \caption{High-precision and high-stability attitude control process.}
   \label{fig:5-11}
   \end{figure}

The payload requires satellite pointing accurate of $(0.07$ arcsecond) and attitude stability attitude control of $(0.0036$ arcsecond $/ 0.02 \mathrm{~sec}$ ). To achieve this high precision and high stability pointing, the satellite attitude pointing information is first measured using a high precision star sensor and gyroscope, and the satellite attitude is controlled using a $20 \mathrm{mN}$ cold gas thruster to achieve the initial target pointing capture stable state of the telescope. Then, the reference star positioning information output from the telescope is used as the tracking target, and the $\mu \mathrm{N}$ level thruster is used as the actuating component for high-precision attitude pointing stability control to achieve the attitude accuracy and stability requirements for the final requirements of multiple mission observations of the same target. The control flow is shown in Figure \ref{fig:5-11}.

5.3.2.3 Thermal Control

\begin{figure}
   \centering
   \includegraphics[width=10cm]{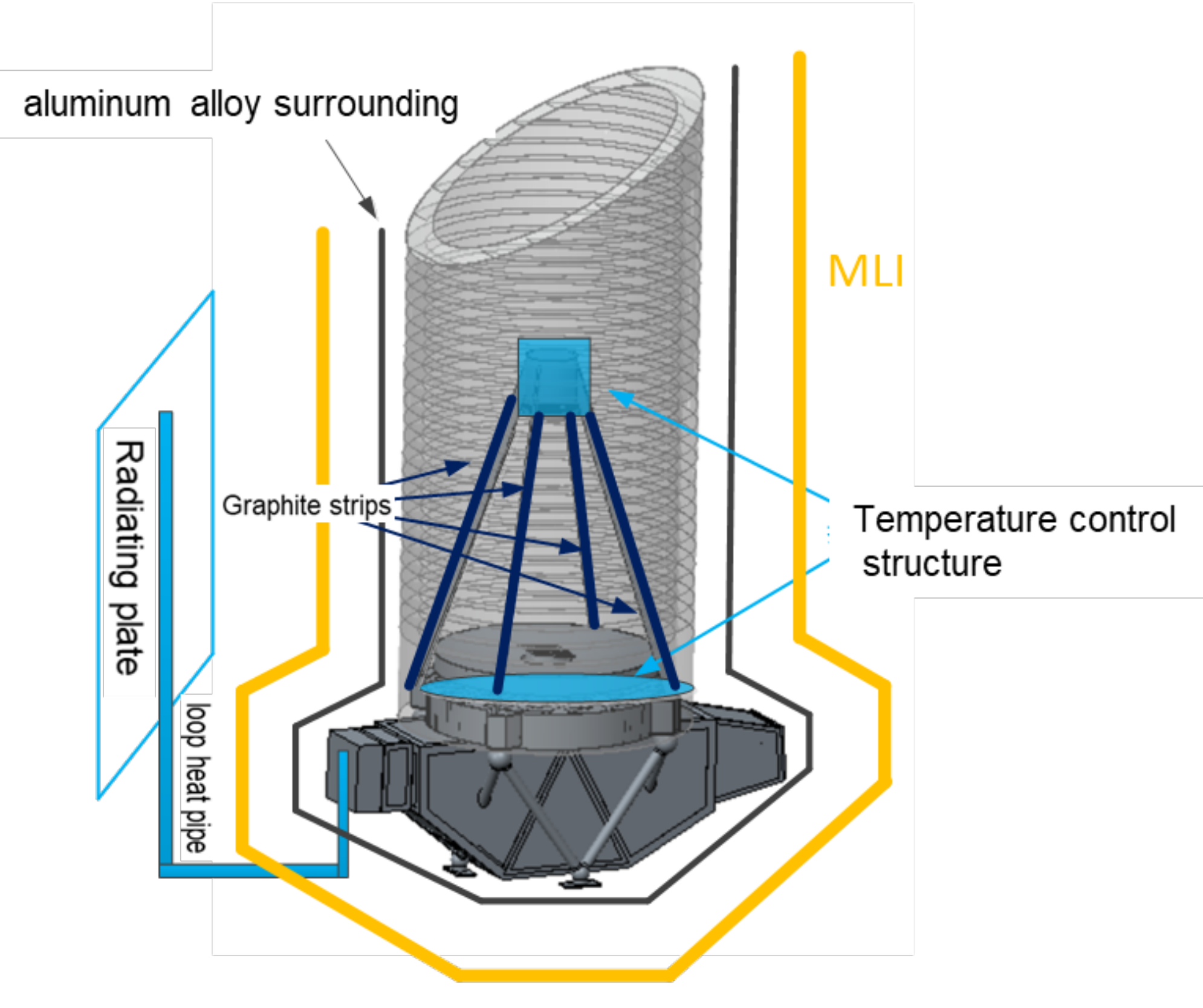}
   \caption{Telescope's thermal control design diagram.}
   \label{fig:5-12}
\end{figure}

The satellite adopts a thermal control scheme combining active thermal control and passive thermal control. The temperature of the satellite platform unit is controlled in the range of $-15 \sim 45^{\circ} \mathrm{C}$. The unit layout is designed to keep the heat-generating unit as far away from the telescope's heat-sensitive components as possible, and the nearby units should operate at a constant power consumption as much as possible during the load observation period. The telescope has an extremely high temperature control requirement, and is installed with the star insulated with a special thermal control design (Figure \ref{fig:5-12}), the scheme is as follows.

(1) Telescope external temperature control scheme. The telescope is designed with a $3 \mathrm{~mm}$ aluminum alloy surrounding structure on the outside of the telescope as a secondary temperature control object, and the temperature is controlled by a heater with a temperature control stability of $\pm 0.5^{\circ} \mathrm{C}$. As a first-level temperature control object, the telescope body is heat exchanged with the surrounding structure by radiation.

(2) The internal temperature control scheme of the telescope. A radiation plate is added between the two primary mirror components to compensate for the heat dissipation from the primary mirror to the deep cold space. A second-level temperature control structure is installed on the outside of the secondary mirror to compensate for the radiation temperature of the secondary mirror. Graphite strips are attached to the truss supporting the secondary mirror to reduce the temperature gradient of the truss.

(3) Focal plane detector box temperature control scheme. The focal box is equipped with a large heat sink to reduce the temperature variation; the focal box is insulated from the telescope body, and a loop heat pipe is used to connect the external radiation heat sink to dissipate the heat, and a loop heat pipe with a heat transfer power of more than $800 \mathrm{~W}$ is selected.

Thanks to the solar-terrestrial Lagrange 2 halo orbit and the anti-sun pointing, the external flux is quite stable after shielding from the solar heat flux. Heat leak from the aperture is certain and is compensated by the heaters. The telescope structure is heat insulated and used as the second stage control unit, including the telescope cavity and the lower part of the baffle, to remain the temperature $\pm 0.2^{\circ} \mathrm{C}$. The mirror is heated through radiation by the bracing structure, which is the first stage control unit. Also there are heaters on the mirror with power level of milliwatt to keep a uniform temperature. The essential way of temperature control is multi-stage technology and we have made some improvement from Taiji-1 according to the CHES telescope.

5.3.2.4 Energy

The satellite adopts fully regulated bus with voltage range of $30 \pm 0.5 \mathrm{~V}$. The solar array adopts triple junction GaAs solar cell with solar sail area of $11.8 \mathrm{~m}^{2}$ and final output power of $1493 \mathrm{~W} .120 \mathrm{Ah}$ battery pack is selected for the Li-ion battery pack.

5.3.2.5 Communications

CHES satellite deep space communication adopts X-band measurement and control/digital transmission integrated communication scheme. Since the CHES satellite has extremely high requirements for attitude stability, a phased array antenna is selected for data transmission in order to avoid the impact of satellite attitude adjustment or vibration generated by the turntable on the load observation. The measurement and control are carried out by a wide-beam omnidirectional antenna with a multi-tone system and a function of sending DOR beacon signals.

5.3.2.6 SMU

SMU consists of hardware and software, the hardware is the house-keeping computer and the software is the housekeeping software. The house-keeping computer uses a dual processing system, a centralized management scheme, and keeps dual cold backup. The software adopts the embedded real-time operating system VxWorks, with different application software.

\subsubsection{Budgets}
\quad(1) Mass budgets

The CHES satellite mass allocation is shown in Table \ref{Tab5.3}.
CHES satellite's dry mass is $1558 \mathrm{~kg}$, including propellant $990 \mathrm{~kg}$, total mass is $2930 \mathrm{~kg}$ (including $382 \mathrm{~kg}$ margin).

\begin{table}
\begin{center}
\caption[]{CHES mass allocation.}\label{Tab5.3}
\begin{tabular}{|c|c|c|}
\hline No. & System & Mass $/ \mathrm{kg}$ \\
\hline 1 & Payload & 765 \\
\hline 2 & Structure and mechanism & 348 \\
\hline 3 & Thermal control & 30 \\
\hline 4 & Electrical and power & 61 \\
\hline 5 & AOCS & 267 \\
\hline 6 & SMU & 25 \\
\hline 7 & TT\&C & 32 \\
\hline 8 & Data transmission & 30 \\
\hline 9 & S/C dry mass & 1558 \\
\hline 10 & Propulsion module propellant & 840 \\
\hline 11 & Service module propellant & 150 \\
\hline 12 & Margin & 382 \\
\hline & Total & 2930 \\
\hline
\end{tabular}
\end{center}
\end{table}
(2) Power budgets

The CHES satellite power allocation lists in Table \ref{Tab5.4}.

\begin{table}
\begin{center}
\caption[]{CHES power allocation.}\label{Tab5.4}
\begin{tabular}{|r|c|c|c|c|}
\hline \multirow{2}{*}{No.} & \multirow{2}{*}{System} & \multirow{2}{*}{Orbit entry} & \multirow{2}{*}{Mission phase} & \multirow{2}{*}{Peak power/W} \\
& & phase power & constant power /W & \\
\hline 1 & Payload & 0 & 650 & 700 \\
\hline \multirow{2}{*}{2} & \multirow{2}{*}{Thermal control} & \multirow{2}{*}{100} & \multirow{2}{*}{0} & \multirow{2}{*}{200} \\
& (Propulsion module ) & & &\\
\hline \multirow{2}{*}{3} & \multirow{2}{*}{Thermal control} & \multirow{2}{*}{30} & \multirow{2}{*}{50} & \multirow{2}{*}{80} \\
& (except propulsion module ) & & &\\
\hline 4 & Energy & 25 & 25 & 25 \\
\hline 5 & AOCS & 60 & 120 & 820 \\
\hline 6 & SMU & 45 & 68 & 68 \\
\hline 7 & TT\&C & 35 & 25 & 95 \\
\hline 8 & Data transmission & 0 & 195 & 195 \\
\hline 9 & Margin & 30 & 110 & 210 \\
\hline & Total & 325 & 1243 & 2393 $(1493)$ \\
\hline
\end{tabular}
\end{center}
\end{table}
Since the payload, thermal control, TT\&C, data transmission, and AOCS are not be in the peak working state at the same time, the satellite peak power is $1493 \mathrm{~W}$ instead of 2393W.

\subsubsection{Key design challenges evaluation}
5.3.4.1 High stability attitude control

Satellite high stability attitude control is the key technology of the satellite. Technology inheritance: (1) precision guidance technology, based on the design of SVOM satellite FGS, SVOM satellite has completed development of Gaia-based FGS electronic star map simulator, and implemented semi-physical simulation verification test.

SVOM is planned to be launched in December 2023. (2) Highly refined closed-loop control technology, based on the development of $\mu \mathrm{N}$ level thrusters and the study of high stability control algorithms already conducted on the Taiji-1 satellite launched in September $2019 .$

Considering that the current technology maturity is five, further measures to improve the technology maturity: Combining the operating mode of CHES payload and the characteristics of the output stellar position information, develop the FGS simulator to form a full-link attitude control semi-physical closed-loop simulation system to verify the attitude control index.

5.3.4.2 High-precision thermal Control

The telescope high-precision thermal control technology is the key technology of the whole satellite. In terms of technology inheritance: (1) multi-stage thermal control technology, based on the Taiji-1 satellite launched in September 2019, which adopts three-stage temperature control technology, with an in-orbit measured core area temperature control accuracy of 6 $\mathrm{mK}$. (2) ultra-high precision thermal control technology: based on the gravitational wave detection project. a. High-resolution low-noise temperature measurement: completed the prototype development and test verification of the thermometer, with a temperature resolution of 0.01 $\mathrm{mK}$; b. High precision high thermal stability temperature control: the principle prototype design was completed; c. High thermal stability active heating temperature control algorithm research: the actual physical model and temperature control algorithm coupling under the $m K$ level active temperature control was realized. The ultra-high-precision thermal control technology has completed the key technology research and review, part of which has developed a stand-alone machine, not yet in orbit flight verification, the relevant technology is to be used in the follow-up detection mission of the Taiji project.

Comprehensive consideration of the current stage of technical maturity for the five levels, to further improve the technical maturity measures: on the basis of simulation analysis, put into production load thermal control for vacuum thermal tests to verify the thermal control scheme.

In summary, the CHES mission will discover Earth-like planets in the habitable zone around the nearby solar-type stars via micro-arcsecond accuracy relative astrometry. CHES will offer the first direct measurements of true masses and three-dimensional orbits of \emph{Earth Twins} and super-Earths orbiting our neighboring stars based on ultra-high-precision astrometry from space. In addition, CHES will further conduct a comprehensive survey and extensively characterize the nearby planetary systems. As to scientific additional benefits, CHES will partly offer informative clues to cosmology, dark matter and compact objects. This will definitely enhance our understanding of the formation of diverse nearby planetary systems and the emergence of other worlds for nearby solar-type stars, and finally to give directions to the evolution of our solar system.

\begin{acknowledgements}
We thank the referee for constructive comments and suggestions to improve the manuscript.
This work is financially supported by the Strategic Priority Research Program on Space Science of the Chinese Academy of Sciences (Grant No. XDA 15020800), the National Natural Science Foundation of China (Grant Nos. 12033010, 41604152, U1938111), Foundation of Minor Planets of the Purple Mountain Observatory, and Youth Innovation Promotion Association CAS (Grant Nos. 2018178).

\end{acknowledgements}

\bibliography{refs}{}
\bibliographystyle{raa}
\label{lastpage}

\end{document}